\documentclass[aps,prx,reprint,groupedaddress,superscriptaddress,floatfix]{revtex4-2}

\usepackage[T1]{fontenc}
\usepackage{newtxtext}
\usepackage{newtxmath}

\usepackage{titlesec}

\usepackage{amsmath}
\usepackage{mathtools}
\usepackage{amsfonts}
\usepackage{bbold}
\usepackage{physics}
\usepackage{color}
\usepackage{xcolor}
\usepackage[colorlinks,citecolor=blue,linkcolor=blue,urlcolor=blue]{hyperref}

\usepackage{graphicx}
\usepackage[all]{hypcap}

\usepackage{enumitem}

\usepackage{tabularx}
\usepackage{makecell}

\begin{document}

\title{Signatures of Integrability and Exactly Solvable Dynamics in an Infinite-Range Many-Body Floquet Spin System}

\author{Harshit Sharma}
\email{ds21phy007@students.vnit.ac.in}
\affiliation{Department of Physics, National Institute of Technology, Nagpur 440010, India}

\author{Udaysinh T. Bhosale}
\email{udaysinhbhosale@phy.vnit.ac.in}
\affiliation{Department of Physics, National Institute of Technology, Nagpur 440010, India}

\date{\today}
\begin{abstract}
In a recent work Sharma and Bhosale [\href{https://journals.aps.org/prb/abstract/10.1103/PhysRevB.109.014412}{Phys. Rev. B \textbf{109}, 014412 (2024)}],
$N$-spin  Floquet model having infinite range Ising interaction was introduced. In this paper, we generalized the strength of
interaction to $J$, such that $J=1$ case reduces to the aforementioned work. We show that for $J=1/2$ the model still exhibits
integrability for an even number of qubits only. We analytically solve the cases of $6$, $8$, $10$, and $12$ qubits, finding its
eigensystem,  dynamics of  the entanglement for various initial states, and the unitary evolution operator. These quantities exhibit
the signature of quantum  integrability (QI). For the general case of even-$N > 12$ qubits, we conjuncture the presence of QI
using the numerical evidences such as spectrum degeneracy, and the exact periodic nature of both the entanglement dynamics and
the time-evolved unitary operator. We numerically show the  absence of QI for odd $N$  by observing a violation of the signatures
of QI. We  analytically and numerically find that the maximum value of time-evolved concurrence ($C_{\mbox{max}}$) decreases with
$N$, indicating the multipartite nature of entanglement. Possible experiments to verify our results are discussed.
\end{abstract}

\maketitle

\section{ Introduction}
Long-range interactions are encountered in various scientific domains, including statistical physics \cite{dauxois2002dynamics,sutherland2004beautiful,campa2014physics}, quantum mechanics \cite{wormer1977quantum,salam2009molecular}, cosmology \cite{dolgov1999long,nusser2005structure,van2012dark,esteban2021long}, atomic and nuclear physics \cite{chomaz2002phase}, plasma physics \cite{elskens2019microscopic}, hydrodynamics \cite{miller1990statistical,schuckert2020nonlocal,morningstar2023hydrodynamics}  and condensed matter physics \cite{defenu2023long}. This interaction  can now be replicated  in  quantum simulator \cite{buluta2009quantum,georgescu2014quantum,gross2017quantum} with artificial ion crystals \cite{deng2005effective,islam2011onset,bermudez2017long}, cold atoms
in cavities \cite{baumann2010dicke}, polar molecules \cite{yan2013observation}, dipolar quantum gases \cite{lahaye2009physics,chomaz2022dipolar,smith2023supersolidity}, Rydberg atoms \cite{saffman2010quantum}, magnetic atoms \cite{griesmaier2005bose,beaufils2008all,baier2016extended}, nonlinear optical media \cite{firstenberg2013attractive}, and solid-
state defects \cite{childress2006coherent}. They are also present between genomic elements which can be detected by using the 3C method (chromosome conformation capture) \cite{sanyal2012long} and in adatoms of graphene \cite{shytov2009long}. Long-range interactions are pervasive and give rise to qualitatively new physics, e.g. emergence of novel quantum phase and dynamical behaviors \cite{kastner2011diverging,gong2016kaleidoscope}. Additionally, they play a crucial role in enabling speed up in quantum information processing  \cite{avellino2006quantum,eldredge2016fast}. Long-range interactions are proven to be very useful in quantum technology applications like quantum heat engine \cite{solfanelli2023quantum}, quantum computing \cite{lewis2021optimal,lewis2023ion}, quantum metrology \cite{pezze2018quantum} and ion trap \cite{gambetta2020long}.

Entanglement stands out as perhaps the most distinctive feature of quantum mechanics, giving rise to unique nonlocal correlations that find no counterpart in the classical domain. Long-range interactions have the potential to profoundly influence the dynamics of correlated systems \cite {dauxois2002dynamics}. Due to the breakdown of quasi-locality in long-range interaction various phenomena arise such as it is capable of faster entanglement generation  \cite{hauke2013spread,eldredge2017fast,colmenarez2020lieb} than the Lieb-Robinson bound   \cite{bravyi2006lieb}, and dynamical phase transitions \cite{zhang2017observation,jurcevic2017direct}. Numerous studies have delved into the examination of entanglement in such system \cite{dur2005entanglement,kuzmak2018entanglement,deng2020universal,schneider2021spreading,
richter2023transport,amghar2023geometrical,kloss2020spin,song2023quantum,duha2024twomode}.
The time evolution of multipartite entanglement measured using quantum Fisher information (QFI) and scrambling in a spin chain has been studied having these interactions \cite{pappalardi2018scrambling}. They have been   measured in experiments too   \cite{garttner2017measuring}.

The long-range interaction decays with distance $r$ according to a power law ($1/r^{\alpha}$). Corresponding to different values of  $\alpha$, these interactions in the natural system fall into categories such as Rydberg atoms ($\alpha=6$, van der Waals interactions), magnetic atoms ($\alpha=3$, dipole-
dipole interaction), Coulomb interactions ($\alpha=1$), atoms coupled to cavities ($\alpha=0$), etc. The case  $\alpha=0$ corresponds to a category of infinite-range or all-to-all interactions \cite{britton2012engineered,schauss2012observation,peter2012anomalous,yan2013observation,
jurcevic2014quasiparticle,hazzard2014many,richerme2014non,douglas2015quantum,gil2016nonequilibrium,Marino2019,sauerwein2023engineering,liu2024signature,vzunkovivc2024mean}.
Models  with $\alpha < d$ are categorized as long-range interaction, where $d$ is the physical dimension of the system \cite{dauxois2002dynamics,defenu2023long}. In such interactions, the energy is not extensive \cite{campa2014physics,dauxois2002dynamics,defenu2023long}. There are  models corresponding to infinite range interaction that are integrable, such as the  LMG model \cite{kumari2022eigenstate} and the model with Ising interaction in a transverse field \cite{sharma2024exactly}. The main theme of this paper revolves around integrability in one such recently introduced  model \cite{sharma2024exactly}.

From the perspective of classical mechanics, integrability can be understood by the connection between the degree of freedom and a sufficient number of constants of motion \cite{babelon2003introduction,retore2022introduction}, whereas  the quantum integrability \cite{owusu2008link,doikou2010introduction,SantosLea12,yuzbashyan2013quantum,
gombor2021integrable,vernier2024strong} is typically associated with an exact solution of models, for example, based on the solution of the Yang-Baxter equation   by obtaining transfer matrix \cite{wadati1993quantum,doikou2010introduction,lambe2013introduction,baxter2016exactly,gaudin2014bethe,zheng2024exact},  and techniques like the Bethe ansatz \cite{bethe1931theorie,faddeev1995algebraic,pan1999analytical,bargheer2008boosting}. Alternatively, it can be identified through other features, for instance, the existence of a set of infinite number of conserved quantities and/or  Poissonian-level statistics  after taking conserved quantities into account \cite{berry1977level,bhosale2021superposition}.  In various studies, the indication of integrability in systems is revealed through signatures like exact periodicity of entanglement dynamics and time evolution of Floquet operator and degenerated spectra or level crossing \cite{LevelCrossingsYuzbashyan2008,yuzbashyan2013quantum,BogomolnyDistribution2013,mishra2015protocol,doikou2010introduction,pal2018entangling,naik2019controlled}.

 In a very recent work \cite{sharma2024exactly}, a many-body  Floquet spin model having an infinite range Ising interaction was introduced. They have shown that it exhibits quantum integrability. In this Ref.~\cite{sharma2024exactly}, they have analytically calculated the eigensystem, reduced density matrix, time evolution of the unitary operator, and entanglement dynamics for $5$ to $11$ qubits. They have measured the entanglement dynamics using linear entropy and concurrence.  For the general case  $N > 11$ qubits, they resort to numerical methods due to the complexity and fairly large calculation, as analytical solutions pose mathematical challenges. The Hamiltonian was defined as follows:  $$H(t)=\sum_{l< l'} \sigma^z_{l} \sigma^z_{l'}+\sum_{n = -\infty}^{ \infty} \delta(n-t/\tau)\sum_{l=1}^{N}\sigma^y_l,$$
 where the first term represents the Ising interaction with unit strength,
while in the second term, $\tau$ is the period at which the magnetic field is applied periodically along the $y$ axis. In various studies, the integrability in system was identified through the signatures, such as the periodicity of entanglement dynamics \cite{ArulLakshminarayan2005,naik2019controlled,mishra2015protocol,
pal2018entangling} and that of the Floquet operator dynamics \cite{pal2018entangling}, and a highly degenerated spectra \cite{yuzbashyan2013quantum,naik2019controlled}. In Ref. \cite{sharma2024exactly} they have used the same signatures to show integrability in their model for any value of $N$.
In their study, the authors reported that the integrability in the system is found  for a specific value of the parameter  $\tau=\pi/4$, and for $\tau\neq\pi/4$ it wasn't. Their findings revealed that the pairwise entanglement using concurrence remains zero for various initial states, indicating the multipartite nature of the entanglement. Furthermore, the signatures of integrability show the same behavior (depending on the parity of $N$) for any value of $N$. The authors have shown that the special case of their model is connected to the one with nearest-neighbor Ising interaction model \cite{ArulLakshminarayan2005,mishra2015protocol,apollaro2016entanglement,bertini2019entanglement,naik2019controlled,shukla2022characteristic} as well as to the well-known quantum chaotic kicked top (QKT) model for the specific value of parameters \cite{Hakke87}. Once it was mapped with  QKT, the integrability in the system was limited only up to four qubits \cite{ExactDogra2019}, whereas, in their work, they generalized it to any $N$ using these signatures \cite{sharma2024exactly}.

In this work, we have generalized this model to include any value of the strength of  Ising interaction to $J$. In the context
of integrability, we  utilize the same signatures, and for
the same value of the $\tau$, as reported in the previous work \cite{sharma2024exactly}.
We analytically obtained the eigenvalues and eigenvectors for an even number of qubits ranging from $6$ to $12$. We explicitly derive the expression for entanglement measures, such as linear entropy and entanglement entropy for various initial unentangled states. Here, we also find that these quantities exhibit a periodic nature for specific values of $J$, other than one \cite{sharma2024exactly}. We also observe the periodic behavior of the time evolution of the unitary operator for the aforementioned cases. However, for cases involving $5$, $7$, $9$, and $11$ qubits,  we have numerically demonstrated the absence of these signatures. For the general case, $N > 12$,  we provide sufficient numerical evidences of Integrability for any even number of qubits and the absence of quantum integrability for any odd qubits by using the absence of the same signatures \cite{sharma2024exactly}. We also observe the decay of the maximum value of concurrence with $N$, indicating the increasing multipartite nature of entanglement.

The rest of this paper is organized as follows. In Sec. \ref{sec:example-section2}, we give a brief introduction to the model investigated. In Sec. \ref{sec:example-section3}, we have given an  exact analytical solution for entanglement measures such as linear entropy, entanglement entropy, and concurrence for six qubits for various initial unentangled states. In  Sec. \ref{sec:example-section4}, the results on  the analytical expression of entanglement and eigenvalues of time evolved unitary operator for $ N= 8,10$ and $12$ qubits are presented. In Sec. \ref{sec:example-section5}, we provide sufficient numerical evidence of the signature of quantum integrability for the general case of even-$N > 12$ qubits and the absence of the QI for odd-$N > 12$ qubits. In Sec. \ref{sec:example-section6},  a summary of the results and  conclusions is given.

\section{Model} \label{sec:example-section2}
The generalization of the  Hamiltonian model from the  Ref. \cite{sharma2024exactly} is given as  follows:
\begin{equation}
\label{Eq:QKT}
H(t)= H_I(h)+\sum_{n = -\infty}^{ \infty} \delta(n-t/\tau)~ H_k,
\end{equation}
where $\delta(t)$ is Dirac delta function and we define,
\begin{eqnarray}
H_I={J} \sum_{ l< l'}\sigma^z_{l} \sigma^z_{l'}~~ \mbox{and} ~~
H_k= \sum_{l=1}^{N}\sigma^y_l.
\end{eqnarray}
Here, the first term represents the ising interaction with a field strength  ${J}$,
while the second term corresponds to the periodically applied magnetic field along the $y$-axis, with a period of $\tau$. The case $J=1$ corresponds to the Hamiltonian from the  Ref. \cite{sharma2024exactly}.
The corresponding Floquet operator is given  as follows:
\begin{eqnarray}\nonumber \label{Eq:QKT1}
\mathcal{U} &=& \exp\left [-i~ \tau H_I(h) \right]\exp\left[-i~ \tau H_k\right]\\
  &=& \exp\left (-i~ J \tau \sum_{ l< l'} \sigma^z_{l} \sigma^z_{l'}   \right)  \exp\left(-i~ \tau \sum_{l=1}^{N}\sigma^y_l \right).
\end{eqnarray}
 The Ising interaction in this model is uniform and all-to-all.  The nearest-neighbor  (NN) interaction model, a special case of our model, has been extensively studied \cite{ArulLakshminarayan2005,mishra2015protocol,apollaro2016entanglement,pal2018entangling,bertini2019entanglement,naik2019controlled,shukla2022characteristic}.   It exhibits permutation symmetry under the exchange of spins.  Due to the presence of symmetries in the model, its effective Hilbert dimension reduces from $2^N$ to $N+1$. Scrambling in models similar to ours has been extensively explored in the Refs.  \cite{belyansky2020minimal,li2020fast,yin2020bound,li2021long,li2022fast,wanisch2023information}. In Ref. \cite{sharma2024exactly}, they have shown  that the model exhibits signatures of quantum integrability  for the parameter  $\tau=\pi/4$ and ${J}=1$. The main objective of this work is to find the values of parameters in which the model shows quantum integrability.  From that perspective, we restricted ourselves to the same $\tau$, while varying the field strength ${J}$.  We find that the model shows identical signatures for the parameter ${J}=1/2$. The Floquet operator corresponding to the Eq.~(\ref{Eq:QKT}) with $\tau=\pi/4$ and ${J}=1/2$ is given as follows:
\begin{equation}\label{Eq:Un1}
{\mathcal U}=\exp\left(-i \frac{\pi}{8}  \sum_{ l< l'} \sigma^z_{l} \sigma^z_{l'}\right)
\exp\left( -i \frac{\pi}{4} \sum_{l=1}^{N}\sigma^y_l \right).
\end{equation}

Throughout this paper, we consistently employ the parameter values $J=1/2$ and $\tau=\pi/4$ unless otherwise stated. In their study \cite{sharma2024exactly}, the authors reported that this model has a close  connection to the very well-known QKT model \cite{Hakke87,Haakebook,UdaysinhPeriodicity2018} and shows quantum integrability up to $4$ qubits. In the Refs. \cite{UdaysinhPeriodicity2018, ExactDogra2019},  this model is extensively studied for a smaller number of qubits ($N = 2$, $3$ and $4$) for these parameters. Its Hamiltonian is given by,
\begin{equation}
\label{Eq:QKT2}
H_{QKT}(t)= \frac{p}{\tau'} \, {\tilde{J}_y} + \frac{k}{2j}{\tilde{J}_z}^2 \sum_{n = -\infty}^{ \infty} \delta(t-n\tau').
\end{equation}
The  Floquet operator in the Eq.~(\ref{Eq:Un1}) also has connection with this model, for the  parameters $p=\pi/2$, $k=j\pi/2~ (=N\pi/4)$, $\tau'=1$, and using many-qubit transformation
$\tilde{J}_{x,y,z}=\sum_{l=1}^{2j}\sigma_l^{x,y,z}/2$, where $\sigma_l^{x,y,z}$ are the standard Pauli matrices. After this transformation Eqs.~(\ref{Eq:Un1}) and (\ref{Eq:QKT2})  are related by $\exp\left(-i~ H_{QKT}\right)=\exp\left(-i~ N {J} \tau\right)\times \mathcal{U}$. The overall phase (global phase) $\exp\left(-i~ N {J} \tau\right)=\exp\left(-i~N\pi/8\right)$  for $J=1/2$ and $\tau=\pi/4$, does not alter the entanglement dynamics. However, the effect of this phase can be observed on the time period of operator dynamics, when it is periodic.

In this work, we will be studying the time evolution from the initial states, that are, localized in spherical phase space. They lie on the unit sphere with spherical coordinates,
$(\theta_0,\phi_0)$. These states are the  standard SU(2) coherent states and given as follows  \cite{Glauber1976,PuriBook}:
\begin{equation}
 |\theta_0,\phi_0\rangle = \otimes^{N} \left(\cos(\theta_0/2) |0\rangle + e^{-i \phi_0} \sin(\theta_0/2) |1\rangle\right).
\end{equation}
The entanglement dynamics as a function of time  can be studied, by evolving these initial states using  the Floquet operator $\mathcal{U}$. Our study utilizes various measures such as linear entropy \cite{buscemi2007linear}, von Neumann entropy \cite{nielsenbook,benenti2004principles} and concurrence \cite{wootters1998entanglement,wootters2001entanglement} to quantify the entanglement in the system. The commutation relation $[\mathcal{U},\otimes_{l=1}^{N}\sigma_l^y ]=0$, implies that the system possesses a symmetry, which is an up-down or parity symmetry (for proof see  the Supplemental Material of the  Ref.  \cite{sharma2024exactly}). In this work, we utilize the same general basis from Ref. \cite{sharma2024exactly} to solve the system  analytically for any $N$ number of qubits. The  basis when $N$ is odd is given as follows:
\begin{equation}\label{Eq:oddBasis1}
\ket{\phi_q^{\pm}}=\frac{1}{\sqrt{2}}\left(\ket{w_q}\pm {i^{\left(N-2q\right)}} \ket{\overline{w_q}}\right),
0\leq q\leq \dfrac{2j-1}{2};
\end{equation}
whereas for even $N$ it is :
\begin{eqnarray}\label{Eq:evenBasis1}
\begin{split}
\ket{\phi_r^{\pm}}&=\frac{1}{\sqrt{2}}\left(\ket{w_r}\pm {(-1)^{\left(j-r\right)}} \ket{\overline{w_r}}\right), 0\leq r\leq j-1\\
\mbox{and} &\;\;
\ket{\phi_j^+}=\left({1}/{\sqrt{\binom{N}{j}}}\right)\sum_{\mathcal{P}}\left(\otimes^{j}\ket{0}\otimes^{j}\ket{1}\right)_\mathcal{P},
\end{split}
\end{eqnarray}
  where
$\ket{w_q}=\left({1}/{\sqrt{\binom{N}{q}}}\right)\sum_\mathcal{P}\left(\otimes^q \ket{1} \otimes^{(N-q)}\ket{0}\right)_\mathcal{P}$ and
$\ket{\overline{w_q}}=\left({1}/{\sqrt{\binom{N}{q}}}\right)\sum_\mathcal{P}\left(\otimes^{q}\ket{0}\otimes^{(N-q)}\ket{1}\right)_\mathcal{P}$,
both being definite particle states \cite{Vikram11}. The $\sum_\mathcal{P}$ denotes the sum over all possible permutations.
These basis states $\ket{\phi_j^{\pm}}$ are  the eigenstate of parity operator having eigenvalues $\pm1$ i.e. $\otimes_{l=1}^{N}\sigma_l^y\ket{\phi_j^{\pm}}=\pm\ket{\phi_j^{\pm}}$.
In this permutation symmetric basis, $\mathcal U$ is  block-diagonalized in two blocks $\mathcal U^{+}$ and $\mathcal U^{-}$. Due to block diagonalization, it becomes easy to compute its $n^{th}$ power, which can simplify further analysis. We analytically calculate the entanglement dynamics for
the coherent states $|\theta_0=0,~\phi_0=0\rangle$ and $|\theta_0=\pi/2,~\phi_0=-\pi/2\rangle$. These states have special importance when the classical phase space of the QKT is considered \cite{ExactDogra2019}.

\section{Exact solution for six qubits} \label{sec:example-section3}
\begin{figure}[t!]\vspace{0.2cm}
\includegraphics[width=0.47\textwidth]{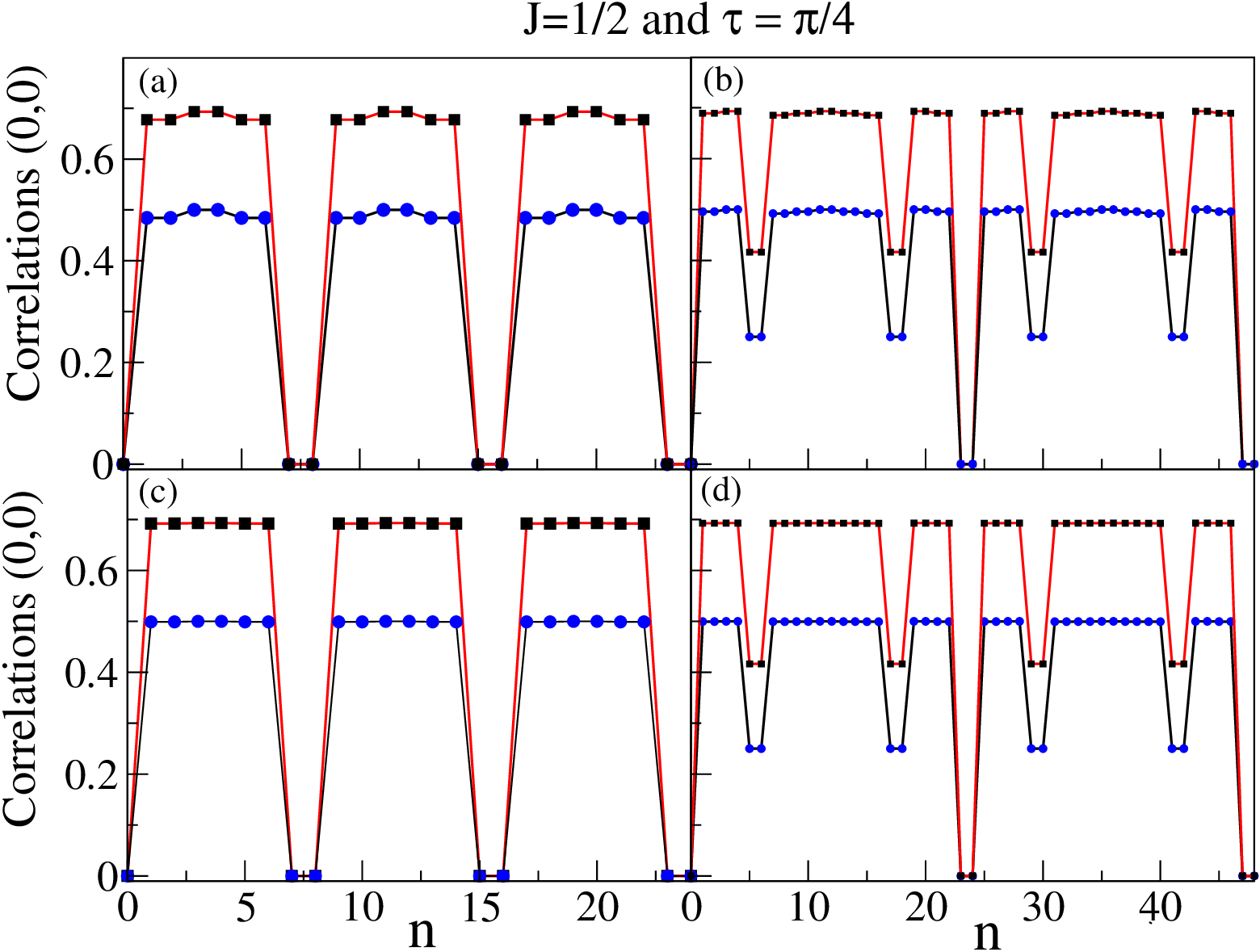}
\caption{Correlations (analytical values) using linear entropy (circles), entanglement entropy (squares)  are plotted for  (a) $6$ qubit
(b) $8$ qubit (c)$10$ qubit and  (d) $12$ qubit  for the  initial state $\otimes ^N {\ket{0}}_y$.}
\label{fig:correlations1}
\end{figure}

By using  the Eq.~(\ref{Eq:evenBasis1}) for $N=6$, the permutation symmetric basis in which $\mathcal{U}$ is block diagonal are given as follows:
\begin{eqnarray}
\ket{\phi_0^{\pm}}&=&\frac{1}{\sqrt{2}}(\ket{w_0} \mp \ket{\overline{w_0}}),\\
\ket{\phi_1^{\pm}}&=&\frac{1}{\sqrt{2}}(\ket{w_1} \pm  \ket{\overline{w_1}}),\\
\ket{\phi_2^{\pm}}&=&\frac{1}{\sqrt{2}}(\ket{w_2} \mp  \ket{\overline{w_2}}),\\
\ket{\phi_3^{+}}&=&\frac{1}{\sqrt{20}}\sum_\mathcal{P}\ket{000111} ,
\end{eqnarray}
where $\ket{w_0}=\ket{000000}$,
$\ket{\overline{w_0}}=\ket{111111}$, $\ket{w_1}=\frac{1}{\sqrt{6}} \sum_\mathcal{P} \ket{000001}_\mathcal{P}$, $\ket{\overline {w_1}}=\frac{1}{\sqrt{6}} \sum_\mathcal{P} \ket{011111}_\mathcal{P}$, $\ket{w_2}$=$\frac{1}{\sqrt{15}} \sum_\mathcal{P} \ket{000011}_\mathcal{P}$ and $\ket{\overline{w_2}}$=$\frac{1}{\sqrt{15}} \sum_\mathcal{P} \ket{001111}_\mathcal{P}$. The seven dimensional ($N+1$) space splits into $4 \oplus 3$ subspace on which operators are $\mathcal{U}_{\pm}$. The unitary operator is given by,
\begin{equation}
 \mathcal{U} = \begin{pmatrix}
            \mathcal{U}_{+} & 0_A \\ 0_B & \mathcal{U}_{-}
            \end{pmatrix},
\end{equation}
where~ $\mathcal{U}_{+}\left(\mathcal{U}_{-}\right)$ are $4\times 4\left(3\times 3 \right)$ dimensional matrices and $0_A$ ($0_B$) is null matrices of the  dimension $4\times 3\left(3\times 4 \right)$. The $\mathcal{U}_{+}$ ($\mathcal{U}_{-}$) are written in the positive (negative) parity subspaces
$\left\lbrace \phi_0^{+}, \phi_1^{+}, \phi_2^{+}, \phi_3^{+}\right\rbrace$ $\left(\left\lbrace \phi_0^{-},\phi_1^{-}, \phi_2^{-}\right\rbrace\right)$
respectively and are obtained as follows:
\begin{equation}
\mathcal{U_+}= \frac{-e^{\frac{i \pi }{8}}}{2\sqrt{2}} \left(
\begin{array}{cccc}
 0 &  \sqrt{3} & 0 & \sqrt{5} \\
 \sqrt{3}~e^{\frac{i \pi }{4}} & 0 & \sqrt{5}~e^{\frac{i \pi }{4}} & 0 \\
 0 &  \sqrt{5} & 0 & - \sqrt{3} \\
 -\sqrt{5}~e^{\frac{i \pi }{4}} & 0 & \sqrt{3}~e^{\frac{i \pi }{4}}  & 0 \\
\end{array}
\right) \mbox{and}
\end{equation}
\begin{equation}
\mathcal{U_-}= \frac{ e^{\frac{i \pi }{8}}}{4} \left(
\begin{array}{ccc}
 1 & 0 &  \sqrt{15} \\
 0 & 4~e^{\frac{i \pi }{4}} & 0 \\
  \sqrt{15} & 0 & -1 \\
\end{array}
\right).
\end{equation}
The eigenvalues of the $\mathcal{U_+} \left( \mathcal{U_-}\right)$ are ${\left\{-(-1)^{1/4},-(-1)^{3/4},(-1)^{1/4}\right\}}\left(\left\{-1,-1,1,1\right\}\right)$ and\\ the eigenvectors are  $\left\lbrace\left[\sqrt{\frac{3}{5}},\sqrt{\frac{3}{5}},\frac{(-1)^{3/8}}{2}  \sqrt{\frac{5}{2}},-\frac{(-1)^{3/8}}{2}  \sqrt{\frac{5}{2}}\right]^T, \right.$\\ $\left. \left[-2
(-1)^{1/8} \sqrt{\frac{2}{5}},2 (-1)^{1/8} \sqrt{\frac{2}{5}},0,0\right]^T, \left[1,1,-\frac{(-1)^{3/8}}{2}  \sqrt{\frac{3}{2}},\right.\right.$\\ $\left. \left.\frac{(-1)^{3/8}}{2}
\sqrt{\frac{3}{2}}\right]^T,\left[0,0,1,1\right]^T\right\rbrace \left(\left\lbrace\left[-\sqrt{\frac{3}{5}},\sqrt{\frac{5}{3}},0\right]^T, \left[0,0,1\right]^T, \right.\right.$\\ $\left. \left.\left[1,1,0\right]^T\right\rbrace\right)$. Thus, the $n$th time evolution of the  blocks $\mathcal{U_{\pm}}$ is given as follows:
\begin{widetext}
\begin{equation}\label{Eq:Apc58}
\mathcal{U}_{+}^n= {\frac{e^{\frac{i n \pi }{4}}}{16}\left[
\begin{array}{cccc}
 {a_n}  \left(3+5 e^{\frac{i n \pi }{2}}\right)  & 2\sqrt{{6}}~b_n{e^{\frac{7i  \pi }{8}}}
  & 4{i \sqrt{15}~ e^{\frac{3i n \pi }{4}}}  \sin\left(\frac{n \pi }{4}\right)\cos\left(\frac{n \pi }{2}\right) & 2\sqrt{{10}}~b_n~{e^{(\frac{ i n \pi }{2}+\frac{3i  \pi }{8})}}   \\
 -2\sqrt{{6}}~b_n~e^{\frac{i  \pi }{8}}   & 8a_n   &-2\sqrt{{10}}~b_n~ {e^{\frac{i  \pi }{8}}}    & 0 \\
 {4i \sqrt{15}~ e^{\frac{3i n \pi }{4}}}  \sin\left(\frac{n \pi }{4}\right)\cos\left(\frac{n \pi }{2}\right) &2\sqrt{{10}}~b_n{e^{\frac{7i  \pi }{8}}}
 & a_n \left(5+3 e^{\frac{i n \pi
}{2}}\right)  & -2\sqrt{{6}}~b_n~{e^{(\frac{ i n \pi }{2}+\frac{3i  \pi }{8})}}    \\
 -2\sqrt{{10}}~b_n~{e^{(\frac{ i n \pi }{2}+\frac{5i  \pi }{8})}}     & 0 & 2\sqrt{{6}}~b_n~{e^{(\frac{ i n \pi }{2}+\frac{i  \pi }{8})}}
  & {8a_n~e^{\frac{i n \pi }{2}}}  \\
\end{array}
\right]}
 \end{equation}
 \begin{equation}\label{Eq:46}
 \mbox{and}~~~~
 \mathcal{U}_{-}^n =\frac{e^{\frac{i n \pi }{8}}}{8}{\left[
\begin{array}{ccc}
 5 + 3~e^{i n \pi } & 0 &  \sqrt{15}\left( 1-e^{{ i n \pi }}\right) \\
 0 & 8 ~ e^{\frac{ i n \pi }{4}} & 0 \\
 \sqrt{15}\left( 1-e^{{ i n \pi }}\right) & 0 & 3 + 5~ e^{i n \pi } \\
\end{array}
\right]},
 \end{equation}
where $a_n$=$1+e^{i n \pi }$ and $b_n$=$1-e^{i n \pi }$. From the Eqs. (\ref{Eq:Apc58}) and (\ref{Eq:46}), we  can observe the  periodic
nature of $\mathcal{U}^n_+ (\mathcal{U}^n_-)$ with period $8(16)$.
Hence, the unitary operator shows periodicity with period $16$ i.e. $\mathcal{U}^n=\mathcal{U}^{n+16}$. In
Refs. \cite{pal2018entangling,sharma2024exactly} it was found that when the Hamiltonian is integrable then the corresponding
unitary operator shows a periodic nature.
Now, it is straightforward to perform the time evolution of any initial state and consequently study various quantum correlations.
We further conducted a detailed analysis on two states, $\ket{0,0}$ and $\ket{\pi/2,-\pi/2}$, deriving exact expressions for the linear
entropy, von Neumann entropy of single-qubit reduced density matrix (RDM), and the concurrence between any two qubits.

\subsection{Initial state $\ket{000000}=\ket{\theta_0 =0,\phi_0 =0}$}
 Let us first start with  the state $|000000\rangle$. The  $n$th time
evolution of the unitary operator on this state  is given by:
\begin{eqnarray}
\ket{\psi_n}&=&\mathcal{U}^n |00000\rangle=\mathcal{U}^n |w_0\rangle =\mathcal{U}^n \left(|\phi_{1}^{+}\rangle+|\phi_{1}^{-}\rangle\right)/\sqrt{2}\nonumber \\ \nonumber
&=&\left(\mathcal{U}_{+}^n |\phi_{1}^{+} \rangle + \mathcal{U}_{-}^n |\phi_{1}^{-} \rangle \right)/\sqrt{2}\\\nonumber
&=& \frac{1}{\sqrt{2}}\left(a_1\ket{\phi_0^+}+a_2\ket{\phi_1^+}+a_3\ket{\phi_2^+}+a_4\ket{\phi_3^+} +a_5\ket{\phi_0^-}+a_6\ket{\phi_2^-}\right),\nonumber
 \end{eqnarray}
where  ${a}_1$=$\left[\frac{e^{\frac{i n \pi }{4}}}{16}  \left(3+5~ e^{\frac{i n \pi }{2}}\right) \left(1+e^{i n \pi }\right)\right]$,
 ${a}_2$ = $\frac{e^{\frac{i \pi }{8}+\frac{i n \pi}{4}}}{4} \sqrt{\frac{3}{2}}  \left(-1+e^{i n \pi }\right)$, ${a}_3=\frac{i \sqrt{15}~ e^{i n \pi }}{8}  \left(\sin\left(\frac{n \pi }{4}\right)-\sin\left(\frac{3 n \pi }{4}\right)\right)$,
 ${a}_4=\frac{ e^{\left(\frac{5i \pi }{8}+\frac{3 i n \pi}{4} \right)}}{4} \sqrt{\frac{5}{2}}  \left(-1+e^{i n \pi }\right)$,~ ${a}_5=\frac{e^{\frac{i n \pi }{8}}}{8} \left(5+3 e^{{i n \pi }}\right)$ and ${a}_6= \sqrt{15}~ e^{\frac{i n \pi }{8}} \left(1-e^{{i n \pi }}\right)/8$. From  $\ket{\psi_n}$, we  will obtain the RDM of a single qubit
$\left(\rho_1(n)=\mbox{Tr}_{\neq1}\left(\ket{\psi_n}\bra{\psi_n}\right)\right)$ and the two qubits
$\left(\rho_{12}(n)=\mbox{Tr}_{\neq1,2}\left(\ket{\psi_n}\bra{\psi_n}\right)\right)$.
Thus we get $\rho_1(n)$, which  is given as follows:\\
\begin{equation}
\rho_1(n)=\frac{1}{4}\left(
\begin{array}{cc}
 2+a_n & w_n \\
 w_n^* & 2-a_n \\
\end{array}
\right),
\end{equation}
\begin{eqnarray}\nonumber
\mbox{where}~ a_n&=&\frac{1}{4} \cos\left(\frac{n \pi }{8}\right) \cos\left(\frac{n \pi }{2}\right) \left[2-4 \cos\left(\frac{n \pi }{4}\right)+5
\cos\left(\frac{n \pi }{2}\right)+5 \cos(n \pi )\right] \mbox{ and} \\ \nonumber
w_n&=&{(-1)^{1/8}~ e^{-\frac{13}{8} i n \pi }} \left(-1+e^{i n \pi }\right)
\left[(5+5 i) \sqrt{2}+5 \left(-1+(-1)^{1/4}\right) e^{\frac{i n \pi }{8}}+5~i \left((-1+i)+\sqrt{2}\right) e^{\frac{9 i n \pi }{8}}\right.\\ \nonumber && \left. -2~e^{\frac{3 i n \pi }{4}}-10 (-1)^{1/4} e^{i n \pi }-\right.
\left.10~ i e^{\frac{5 i n \pi }{4}}-2~ (-1)^{3/4} e^{\frac{3 i n \pi }{2}}+\left((1+5 i)-(2+3 i) \sqrt{2}\right)
\left(e^{\frac{5 i n \pi }{8}}+e^{\frac{13 i n \pi }{8}}\right)\right.\\ \nonumber && \left.+10~
e^{\frac{7 i n \pi }{4}}+5~\left(-i+(-1)^{3/4}\right) e^{\frac{17 i n \pi }{8}}+10~ (-1)^{3/4} e^{\frac{i n \pi }{2}}+10~ i ~e^{\frac{9 i n \pi }{4}}\right]\Big{/}{64}.
\end{eqnarray}
\end{widetext}
The eigenvalues of  $\rho_1(n)$ are $\lambda_n$ and $1-\lambda_n$ where $\lambda_n$=\\$\frac{1}{2}-\frac{1}{16} \left\lbrace 17+8 \left(2+\sqrt{2}\right) \cos\left(\frac{n \pi }{4}\right)+8 \left(2-\sqrt{2}\right) \cos\left(\frac{3 n \pi }{4}\right)+\right.$\\ $\left.15 \left[\cos\left(\frac{n
\pi }{2}\right)- \sin\left(\frac{n
\pi }{2}\right)\right)-8 \sqrt{2} \left(\sin\left(\frac{n \pi }{4}\right)+  \sin\left(\frac{3 n \pi }{4}\right)\right]\right\rbrace^\frac{1}{2}$.
Using these eigenvalues we can obtain the linear entropy \cite{buscemi2007linear} given by $2\lambda_n(1-\lambda_n)$ and is plotted in the Fig.~\ref{fig:correlations1}. The entanglement entropy \cite{nielsenbook,benenti2004principles} can also  be calculated  using $-(\lambda_n\log\lambda_n +(1-\lambda_n)\log(1-\lambda_n) )$ and is  plotted in the same figure.
It can be shown from the expressions and the figure that both,  the linear and entanglement entropy have a periodic nature with period eight i.e. $S(n+8)=S(n)$. This periodic nature has been previously observed in  integrable systems, particularly those involving
periodically kicked spin chains Refs.  \cite{ArulLakshminarayan2005,mishra2015protocol,naik2019controlled,sharma2024exactly}. For this particular state, we provided analytical proof that the entanglement content remains unchanged for consecutive odd and even values of  $n$ \cite{supplementry2023}
i.e. $S_{(0,0)}^{(6)}(2n-1) = S_{(0,0)}^{(6)}(2n)$, which is shown in the
Fig. \ref{fig:correlations1}. The linear and entanglement entropy attains its maximum upper bound value $0.5$ and $\ln2 \approx 0.6932$ for the parameter $\tau=\pi/4$ and ${J}=1/2$, which is also observed in the previous study for this  model \cite{sharma2024exactly} with same $\tau$ but ${J}=1$. We can observe from the figure that the entanglement content takes more time ($n=3$) to reach the maximum upper bound value compared to the previous work ($n=1$) in
Ref. \cite{sharma2024exactly}.
\subsubsection{Concurrence}
The linear entropy measures the entanglement between a single qubit with  the rest of the qubits  in a pure state, while the concurrence measures the entanglement between any pair of qubits within the system (pure or mixed). Regardless of which particular qubits are chosen, the presence of permutation symmetry in the state results in only one concurrence value \cite{wootters1998entanglement,wootters2001entanglement}. The concurrence is given by,
\begin{equation}
\label{Eq:App5Q3}
\mathcal{C}(\rho_{12})=\text{max}\left(0, \sqrt{\lambda_1}-\sqrt{\lambda_2}-\sqrt{\lambda_3}-\sqrt{\lambda_4} \right),
\end{equation}
where $\lambda_l$ are eigenvalues in decreasing order of
$(\sigma_y \otimes \sigma_y)\rho_{12} (\sigma_y \otimes \sigma_y) \rho_{12}^*$, where $\rho_{12}^*$ is complex conjugation in
the standard ($\sigma_z$) basis. The two-qubit RDM, $\rho_{12}$, is  given as follows:
\begin{equation}
\rho_{12}(n)={\frac{1}{4}\left(
\begin{array}{cccc}
 b_{m} & a_{m} & a_{m} & d_{m}^* \\
 a_{m}^* & e_{m} & e_{m} & a_{m}^*  \\
 a_{m}^* & e_{m} & e_{m} & a_{m}^*  \\
 d_{m} & a_{m} & a_{m} & f_{m} \\
\end{array}
\right)},
\end{equation}
where the coefficients are,
\begin{eqnarray}\nonumber
b_{m}&=&\frac{1}{16} \left[22+6 \cos\left(\frac{n \pi }{8}\right)+10 \cos\left(\frac{3 n \pi }{8}\right)+4 \cos\left(\frac{n\pi }{2}\right)\right.\\ &&\left. \nonumber +10 \cos\left(\frac{5 n \pi }{8}\right)+6\cos\left(\frac{7 n \pi }{8}\right)+6 \cos(n \pi )\right],  \\ \nonumber
f_{m}&=&\frac{1}{4} \left[58+94 \cos\left(\frac{n \pi }{8}\right)+78 \cos\left(\frac{n \pi }{4}\right)+62 \cos\left(\frac{3
n \pi }{8}\right)\right.\\ \nonumber  && \left.+56 \cos\left(\frac{n \pi }{2}\right)+46\cos\left(\frac{5 n \pi }{8}\right)+46 \cos\left(\frac{3 n \pi }{4}\right)\right.\\ \nonumber  && \left.+26 \cos(n \pi )+46\cos\left(\frac{7 n \pi }{8}\right)\right] \sin^2\left(\frac{n \pi }{16}\right),
\\ \nonumber
d_{m}&=&\frac{1}{16} \left[4 \cos\left(\frac{n \pi }{2}\right)-6 \cos(n \pi )-{2~i}
 \left( i -4~ i \sin\left(\frac{n \pi }{2}\right)+\right.\right.\\ \nonumber  && \left.\left.\sin\left(\frac{n \pi }{8}\right)+ \sin\left(\frac{3 n \pi }{8}\right)- \sin\left(\frac{5
n \pi }{8}\right)-\sin\left(\frac{7 n \pi }{8}\right)\right)\right],\\ \nonumber
 e_{m}&=&\frac{1}{8} \left[3 \sin\left(\frac{n \pi }{4}\right)+\sin\left(\frac{3 n \pi }{4}\right)\right]^2  ~~\mbox{and}\\ \nonumber
a_{m}&=&(-1)^{3/8} e^{-\frac{13}{8} i n \pi } \left(-1+e^{i n \pi }\right)
\left[6 \sqrt{2}+e^{\frac{i n \pi }{8}} \left((2-2 i)\right.\right.\\ \nonumber  && \left.\left. -3 \sqrt{2} +6 ~i \sqrt{2} e^{\frac{3 i n \pi }{8}}+4 \left(-2+\sqrt{4-3 ~i}\right)\cos\left(\frac{n \pi }{2}\right)\right.\right.\\ \nonumber
&&\left.\left.-(4-4~ i)~\left( e^{\frac{5 i n \pi }{8}}-e^{\frac{-3 i n \pi }{8}}\right)+\left(4-6 e^{\frac{i  \pi }{4}}\right)~ e^{i n \pi }\right.\right.\\ \nonumber
&&\left.\left.-6 \sqrt{2} ~e^{\frac{7 i n \pi }{8}}-(4+4 ~i)~ \left(e^{\frac{9 i n
\pi }{8}}-e^{\frac{ i n \pi }{8}}\right)+2~ i \sqrt{2}~  \right.\right.\\ \nonumber
&&\left.\left.  e^{\frac{-5 i n \pi }{8}}+i(-2+2~
i)-3~ i \sqrt{2} \right)\right]\Big{/}{64 \sqrt{2}}.\\ \nonumber
\end{eqnarray}
\begin{figure}[t!]\vspace{0.2cm}
\includegraphics[width=0.47\textwidth]{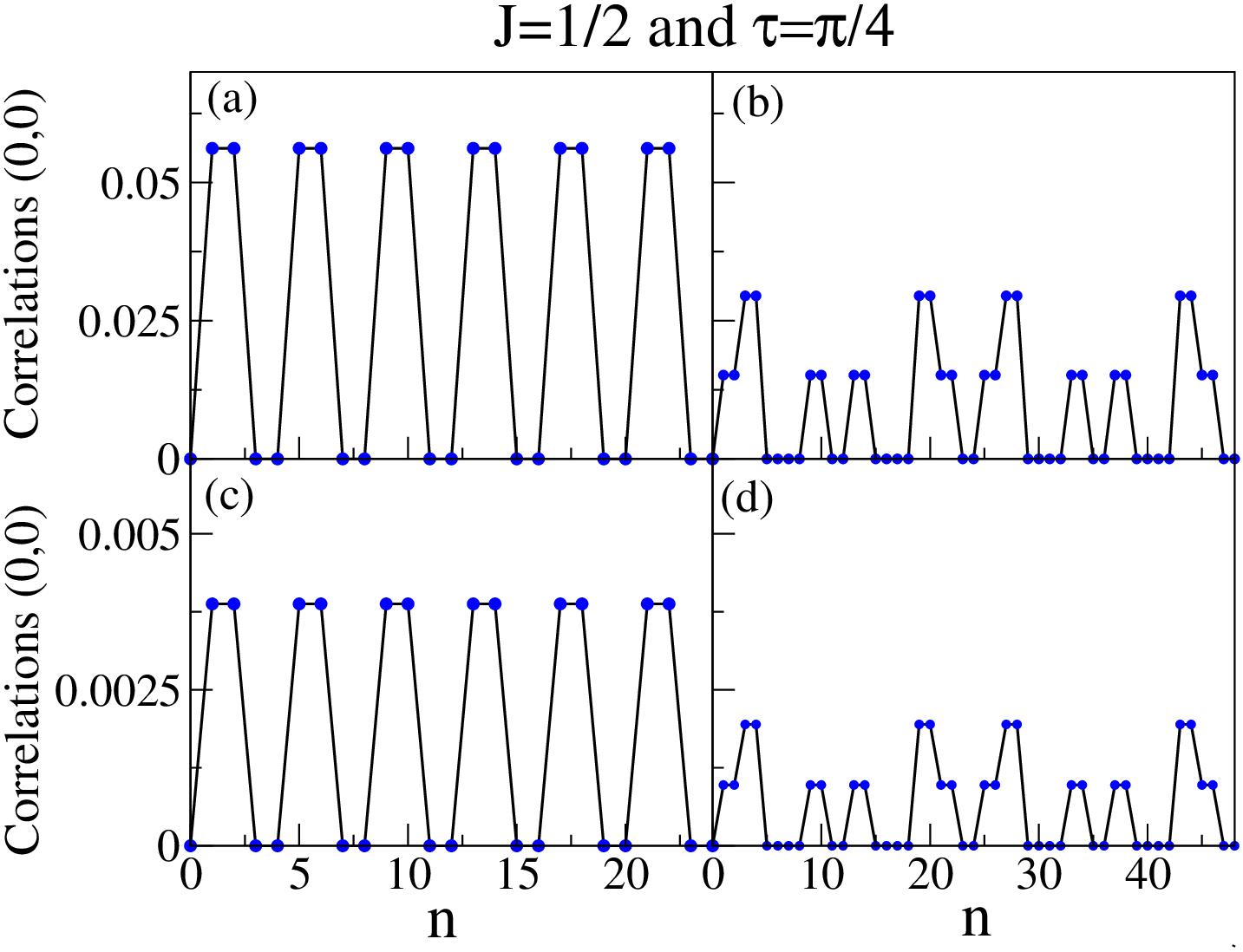}
\caption{The concurrence (circles) are plotted for  (a) $6$ qubit  (b) $8$ qubit
(c) $10$ qubit and  (d) $12$ qubit   for the  initial state $\otimes ^N {\ket{0}}_y$.}
\label{fig:correlations15}
\end{figure}
By evaluating the eigenvalues of $(\sigma_y \otimes \sigma_y)\rho_{12} (\sigma_y \otimes \sigma_y) \rho_{12}^*$, we obtain  fairly long expressions (please refer supplemental material \cite{supplementry2023} for their numerical values). Using these values in the Eq. (\ref{Eq:App5Q3}), the computed concurrence values are plotted in Fig. \ref{fig:correlations15}.
 Similarly, concurrence is also periodic in nature and it  remains the same for consecutive odd and even values of $n$
 \cite{supplementry2023}, which can be seen in Fig.~\ref{fig:correlations15}, whereas
 in Ref. \cite{sharma2024exactly} the pairwise concurrence remains zero for all the values of $n$. Thus all the entanglement measure quantities  are periodic in nature which is the signature of quantum integrability \cite{ArulLakshminarayan2005,naik2019controlled,mishra2015protocol,pal2018entangling,sharma2024exactly}.
\subsection{ Initial state $\ket{++++++}$= $\ket{\theta_0 =\pi/2,\phi_0 =-\pi/2}$.}
 Let us  now focus on another state  $\ket{++++++}_y$, where $|+\rangle_y=\frac{1}{\sqrt{2}}(|0\rangle+i|1\rangle)$ is an eigenstate of $\sigma_y$ with eigenvalue $+1$. The evolution of this state is entirely confined to the positive parity subspace of the seven-dimensional permutation symmetric space of six-qubits hilbert space. It can be expressed as:
 \begin{equation}
  \otimes ^6 {\ket{+}}_y=\frac{1}{4\sqrt{2}} \ket{\phi_0^+}+\frac{i\sqrt{3}}{4}\ket{\phi_1^+}- \frac{\sqrt{15}}{4\sqrt{2}} \ket{\phi_2^+}-\frac{i\sqrt{5}}{4}\ket{\phi_3^+}.
 \end{equation}
 The state $\ket{\psi_n}$ can than  be obtained  by applying  the $n$ iterations  of unitary operator $\mathcal{U}$  on it as follows:
\begin{eqnarray}
\ket{\psi_n}&=&\mathcal{U}_{+}^n\ket{++++++}\\  \nonumber
&=&e^{{\frac{ in \pi }{4}}}\left({\alpha}_{n}^\prime \ket{\phi_0^+}+{\beta}_{n}^\prime  \ket{\phi_1^+}+{\gamma}_{n}^\prime  \ket{\phi_2^+}+\zeta_n  \ket{\phi_3^+}\right),
\end{eqnarray}
where   the coefficients can be expressed as,
\begin{eqnarray} \nonumber
 {\alpha}_{n}^\prime &=&\frac{e^{-\frac{3}{4} i n \pi }}{{16 \sqrt{2}}} \left[3 \left(-1+e^{\frac{3i \pi }{8}}\right)+5 \left(1+e^{\frac{7i \pi }{8}}\right) e^{\frac{i
n \pi }{2}}\right.\\ \nonumber && \left.-3 \left(1+e^{\frac{3i \pi }{8}}\right) e^{i n \pi }-5 \left(-1+e^{\frac{7i \pi }{8}}\right) e^{\frac{3 i n \pi }{2}}\right],\\ \nonumber  ~{\beta}_{n}^\prime &=&\frac{\sqrt{3}~ e^{-\frac{3}{4} i n \pi }}{8}  \left[i-e^{\frac{i\text{  }\pi }{8}}+\left(i+e^{\frac{i\pi }{8}}\right)
e^{i n \pi }\right],\\ \nonumber
{\gamma}_{n}^\prime &=&\frac{i~e^{-\frac{3}{4} i n \pi }}{16}  \sqrt{\frac{15}{2}}  \left[1-e^{\frac{3i\text{  }\pi }{8}}+\left(i+e^{\frac{3i\text{
 }\pi }{8}}\right) e^{\frac{i n \pi }{2}}\right]\\ \nonumber && \left[i+(1+i) e^{\frac{i n \pi }{2}}+e^{i n \pi }\right] ~~\mbox{and}\\ \nonumber~ \zeta_n &=&\frac{\sqrt{5}~ e^{-\frac{1}{4} i n \pi }}{8}  \left[-i+e^{\frac{5i\text{  }\pi }{8}}-\left(i+e^{\frac{5i\text{  }\pi
}{8}}\right) e^{i n \pi }\right]. \\ \nonumber
\end{eqnarray}
The  $\rho_1(n)$, is given as follows:
\begin{equation}
\rho_1(n)={\frac{1}{2}\left[
\begin{array}{cc}
 1 & Z_n \\
 Z_n^* & 1 \\
\end{array}
\right]},
\end{equation}
where~~$Z_n={-i} \left[3(1+\cos(n\pi)) +\sqrt{2}\left(1-\cos(n\pi)\right)+10\right.$\\ $  \left.\cos\left(\frac{n \pi }{2}\right)\right]\Big{/}{16}$. The eigenvalues of  $\rho_1(n)$ are $\lambda_n$ and $1-\lambda_n$\\ where $\lambda_n=
 \left[8-\sqrt{18+30\cos\left(\frac{n \pi }{2}\right)+16 \cos(n \pi )}\right]\Big{/}16$.
 \begin{figure}[t!]\vspace{0.4cm}
\includegraphics[width=0.47\textwidth]{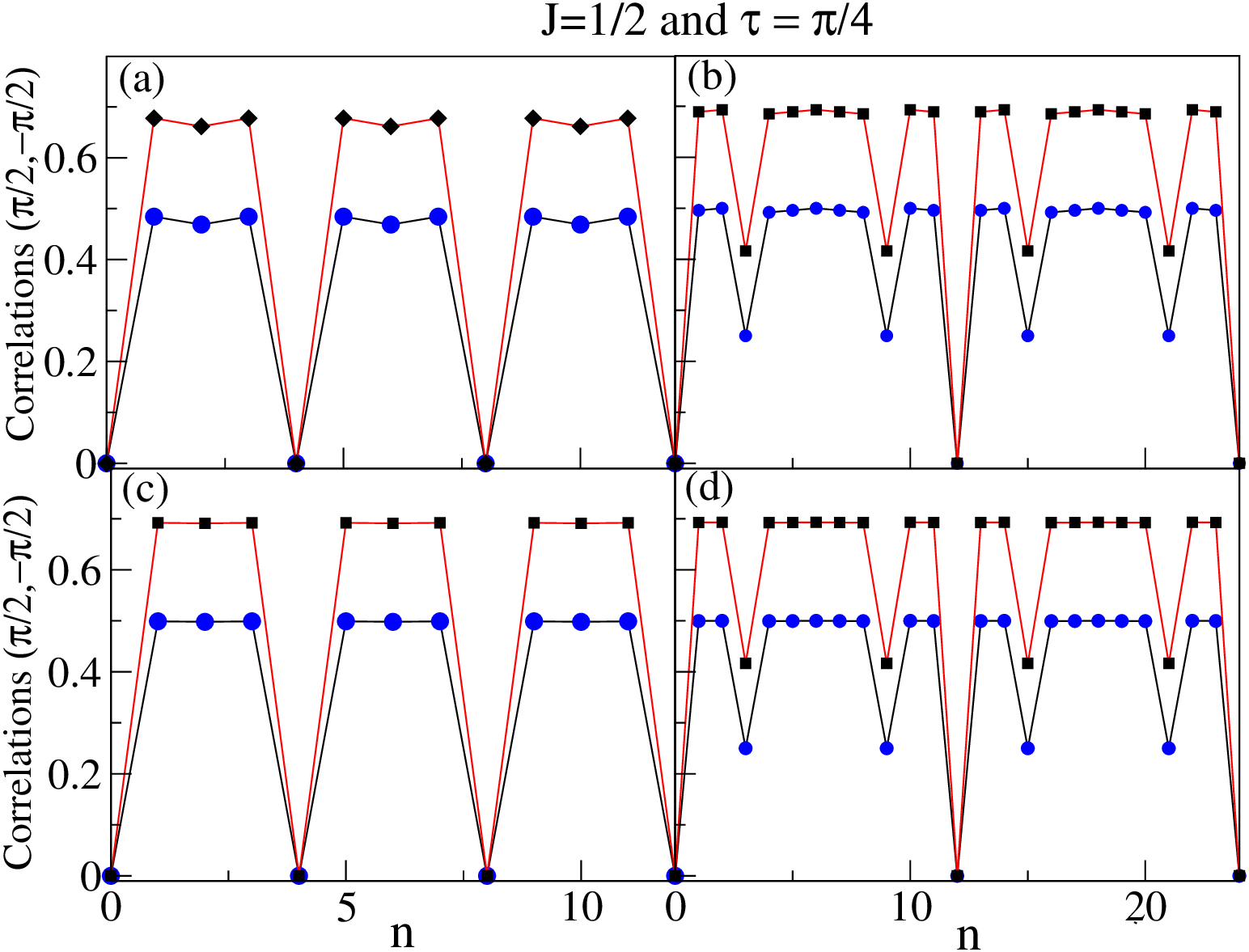}
\caption{Correlations using linear entropy (circles), and entanglement entropy (squares)  are plotted for  (a) $6$ qubit (b) $8$ qubit
(c) $10$ qubit and  (d) $12$ qubit   for the  initial state $\otimes ^N {\ket{+}}_y$.}
\label{fig:correlations2}
\end{figure}
The linear entropy and entanglement entropy can be calculated by using  eigenvalues ($\lambda_n$) and  are   plotted in Fig. \ref{fig:correlations2}. The entanglement content is  periodic in nature  having a period of $4$.
\subsubsection{Concurrence}
The  $\rho_{12}(n)$, for this state  is given as follows:
\begin{equation}
\rho_{12}(n)={\frac{1}{2}\left(
\begin{array}{cccc}
 h_1 &h_4  & h_4 & h_2 \\
 h_4^* & h_3 & h_3 & h_6 \\
 h_4^* & h_3 & h_3 & h_6 \\
 h_2^* & h_6^* & h_6^* & h_1 \\
\end{array}
\right)},
\end{equation}
where  the coefficients are,\\
\begin{eqnarray} \nonumber
h_1&=&\frac{1}{8} \left[5- \left(\cos\left(\frac{n \pi }{2}\right)+\sin\left(\frac{n \pi }{2}\right)\right)\right],\\ \nonumber
h_2&=&\frac{-1}{8} \left[1+ 3\cos\left(\frac{n \pi }{2}\right)-\sin\left(\frac{n \pi }{2}\right)\right],  \\ \nonumber
h_3&=&\frac{1}{8} \left[3+\cos\left(\frac{n \pi }{2}\right)+\sin\left(\frac{n \pi }{2}\right)\right],  \\ \nonumber
h_6&=&\frac{(1-i)}{64} \left[3(1+ i)(1+\cos(n\pi))+4 \sqrt{2}(1-\cos(n\pi))\right.\\ \nonumber && \left. +20\left(1+ i\right) \cos\left(\frac{n
\pi }{2}\right)\right]\mbox{and}\\ \nonumber
h_4&=&\frac{-(1+i)}{64} \left[3(1+ i)(1+\cos(n\pi))+4 \sqrt{2}(1-\cos(n\pi))\right.\\ \nonumber && \left. +20\left(1+ i\right) \cos\left(\frac{n
\pi }{2}\right)\right].\nonumber
\end{eqnarray}
 \begin{figure}[t!]\vspace{0.4cm}
\includegraphics[width=0.47\textwidth]{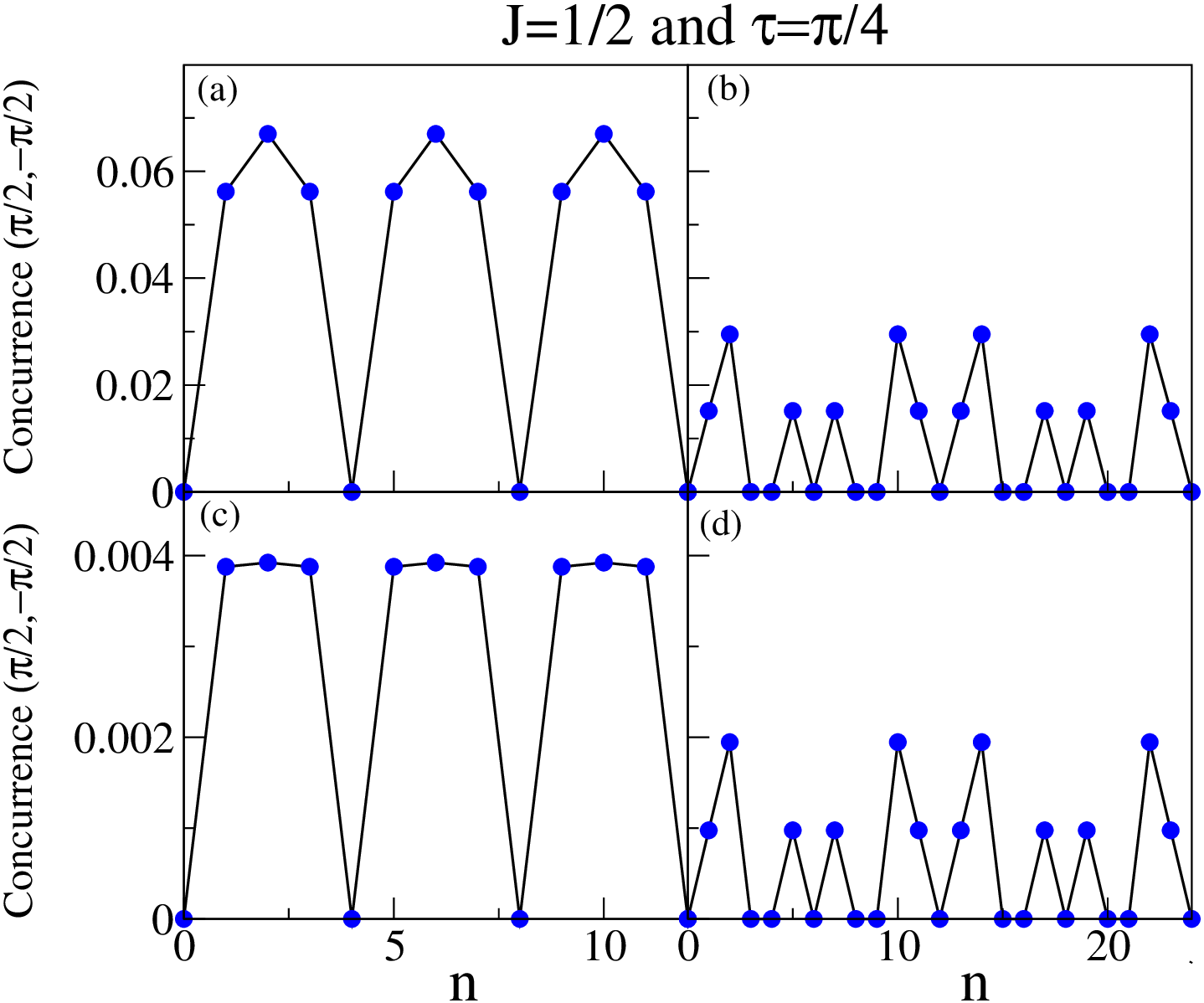}
\caption{The concurrence (circles) are plotted for  (a) $6$ qubit and (b) $8$ qubit
(c) $10$ qubit and  (d) $12$ qubit   for  the  initial state $\otimes ^N {\ket{+}}_y$.}
\label{fig:correlations18}
\end{figure}
The  eigenvalues of $(\sigma_y \otimes \sigma_y)\rho_{12} (\sigma_y \otimes \sigma_y) \rho_{12}^*$ are $\left\lbrace  \frac{1}{4} \sin^4\left(\frac{n \pi}{4}\right),
 \frac{1}{256} \left[23 - 6 \cos\left(\frac{n\pi}{2}\right) - 17 \cos\left({n\pi}\right)\pm
    4 \right.\right.$\\ $\left.\left.\left[ \cos^2\left(\frac{n\pi}{2}\right)(-5 + \cos\left({n\pi}\right)) \left( (-41 + 16 \sqrt{2})\cos\left({n\pi}\right) -109
        \right.\right.\right.\right.$\\ $\left.\left.\left.\left. + 16 \sqrt{2}+ 2 (-69 + 16 \sqrt{2}) \cos\left(\frac{n\pi}{2}\right) \right) \sin^4\left(\frac{n \pi}{4}\right)\right]^\frac{1}{2}\right],0\right\rbrace$.
        The computed concurrence values  are plotted in the
        Fig. \ref{fig:correlations18}. It can be  observed that it is periodic in nature with a period of $4$, whereas in Ref. \cite{sharma2024exactly} the pairwise concurrence remains zero. The periodic nature of entanglement dynamics and operator dynamics are the signature of integrability \cite{sharma2024exactly}, hence for $6$ qubit the model is integrable  for these parameters as well.
\section{Exact solutions for eight, ten and  twelve  qubits }\label{sec:example-section4}
We followed a similar process from section \ref{sec:example-section3} to solve the analytically cases involving  $N$= $8$, $10$, and $12$ qubits at the particular value of $J$=$1/2$ and  $\tau$=$\pi/4$. We created distinct tables  for the states $\otimes^N|0\rangle$ (refer
Tables \ref{Table:LinearEntropy1} and  \ref{Table:LinearEntropy}) and $\otimes^N|+\rangle$ (refer
Table \ref{Table:LinearEntropy2}) to show  the analytical expression for the eigenvalues and linear entropy for the qubits
$8$, $10$ and $12$. Using these expressions, it can be shown that they exhibit a periodic behavior. For the initial state
$\otimes^N \ket{0}$ and the  cases of  $6$ and $10$ qubits period is $8$, whereas for $8$ and $12$ qubits it is  $24$. This
can be observed in the
Fig. \ref{fig:correlations1}. We  observe that each correlation is  same for consecutive odd and even values of $n$ \cite{ExactDogra2019}. We have plotted the analytically obtained concurrence in Fig.~\ref{fig:correlations15}. From the values and the plot, we observe that it shows a periodic nature. For $N=6$ \mbox{and} $10$ ($8$ \mbox{and} $12$), we find  $T=4 (12)$. The maximum value of concurrence ($C_{max}$) for $N$=~$6$, $8$, $10$ and $12$ is $0.05618622$, $0.029509$, $0.0038762$ and $0.0019455$ respectively (All values are close upto $12$ decimal places). Reduction of  $C_{\mbox{max}}$  with $N$  shows an increased multipartite nature of entanglement. The signatures of QI are obtained by limiting $n$ to positive integers only.
The data points in all the figures of this paper correspond to $n\in \mathcal{N}^+$.

 The Table~\ref{Table:LinearEntropy2} presents the analytical expressions for the eigenvalues and  linear entropy  for the initial  state $\otimes^N|+\rangle$. Using these eigenvalues expression for entanglement entropy can be obtained easily (not shown here). These correlations are plotted in
 Fig.~\ref{fig:correlations2}. It can be seen that they show periodic behavior. For the cases  $6$ and $10$ ($8$ and $12$) qubits the period is $4$ ($12$). The values of  $C_{max}$ for $N$=$6$, $8$, $10$ and $12$ are  $0.0669873$, $0.0295085$, $0.00392163$ and $0.0019455$ respectively. Thus, for this state also $C_{\mbox{max}}$ decreases with  $N$, displaying the multipartite nature of entanglement.

For this case also we study  the dynamics of the floquet operator  in which  we observe that it shows  periodic nature  with time i.e. ${\mathcal U}^{n+T}={\mathcal U}^n $, where $n\geq 1$ and $T$ is the period.
 For qubits $N$= $8$, $10$ and $12$, the period is $48$. Through analytical calculations, we determine the eigenvalues  of $\mathcal{U}$ and identify degeneracy among them (see Supplemental Material \cite{supplementry2023}).
 \begin{figure}[t!]\vspace{0.4cm}
\includegraphics[width=0.47\textwidth]{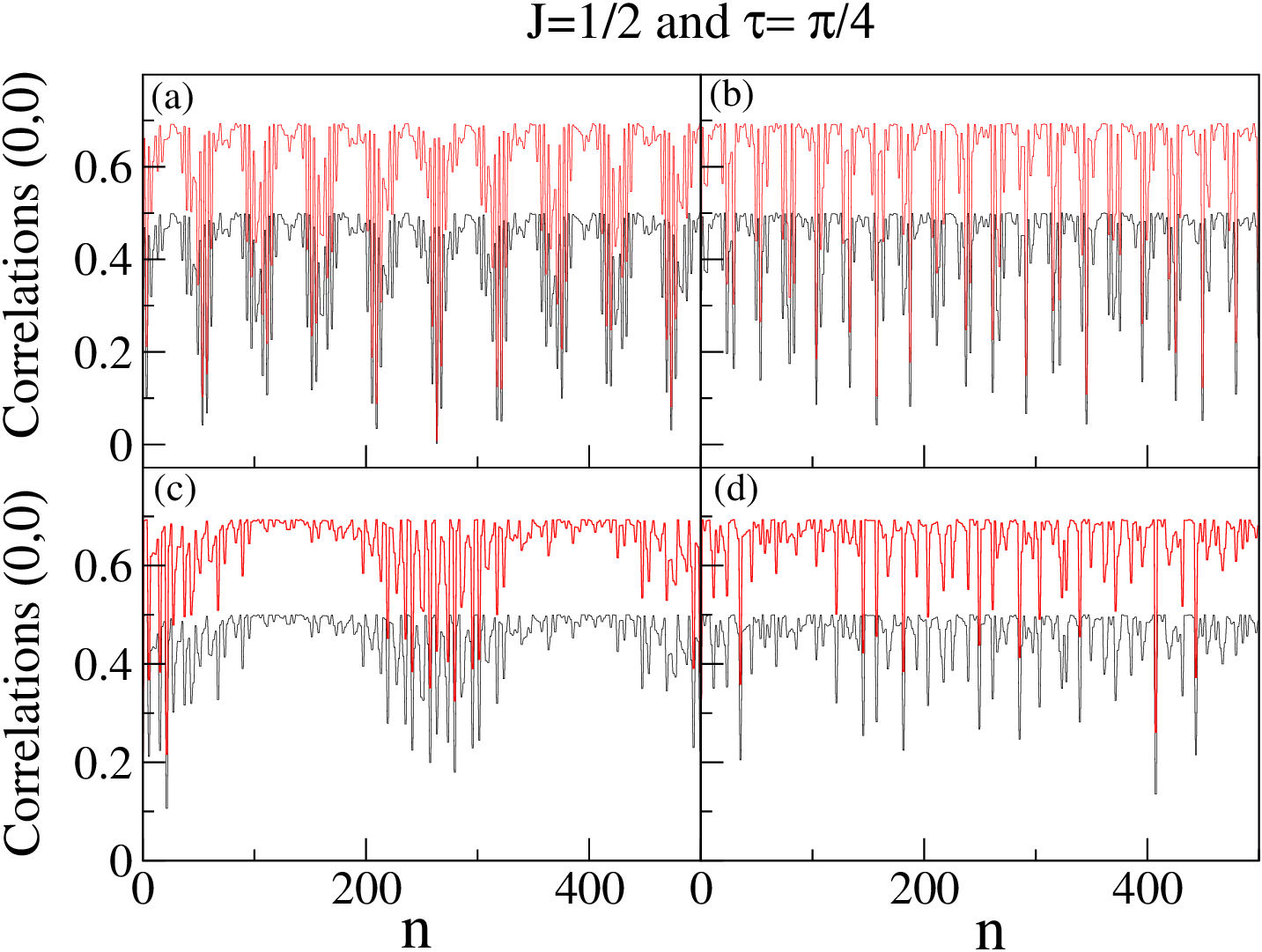}
\caption{(Color online) Correlations using linear entropy (black), and entanglement entropy (red)  are plotted for  (a) $5$ qubit  (b) $7$ qubit
(c) $9$ qubit and  (d) $11$ qubit  for  initial state $\otimes ^N {\ket{0}}_y$.}
\label{fig:correlations182}
\end{figure}
\begin{figure}[t!]\vspace{0.4cm}
\includegraphics[width=0.47\textwidth]{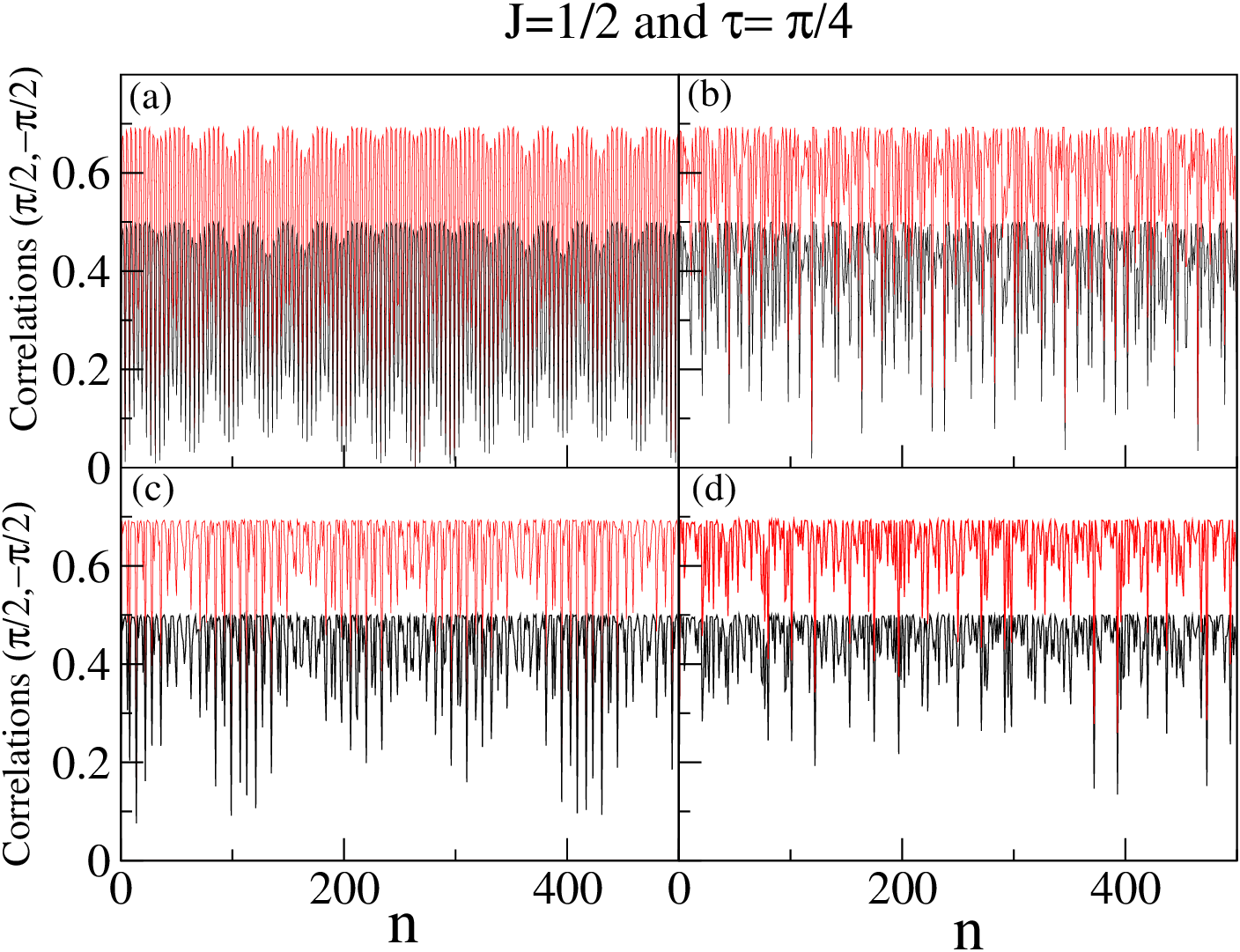}
\caption{(Color online) Correlations using linear entropy (black), and entanglement entropy (red)  are plotted for  (a) $5$ qubit  (b) $7$ qubit
(c) $9$ qubit and  (d) $11$ qubit  for the  initial state $\otimes ^N {\ket{+}}_y$.}
\label{fig:correlations183}
\end{figure}
 \begin{figure}[t!]\vspace{0.4cm}
\includegraphics[width=0.47\textwidth]{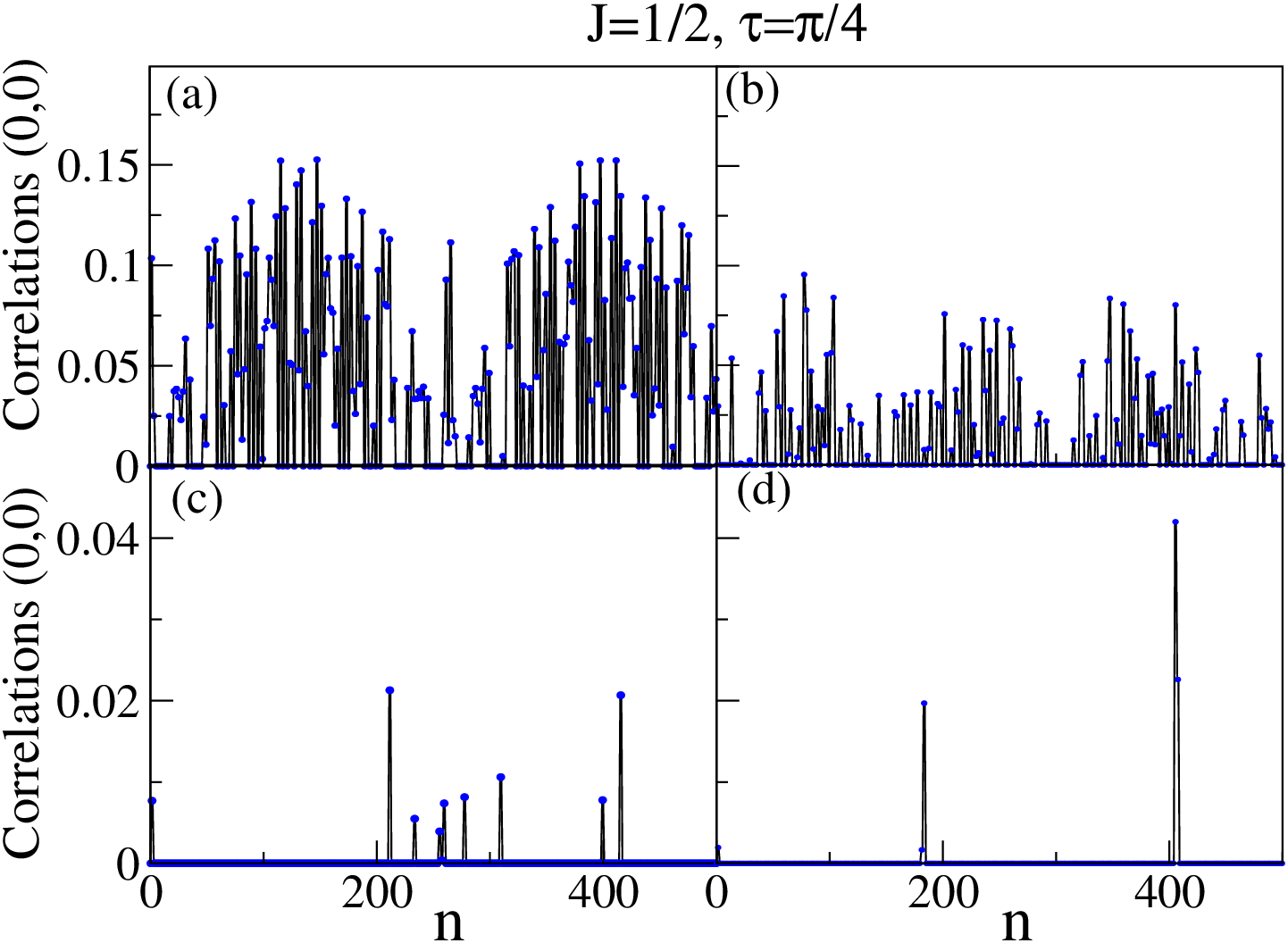}
\caption{The concurrence (circles) are plotted for  (a) $5$ qubit and (b) $7$ qubit
(c) $9$ qubit and  (d) $11$ qubit   for  the  initial state $\otimes ^N {\ket{0}}_y$.}
\label{fig:correlations189}
\end{figure}
 \begin{figure}[t!]\vspace{0.4cm}
\includegraphics[width=0.47\textwidth]{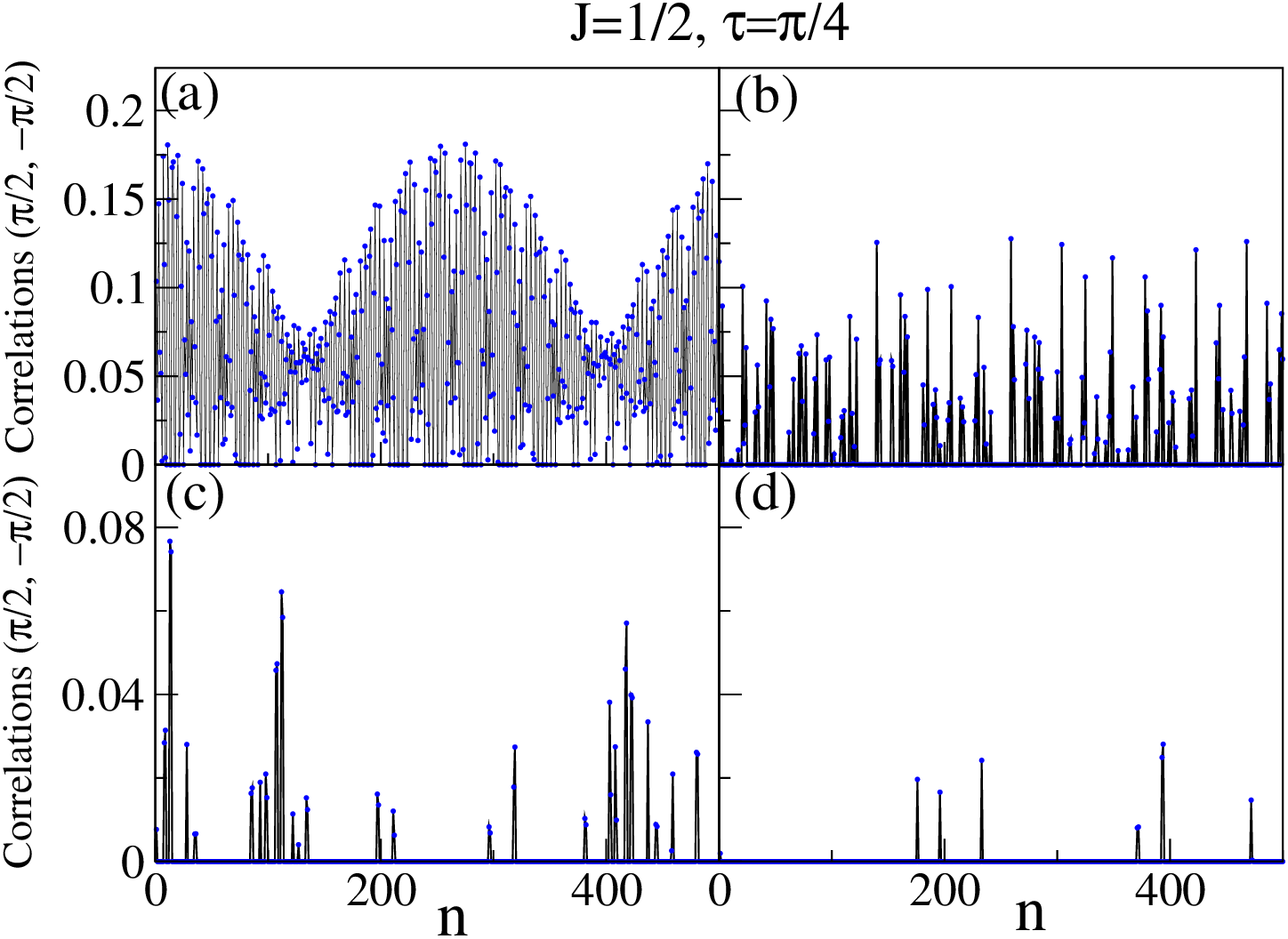}
\caption{The concurrence (circles) are plotted for  (a) $5$ qubit and (b) $7$ qubit
(c) $9$ qubit and  (d) $11$ qubit   for  the  initial state $\otimes ^N {\ket{+}}_y$.}
\label{fig:correlations1899}
\end{figure}

 For the cases  $5$, $7$, $9$, and $11$ qubits,  due to complexity and fairly large expressions, we restrict ourselves to numerical results instead of analytical ones. We observed  that the  entanglement dynamics  for both the initial states do not exhibit periodic behavior (checked for $n$ up to $1000$), which is  shown in Figs. \ref{fig:correlations182}, \ref{fig:correlations183},  \ref{fig:correlations189} and \ref{fig:correlations1899} (plotted  for $n$ up to $500$).
 We also study the  Floquet operator dynamics in which we observe  that it does not exhibit periodic behavior for large values of $n$, as shown in
 Fig. \ref{fig:correlations7}. The periodic behavior of entanglement dynamics, and floquet operator dynamics, and degenerated spectrum are the signature of integrability (as explained in the introduction). In the previous study
 Ref. \cite{sharma2024exactly}, they have shown, that  this model is integrable for the parameters $J=1$ and $\tau=\pi/4$
 using these  signatures,
whereas we observe these signatures only for an even $N$ ranging from $6$ to $12$. Thus, for the parameter  $J=1/2$ and $\tau=\pi/4$, the model shows integrability for $6$, $8$, $10$, and $12$  qubits and absence of integrability for odd $N$ ranging from $5$ to $11$.

 \begin{table*}[t]
  \centering
 \caption{The eigenvalues for the   initial state $\otimes^N\ket{0}$ and various value of  $N$.}
 \renewcommand{\arraystretch}{2.4}
 \begin{tabular}{|p{0.35cm}|p{16.8cm}| }
 \hline
$N$ &\hspace{8cm} Eigenvalues ($\lambda$)  \\
\hline
8 &
$\lambda=\frac{1}{2}\pm\frac{1}{96}
 \left\lbrace \left(5+7 \cos\left(\frac{2 n \pi }{3}\right)\right)^2\left[ \cos^2\left(\frac{3 n \pi }{8}\right) +8 \cos\left(\frac{3
n \pi }{8}\right)  \cos\left(\frac{5 n \pi }{8}\right)+4 \cos^2\left(\frac{5n \pi }{8}\right)\right]+\sin^2\left(\frac{n \pi }{2}\right)
 \left[9 \cos^2\left(\frac{1}{8} (3+4 n) \pi \right)+147\right.\right.$\newline$\left.\left.\left.~~~~( \cos^2\left(\frac{1}{24} (9+5 n)\pi \right)+ \cos^2\left(\frac{1}{24} (9+13 n) \pi \right)\right)+
\sin\left(\frac{1}{24} (9+5 n) \pi \right)\left(-98 \sqrt{3} \cos\left(\frac{1}{24} (9+13n) \pi \right) +49 \sin\left(\frac{1}{24}
(9+5 n) \pi \right)-\right.\right.\right.$\newline$  \left.\left.\left.~~~~280\sin\left(\frac{1}{8} (3+7 n) \pi \right)\right)+49 \sqrt{3} \sin\left(\frac{1}{12} (9+5 n) \pi \right)+
280 \sqrt{3} \cos\left(\frac{1}{24} (9+37 n) \pi \right) \sin\left(\frac{1}{8} (3+7 n) \pi \right)+400 \sin^2\left(\frac{1}{8}
(3+7 n) \pi \right)-\right.\right.$\newline$ \left.\left.~~~~
\sin\left(\frac{1}{24} (9+13 n) \pi \right)\left(98 \sin\left(\frac{1}{24} (9+5 n) \pi \right)-280 \sin\left(\frac{1}{8} (3+7
n) \pi \right)-49 \sin\left(\frac{1}{24} (9+13 n) \pi \right)\right)-
14 \cos\left(\frac{1}{24} (9+5 n) \pi \right)\right.\right.$\newline$ \left.\left. ~~~\left( 21 \cos\left(\frac{1}{24} (9+37 n) \pi \right)+\sqrt{3} \left(20 \sin\left(\frac{1}{8}
(3+7 n) \pi \right)+7 \sin\left(\frac{1}{24} (9+13 n) \pi \right)\right)\right)-49 \sqrt{3} \sin\left(\frac{1}{12} (9+13 n) \pi \right)+
\sin\left(\frac{1}{8} (\pi +4 n \pi )\right)\right.\right.$\newline$\left.\left. ~~~\left(18 \cos\left(\frac{1}{8} (3+4 n) \pi \right) +9 \sin\left(\frac{1}{8}
(\pi +4 n \pi )\right)\right)\right]\right\rbrace^\frac{1}{2}$.
 \\
\hline
10&$\lambda =0.5\pm 0.5 \cos(n\pi) \left[
  0.2509765625 + 0.426776695296637 \cos(\frac{n\pi}{4}) +
   0.073223304703364 \cos(\frac{3n\pi}{4}) - 0.176776695296637 \right.$\newline
 $~~~~~~\left.\left(\sin(\frac{n\pi}{4})+\cos(\frac{3n\pi}{4})\right) +
   0.2490234375 \left(\cos(\frac{n\pi}{2}) - \sin(\frac{n\pi}{2})\right)\right]^{\frac{1}{2}}$.
\\
\hline
12&$\lambda =0.5 \pm
\left[0.041748046875+ 0.040916791185\cos\left(\frac{n\pi}{12}\right)+0.035634186961\left(\cos\frac{n\pi}{4}\right)+0.06243896484375\cos\left(\frac{n\pi}{3}\right)\right.$\newline$~~~~ \left.+0.026199786034\cos\left(\frac{5n\pi}{12}\right)+0.01542619052857\cos\left(\frac{7n\pi}{12}\right)+0.02081298828125\cos\left(\frac{2n\pi}{3}\right)
+0.0061138599142\cos\left(\frac{3n\pi}{4}\right)\right.$\newline$ ~~~~\left.+0.00070918537814\cos\left(\frac{11n\pi}{12}\right)-0.0053867977527\left(\sin\left(\frac{n\pi}{12}\right)+\sin\left(\frac{11n\pi}{12}\right)\right)-0.0147601635233\left(\sin\left(\frac{n\pi}{4}\right)+\sin\left(\frac{3n\pi}{4}\right)\right)\right.$\newline$~~~~\left.-0.036049153161\left(\sin\left(\frac{n\pi}{3}\right)+\sin\left(\frac{2n\pi}{3}\right)\right)-0.020103802903\left(\sin\left(\frac{5n\pi}{12}\right)+\sin\left(\frac{7n\pi}{12}\right)\right)\right]^\frac{1}{2}$.
\\
\hline
\end{tabular}
  \label{Table:LinearEntropy1}
\end{table*}
\begin{table*}[t]
  \centering
 \caption{The linear entropy ($S(n)$) for the initial state $\otimes^N\ket{0}$ and various values of  $N$.}
 \renewcommand{\arraystretch}{2.4}
 \begin{tabular}{|p{0.35cm}|p{17.3cm}| }
 \hline
$N$ &\hspace{8cm} $S(n)$   \\
\hline
8 &
$S_{(0,0)}^{8}(n)=\frac{1}{2}
\left\lbrace 1-\frac{1}{576} \left[5+7 \cos\left(\frac{2 n \pi }{3}\right)\right]^2 \left[\cos\left(\frac{3 n \pi }{8}\right)+\cos\left(\frac{11
n \pi }{8}\right)\right]^2+\sin^2\left(\frac{n \pi }{2}\right) \left[-\left(18-9\sqrt{2}\right) \cos(n \pi )+378 \sin\left(\frac{1}{12} (3+5 n) \pi \right) \right. \right.$\newline$ \left.\left.\hspace{1.5cm} +560 \left(\cos\left(\frac{2 n \pi }{3}\right)+\sqrt{3} \sin\left(\frac{2 n \pi }{3}\right)\right)+196 \left(\cos\left(\frac{4n \pi }{3}\right) - \sqrt{3} \sin\left(\frac{4n \pi }{3}\right)\right)+378 \sqrt{3} \sin\left(\frac{1}{12} (9+13 n) \pi \right)+9 \left(-90+\sqrt{2}\right)\right.\right.$\newline$ \left.\left.
\hspace{1.5cm}-378 \sqrt{3} \sin\left(\frac{1}{12} (9+5 n) \pi \right)+378 \sin\left(\frac{1}{12}(3+13 n) \pi \right) -792 \sin\left(\frac{1}{4} (\pi +7 n \pi )\right)\right]\right\rbrace$
 \\
\hline
10&$ S_{(0,0)}^{(10)}(n)=  0.37451171875-(0.21338834765) \cos\left(\frac{n \pi }{4}\right)-0.036611652352 \cos\left(\frac{3
n \pi }{4}\right)+0.088388347649\left(\sin\left(\frac{ n \pi }{4}\right)+\sin\left(\frac{3
n \pi }{4}\right)\right)$\newline$~~~~~~~~~~~~~~~ -0.12451171875\left( \cos\left(\frac{n \pi }{2}\right)-\sin\left(\frac{n \pi }{2}\right)\right)$.
\\
\hline
12&$ S_{(0,0)}^{(12)}(n)=  \frac{1}{16} \left\lbrace 8-0.1953125 e^{-i n \pi } \left(1\, +e^{i n \pi }\right)^2 \left[\cos\left(\frac{3 n \pi }{8}\right)+1.1\cos\left(\frac{25
n \pi }{24}\right)+1.1 \cos\left(\frac{41 n \pi }{24}\right)\right]^2 -0.1667096466\left[\cos\left(\frac{n \pi }{8}\right)-\right.\right.$\newline$\left.
\left.\hspace{1.6cm}\cos\left(\frac{7 n \pi }{8}\right)-0.4142135623731 \left(\sin\left(\frac{n \pi }{8}\right)+\sin\left(\frac{7 n \pi }{8}\right)\right)+ (1-\cos(n\pi))\left( -0.944591414368 \cos\left(\frac{29 n \pi }{24}\right)-0.15540858564 \right.\right.\right.$\newline$
\left.\left.\left. \hspace{1.55cm}\cos\left(\frac{13 n \pi }{24}\right)-1.180445403469\sin\left(\frac{13
n \pi }{24}\right)+0.724810484858 \sin\left(\frac{29 n \pi }{24}\right)\right)\right]^2\right\rbrace$.
\\
\hline
\end{tabular}
  \label{Table:LinearEntropy}
\end{table*}

\begin{table*}[t!]
  \centering
 \caption{The  eigenvalues and linear entropy ($S(n)$) for the initial state $\otimes^N\ket{+}$ and various values of  $N$.}
 \renewcommand{\arraystretch}{2.4}
\begin{tabular}{|p{0.35cm}|p{7.2cm}|p{9.65cm}|  }
 \hline
$N$ &\hspace{2.6cm} Eigenvalues & \hspace{4.5cm} $S(n)$   \\
\hline
8 &$ \frac{1}{2}\pm\frac{1}{48}\left[10 \cos\left(\frac{n \pi }{4}\right)+7 \left(\cos\left(\frac{5 n \pi }{12}\right)+\cos\left(\frac{11
n \pi }{12}\right)\right)\right]$& $S_{(\pi/2,-\pi/2)}^{(8)}(n)=\frac{1}{2}\left[1-\frac{1}{144} \cos^2\left(\frac{n \pi }{4}\right) \left(5+7 \cos\left(\frac{2 n \pi }{3}\right)\right)^2\right]$.
 \\
\hline
10&$ \frac{1}{2}\pm\frac{1}{64}\left[\sqrt{258+510 \cos\left(\frac{n \pi }{2}\right)+256 \cos(n \pi )}\right]$&$S_{(\pi/2,-\pi/2)}^{(10)}(n)=\frac{1}{2}\left\lbrace 1-\frac{1}{4096}\left[17 \left(1+\cos(n \pi )\right)+\sqrt{2} \left(1-\cos(n \pi )\right)\right.\right.$\newline$ \left.\left.\hspace{2cm}+30 \cos\left(\frac{ n \pi
}{2}\right)\right]^2\right\rbrace$.
\\
\hline
12&$\frac{1}{2}\pm\frac{1}{64} \left[11 \cos\left(\frac{n \pi }{12}\right)+10 \cos\left(\frac{3 n \pi }{4}\right)+11
\cos\left(\frac{17 n \pi }{12}\right)\right]$&$S_{(\pi/2,-\pi/2)}^{(12)}(n)=\frac{1}{2}-\frac{1}{2048}\left[11 \cos\left(\frac{n \pi }{12}\right)+10 \cos\left(\frac{3 n \pi }{4}\right)+11 \cos\left(\frac{17
n \pi }{12}\right)\right]^2$.
\\
\hline
\end{tabular}
  \label{Table:LinearEntropy2}
\end{table*}
\section{Results for general $N$ qubits.}
\label{sec:example-section5}  
\begin{figure}[t!]\vspace{0.4cm}
\includegraphics[width=0.47\textwidth]{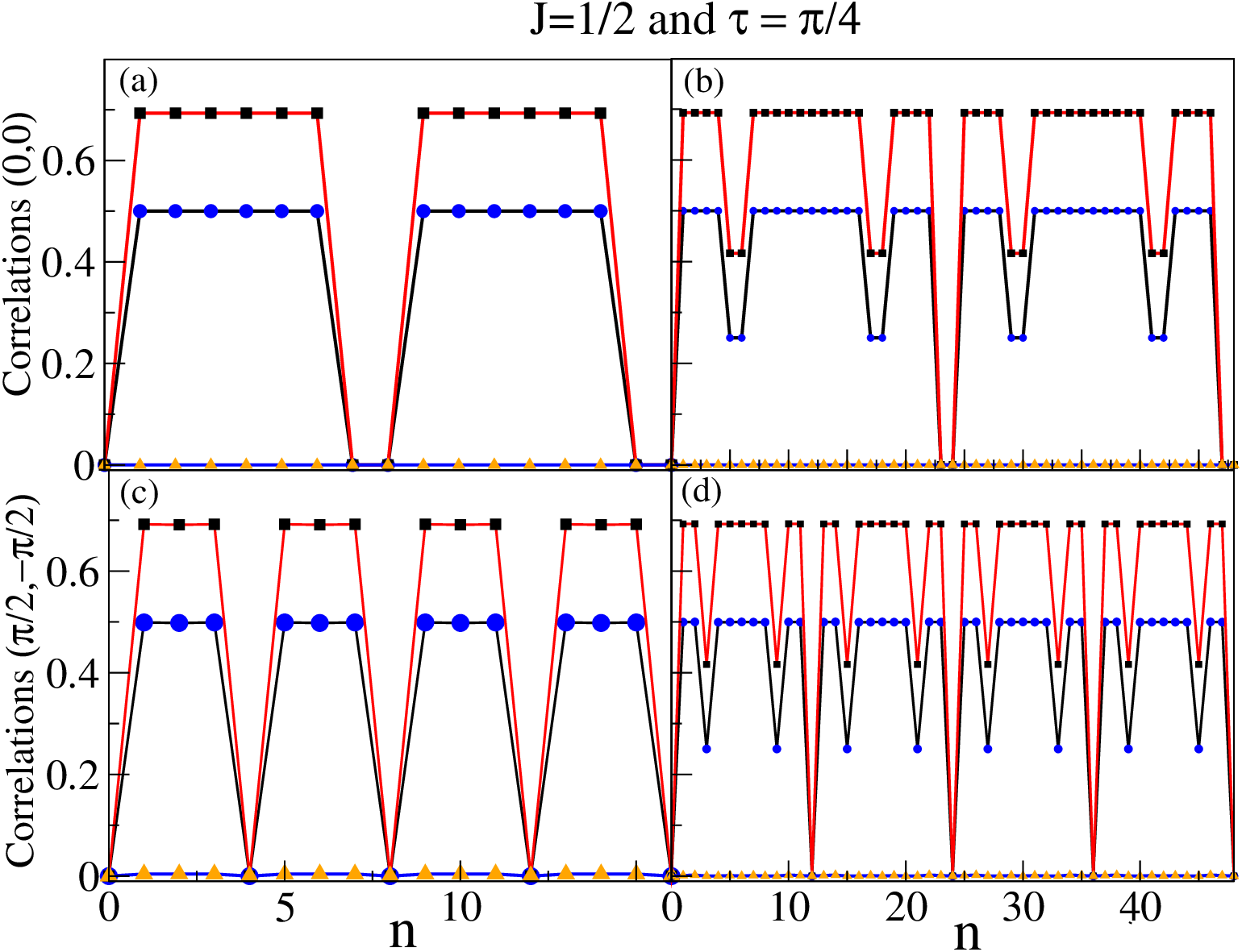}
\caption{Correlations using linear entropy (circles), entanglement entropy (squares), and concurrence (upper triangle) are plotted for  ((a) $1002$ qubit, (b) $1000$ qubit) for the  initial states $\otimes ^N {\ket{0}}_y$ and
((c) $1002$ qubit, (d) $1000$ qubit)   for the  initial states $\otimes ^N {\ket{+}}_y$ }
\label{fig:correlations4}
\end{figure}

\begin{figure}[t!]\vspace{0.4cm}
\includegraphics[width=0.47\textwidth]{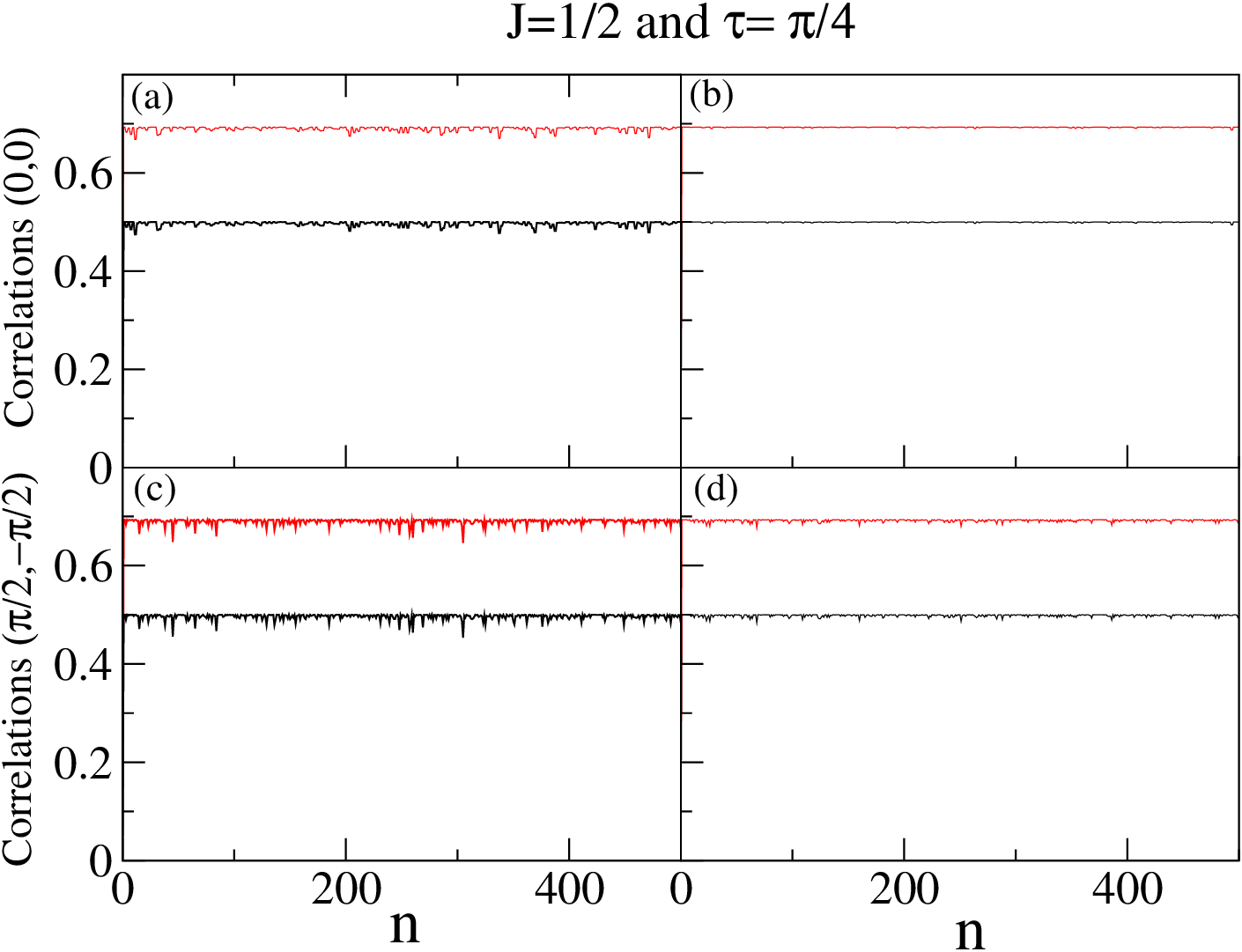}
\caption{(Color online) Correlations using linear entropy (black) and  entanglement entropy (red)  are plotted for
(a)  $101$ qubit, (b) $203$ qubit for  the initial states $\otimes ^N {\ket{0}}_y$ and
(c) $101$ qubit, (d) $203$ qubit  for the initial states $\otimes ^N {\ket{+}}_y$}
\label{fig:correlations6}
\end{figure}

Using our procedure, in principle  one can solve for the case of  any finite $N$, but  various expressions for the eigensystem, floquet operator, and its $n^{th}$ power become more cumbersome. Therefore, for $N>12$
we restrict ourselves  only to  the numerical methods. Here, we use the same signatures used to establish integrability for
$N=6,~8,~10$ and $12$ qubits. Surprisingly, we find exactly the same behavior of these  signatures for the said parameters ($J=1/2$ and $\tau=\pi /4$). Particularly, we find that the  entanglement dynamics is periodic such  that  for the initial states  $\otimes ^N {\ket{0}}_y$  ($\otimes ^N {\ket{+}}_y$)  and  $N=4m+2$ the  period is  $8$  ($4$) respectively, whereas for $N=4m+4$  it is $24$($12$), where $m \in \left\lbrace 0,1,2,...\right\rbrace$. The results are  shown in Fig. \ref{fig:correlations4}.  We numerically find that $C_{\mbox{max}}$ tends to zero  as $N$ increases for both the initial states.  Thus, the entanglement becomes  multipartite in  nature with  $N$. For any odd $N>11$ with the same parameter, we find  that the entanglement dynamics is not periodic at all, which is shown in Fig. \ref{fig:correlations6}. We observed the same non-periodic behavior for $n$ up to $1000$ (results not shown here).

\begin{figure}[t!]\vspace{0.2cm}
\includegraphics[width=0.47\textwidth]{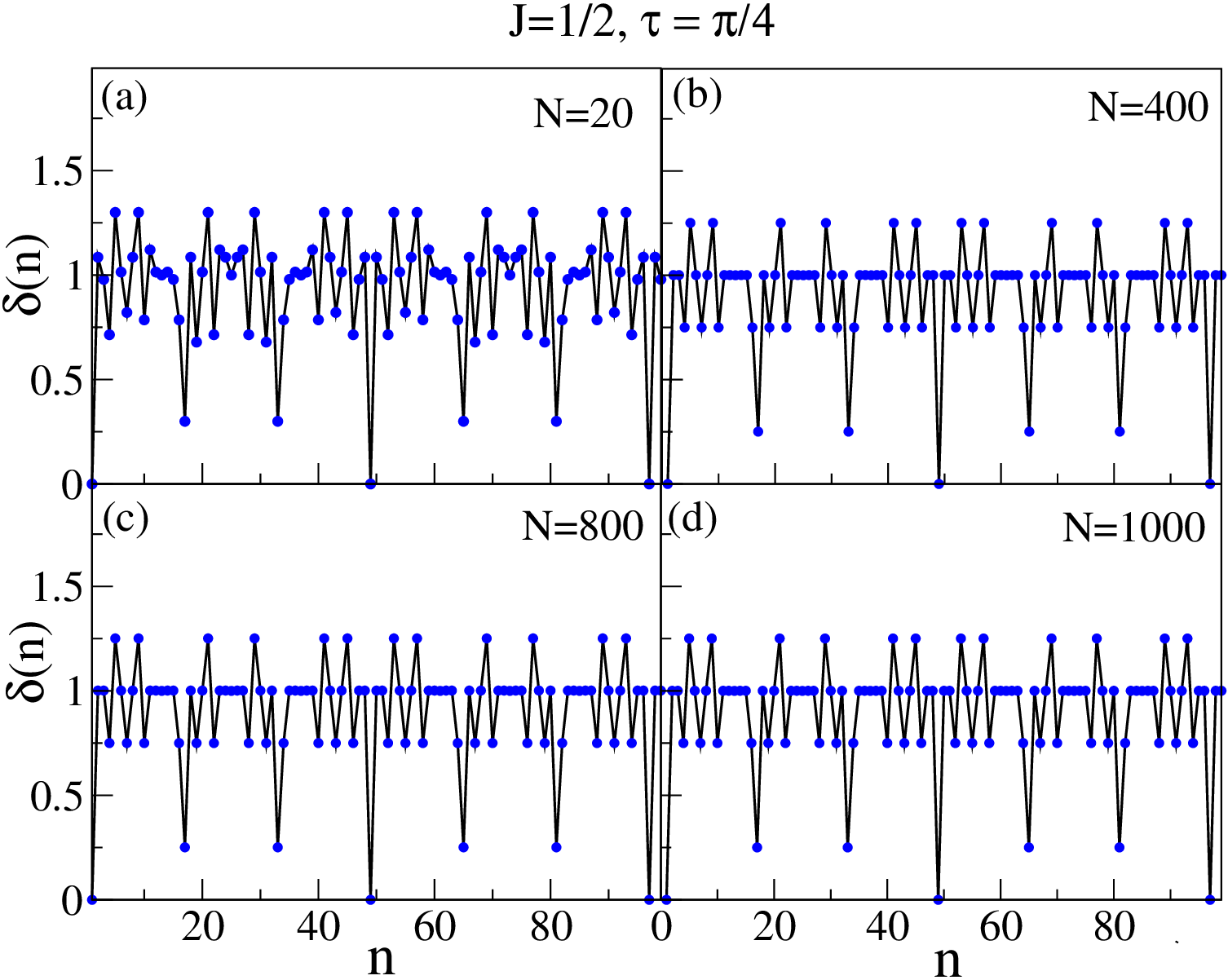}
\caption{The deviation $\delta(n)$ for various values of $N$ (integrable case).}
\label{fig:correlations5}
\end{figure}
\begin{figure}[t!]\vspace{0.2cm}
\includegraphics[width=0.47\textwidth]{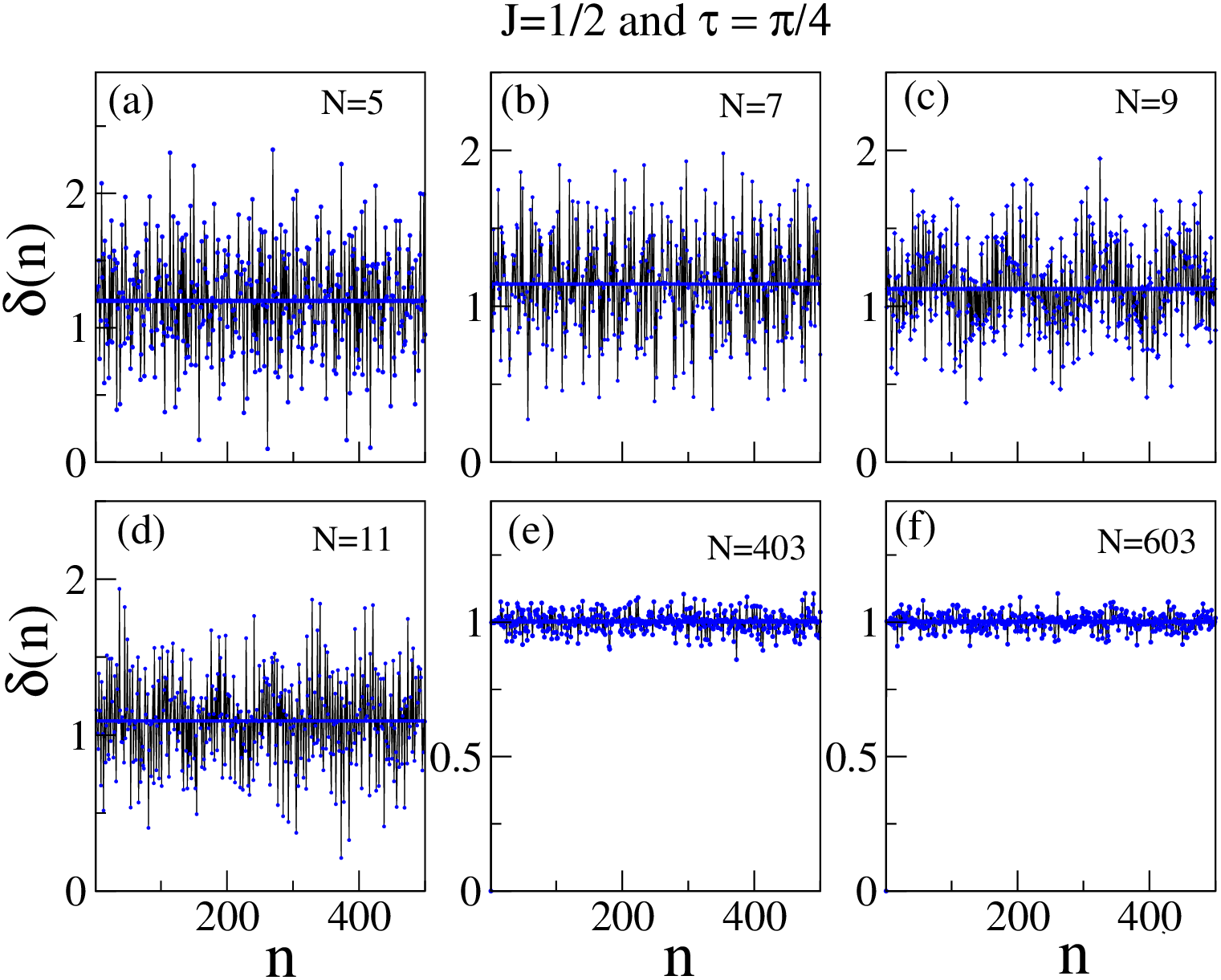}
\caption{The deviation $\delta(n)$ for  various values of $N$ (nonintegrable case).}
\label{fig:correlations7}
\end{figure}

We also study the operator dynamics using a numerical method. We define
$\delta(n)=\sum_{i,j}(A_{i,j} A^*_{i,j})/2N$, where, $A_{i,j}={\mathcal U}^n_{i,j}-{\mathcal U}_{i,j}$ \cite{sharma2024exactly}. Numerically it is observed that the
division by $2N$ ensures the average of $\delta(n)$ is one. For any $n$, $\delta(n)=0$  if and only if ${\mathcal U}^n={\mathcal U}$, which confirms
the periodic nature of $\mathcal{U}$ implying integrability. The definition above  makes sure that $\delta(n)$ is invariant under the global unitary transformation (for proof see Supplemental Material \cite{supplementry2023}). A similar definition in previous work was lacking this invariance, although it can detect the periodic behavior correctly \cite{sharma2024exactly}.
For our parameter results are plotted in Fig. \ref{fig:correlations5}. We have checked periodicity for $N$ as large as $1000$  and   $n$ up to $10000$ (results are not shown here). The observed period is $48$  for all  even $N>6$.  The same periodic nature could not be observed for (a) even $N$ and $J$ $\notin$ $\left\lbrace 1,1/2\right\rbrace$, (b) odd $N$ and $J$ $\neq1$. Which is shown in the
Figs. \ref{fig:correlations5} and \ref{fig:correlations7} (results for odd $N$ are shown in Fig. \ref{fig:correlations7}).
\begin{figure}[t!]\vspace{0.2cm}
\includegraphics[width=0.47\textwidth]{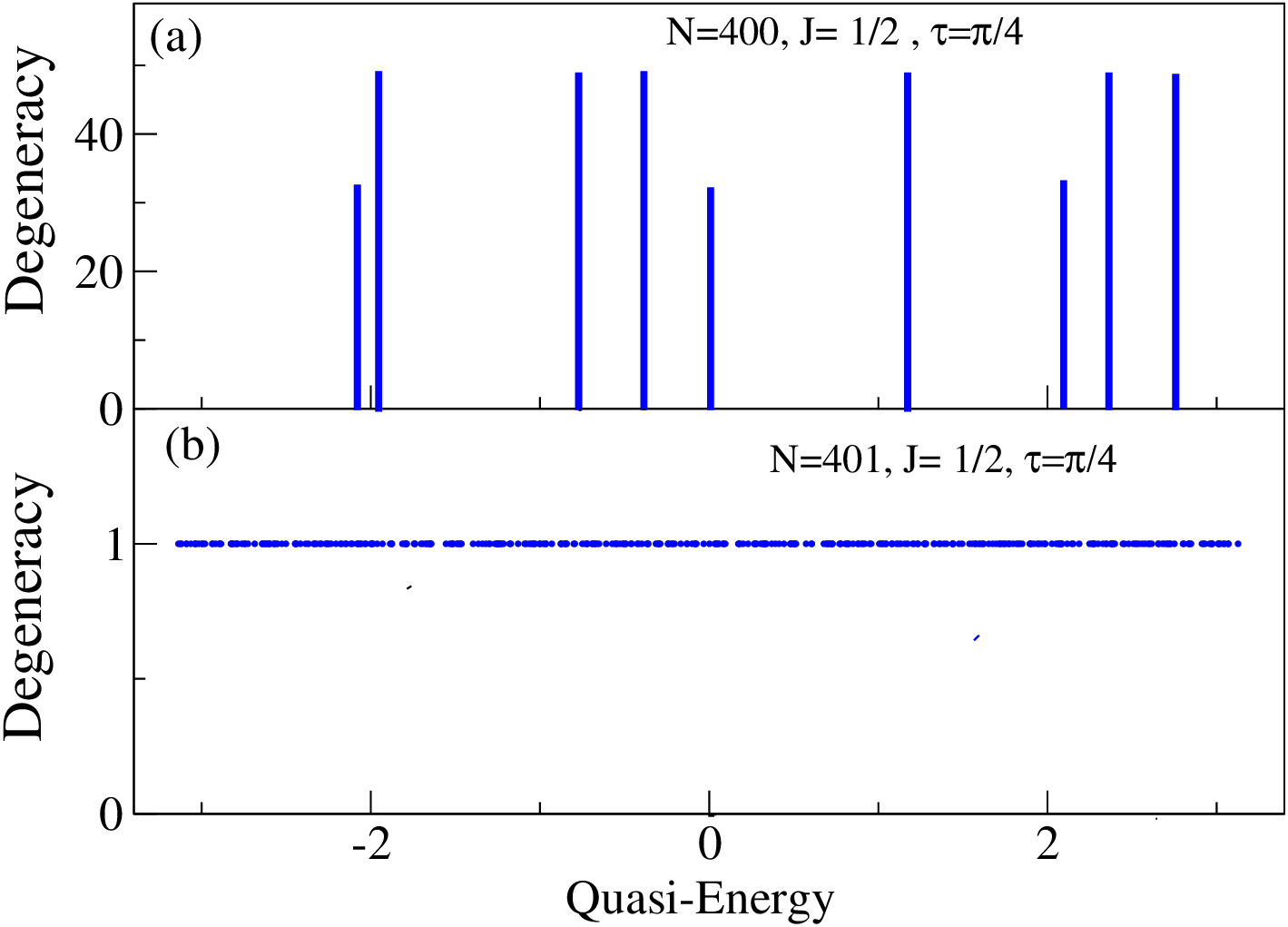}
\caption{Degeneracy of the quasienergies of $\mathcal{U}$ for (a) integrable case and (b) nonintegrable case.}
\label{fig:correlations8}
\end{figure}
\begin{figure}[t!]\vspace{0.2cm}
\includegraphics[width=0.47\textwidth]{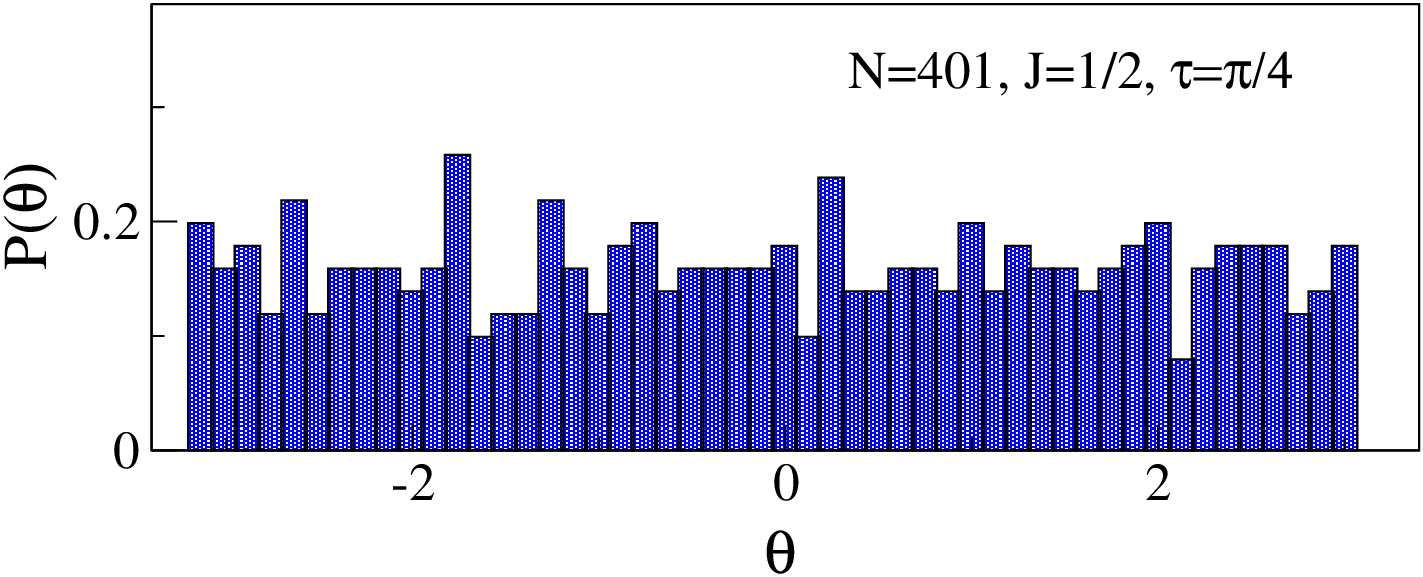}
\caption{ Probability distribution of quasi-energies}
\label{fig:correlations811}
\end{figure}
 Additional signature of integrability can be found from the eigenangle spectrum of $\mathcal U$~\cite{LevelCrossingsYuzbashyan2008,yuzbashyan2013quantum,naik2019controlled}.
 We find that it is  highly degenerate to take values from the finite set
 $\{0,\pm \pi/4, \pm \pi/8, \pm 3 \pi/4,\pm 3\pi/8 ,\pm 5\pi/8,\pm 7\pi/8 \}$ (see (a) of
 Fig. \ref{fig:correlations8}), whereas for odd $N$, $J=1/2$ and $\tau=\pi/4$, we find that degeneracy disappears
 (see (b) Fig.~\ref{fig:correlations8}).
 In fact, we find that the eigenangles are almost uniformly distributed (see
 Fig.~\ref{fig:correlations811}). Thus, with these signatures, we can readily  conjecture that the system is quantum integrable  for any  even-$N > 12$, whereas for an odd $N$, it isn't.

\section{Conclusions}\label{sec:example-section6}
In this paper, we have generalized the recently introduced many-body model \cite{sharma2024exactly}.
The model consisted of qubits with all-to-all ising interaction. Keeping other parameters same, we have generalized this
interaction to $J$, with $J=1$ reducing to the recent case \cite{sharma2024exactly}. We have studied QI in the generalized
version. We have found that apart from $J=1$, QI also exists for $J=1/2$  and {\it only} for an even number of qubits.
We have analytically calculated the eigensystem, time evolution of the unitary operator, reduced density matrix, and
entanglement dynamics for an even number of qubits ranging from $6$ to $12$. We have used linear entropy, von Neumann
entropy, and concurrence to measure the entanglement dynamics for various initial unentangled states. The behavior of
entanglement dynamics and unitary operator dynamics are in accordance with QI systems with ising interaction. Due to
fairly large expressions and complexity, we restricted ourselves to the numerical simulations for any $N>12$ qubits.
We have introduced a quantity, $\delta(n)$, to investigate the periodic nature of the time-evolution operator numerically.
We have shown that it remains invariant under the global unitary transformation. If this quantity is zero then it indicates
the periodic nature of the unitary operator. We have observed that the entanglement dynamics and floquet operator dynamics
show periodic behavior for any even $N$, while for any odd $N$, it does not. We have also observed the high degeneracy in
the spectra for an even number of qubits.  While for odd $N$,  the degeneracy in the spectra disappears. We observed that
its pairwise concurrence tends to zero with $N$, indicating the multipartite nature of entanglement. Thus, we can very well
conjuncture, that the system also exhibits quantum integrability for parameters  $J=1/2$ and $\tau=\pi/4$ for any even number
of qubits, whereas it is absent for any odd number of qubits.

In recent work, a generalization of the Hubbard-Stratonovich
 transformation was employed to get an exact  analytical solution for quantum-strong-long-range Ising chains \cite{roman2023exact}. Our model can be explored further in this direction. The experimental verification of our results, (for the smaller number of qubits) can be conducted in various setups such as
 NMR \cite{Krithika2019}, superconducting qubits \cite{neill2016ergodic} and laser-cooled atoms \cite{Chaudhury09}, where the QKT has been implemented,
whereas for a large number of qubits (of the order of $100$s), one can use ion traps \cite{monroe2021programmable,defenu2023long}.
The validity of our conjecture can  be established in this setup. While our study has successfully identified integrability in this model
for the specified values of parameters  $J=1, 1/2$ and  $\tau=\pi/4$. We hope our work raises several open questions. Few of them are as follows:
1) As the number of qubits increases, entanglement becomes multipartite in nature. One such measure is the Meyer and Wallach Q measure,
which is very much related to linear entropy and we have studied it here. Further analysis using various multipartite
entanglement measures can be an interesting direction to pursue
\cite{hauke2016measuring,szalay2015multipartite,li2024parametrized,beckey2021computable};
~2) Are there other possible values of $J$ that exhibit integrability within this framework?;
~3) Our findings encourage further exploration to search for other integrable systems in which all-to-all interaction is present.

\section{Acknowledgment}
The authors are grateful to the Department of Science and Technology (DST) for their generous financial support, making this research possible through the sanctioned project SR/FST/PSI/2017/5(C) to the Department of Physics of VNIT, Nagpur.

\let\oldaddcontentsline\addcontentsline
\renewcommand{\addcontentsline}[3]{}
\bibliography{reference2023,refrence11,reference22013}
\let\addcontentsline\oldaddcontentsline

\end{document}


\let\oldaddcontentsline\addcontentsline
\renewcommand{\addcontentsline}[3]{}

\let\addcontentsline\oldaddcontentsline

\newpage\clearpage

\onecolumngrid
\setlength{\belowcaptionskip}{0pt}
\setcounter{secnumdepth}{4}
\setcounter{tocdepth}{5}
\setcounter{page}{1}
\setcounter{figure}{0}
\setcounter{table}{0}
\setcounter{equation}{0}
\setcounter{section}{0}
\renewcommand{\thepage}{S\arabic{page}}
\renewcommand{\thesection}{S\arabic{section}}
\renewcommand{\thesection}{S\arabic{section}}
\renewcommand{\thetable}{S\arabic{table}}
\renewcommand{\thefigure}{S\arabic{figure}}
\renewcommand{\theequation}{S\arabic{equation}}
\renewcommand{\bibnumfmt}[1]{[S#1]}
\renewcommand{\citenumfont}[1]{S#1}
\newcounter{mytable}
\makeatletter
\@addtoreset{table}{mytable}
\makeatother
\renewcommand{\thetable}{S\Roman{table}}
\counterwithout*{equation}{section}

\titleformat{\paragraph}[runin]{\normalfont\normalsize\bfseries}{}{0em}{}[.]
\titlespacing*{\paragraph}{0pt}{3.25ex plus 1ex minus .2ex}{1em}

\setlength{\belowcaptionskip}{0pt}

\begin{center}
    \large\textbf{Supplemental Materials for\\``\textit{
    Signatures of Integrability and Exactly Solvable Dynamics in an Infinite-Range Many-Body Floquet Spin System''}}
\end{center}

{\hypersetup{linkcolor=black}\tableofcontents}

\newpage

In this supplement, we discussed the detailed analysis of the analytical solutions for systems consisting of $6$, $8$,
$10$, and $12$ qubits. In the first four sections, we provided  the analytical calculations of the eigensystem, the time evolution
of the unitary operator, and various entanglement measures such as linear entropy, entanglement entropy, and concurrence.
Within these sections, we show the periodic nature of both the unitary operator and entanglement measures. In the final
section (Sec. \ref{sec:example-section}), we presented a proof of the invariance of the quantity $\delta(n)$ under the global
unitary transformations.
\section{Exact analytical solution for six-qubits case}
This section provides  a detailed description of the analytical solutions which has been discussed in the main text. In accordance with
Eq. $4$, from the main text, the unitary operator acting on a system of $6$-qubits ($j=3$) is expressed  as follows:
\begin{eqnarray}\label{Eq:5q87}
\begin{split}
  \mathcal{U} &= \exp\left[-i \frac{\pi}{8}(\sigma_1^z \sigma_2^z +\sigma_1^z \sigma_3^z +\sigma_1^z \sigma_4^z+\sigma_1^z \sigma_5^z
 +\sigma_1^z \sigma_6^z+\sigma_2^z \sigma_3^z+\sigma_2^z \sigma_4^z+\sigma_2^z \sigma_5^z+\sigma_2^z \sigma_6^z+\sigma_3^z \sigma_4^z
 +\sigma_3^z \sigma_5^z \right.\\
 & \hspace{1.4cm}\left. + \sigma_3^z \sigma_6^z+\sigma_4^z \sigma_5^z  +
 \sigma_4^z \sigma_6^z+\sigma_5^z \sigma_6^z) \right]\times
  \exp \left[-i \frac{\pi}{4} (\sigma_1^y+\sigma_2^y+\sigma_3^y+\sigma_4^y+\sigma_5^y+\sigma_6^y ) \right],
 \end{split}
\end{eqnarray}
where  $\sigma_l^z$ and $\sigma_l^y$  are  Pauli operators for the $l^{th}$ qubit. The solution to the 6-qubit case proceeds from the general observation that (for proof, see supplemental material of Ref.~{\cite{sharma2024exactly}}),
\begin{equation}
[\mathcal{U},\otimes_{l=1}^{2j} \sigma^y_l]=0,
\end{equation}
 {\it i.e.,} there is
an ``up-down" or parity symmetry \cite{ExactDogra2019}. Here, we are confined to permutation symmetric subspace  of dimension $(2j+1=7)$ within the  total
$2^{2j}$= $2^{6}$= $64$ dimensional Hilbert space. The basis vectors are given as follows:
\begin{eqnarray}
\ket{\phi_0^{\pm}}&=&\frac{1}{\sqrt{2}}(\ket{w_0} \mp \ket{\overline{w_0}}),\label{Eq:5q91}\\
\ket{\phi_1^{\pm}}&=&\frac{1}{\sqrt{2}}(\ket{w_1} \pm  \ket{\overline{w_1}}),\label{Eq:5q92}\\
\ket{\phi_2^{\pm}}&=&\frac{1}{\sqrt{2}}(\ket{w_2} \mp  \ket{\overline{w_2}}),\label{Eq:5q93}\\
\ket{\phi_3^{+}}&=&\frac{1}{\sqrt{20}}\sum_\mathcal{P}\ket{000111} \label{Eq:5q94},
\end{eqnarray}
where $\ket{w_0}=\ket{000000}$,
$\ket{\overline{w_0}}=\ket{111111}$, $\ket{w_1}=\frac{1}{\sqrt{6}} \sum_\mathcal{P} \ket{000001}_\mathcal{P}$, $\ket{\overline {w_1}}=\frac{1}{\sqrt{6}} \sum_\mathcal{P} \ket{011111}_\mathcal{P}$, $\ket{w_2}$=$\frac{1}{\sqrt{15}} \sum_\mathcal{P} \ket{000011}_\mathcal{P}$ and $\ket{\overline{w_2}}$=$\frac{1}{\sqrt{15}} \sum_\mathcal{P} \ket{001111}_\mathcal{P}$. The unitary operator $\mathcal{U}$ can be expressed in terms of two blocks  ($\mathcal{U}_{\pm}$), each consisting of  matrices with different dimensions and  is given by,
 \begin{equation}
 \mathcal{U} = \begin{pmatrix}
            \mathcal{U}_{+} & 0 \\ 0 & \mathcal{U}_{-}
            \end{pmatrix},
\end{equation}
where the  blocks $\mathcal{U}_{\pm}$ are $4\times 4 \left(3\times3\right)$ dimensional matrices  and $0$ is null  rectangular matrices of  dimensions $3\times 4$. The $\mathcal{U}_{+}\left(\mathcal{U}_{-}\right)$ written in the basis of positive (negative) parity states  $\left\lbrace \phi_0^{+},\phi_1^{+},\phi_2^{+},\phi_3^{+}\right\rbrace\left(\left\lbrace \phi_0^{-},\phi_1^{-},\phi_2^{-}\right\rbrace\right)$ is given as follows:
\begin{equation}\label{Eq:5q88}
\renewcommand{\arraystretch}{1.4}
\mathcal{U}_{+} = \begin{pmatrix}
            \bra{\phi_0^{+}}\mathcal{U}\ket{\phi_0^{+}}& \bra{\phi_0^{+}}\mathcal{U}\ket{\phi_1^{+}}&\bra{\phi_0^{\pm}}\mathcal{U}\ket{\phi_2^{+}}&\bra{\phi_0^{+}}\mathcal{U}\ket{\phi_3^{+}} \\  \bra{\phi_1^{+}}\mathcal{U}\ket{\phi_0^{+}} & \bra{\phi_1^{+}}\mathcal{U}\ket{\phi_1^{+}}&\bra{\phi_1^{+}}\mathcal{U}\ket{\phi_2^{+}}&\bra{\phi_1^{+}}\mathcal{U}\ket{\phi_3^{+}}\\\bra{\phi_2^{+}}\mathcal{U}\ket{\phi_0^{+}}&\bra{\phi_2^{+}}\mathcal{U}\ket{\phi_1^{+}}&\bra{\phi_2^{+}}\mathcal{U}\ket{\phi_2^{+}}&\bra{\phi_2^{+}}\mathcal{U}\ket{\phi_3^{+}}\\\bra{\phi_3^{+}}\mathcal{U}\ket{\phi_0^{+}}&\bra{\phi_3^{+}}\mathcal{U}\ket{\phi_1^{+}}&\bra{\phi_3^{+}}\mathcal{U}\ket{\phi_2^{+}}&\bra{\phi_3^{+}}\mathcal{U}\ket{\phi_3^{+}}
\end{pmatrix}    \mbox{ and}
\end{equation}
\begin{equation}\label{Eq:5q89}
\renewcommand{\arraystretch}{1.4}
\mathcal{U}_{-} = \begin{pmatrix}
            \bra{\phi_0^{-}}\mathcal{U}\ket{\phi_0^{-}}& \bra{\phi_0^{\pm}}\mathcal{U}\ket{\phi_1^{-}}&\bra{\phi_0^{-}}\mathcal{U}\ket{\phi_2^{-}} \\ \bra{\phi_1^{-}}\mathcal{U}\ket{\phi_0^{-}} & \bra{\phi_1^{-}}\mathcal{U}\ket{\phi_1^{-}}&\bra{\phi_1^{-}}\mathcal{U}\ket{\phi_2^{-}}\\ \bra{\phi_2^{-}}\mathcal{U}\ket{\phi_0^{-}}&\bra{\phi_2^{-}}\mathcal{U}\ket{\phi_1^{-}}&\bra{\phi_2^{-}}\mathcal{U}\ket{\phi_2^{-}}\\
            \end{pmatrix}.
\end{equation}
 An unitary operator describes the evolution of states over time. To understand the time evolution of observables, one needs to compute the $n^{th}$ power of the unitary operator. In this context, due to block diagonalization, it becomes easier to calculate its $n^{th}$ power. This can be expressed as follows:
\begin{equation}
\mathcal{U}^n = \begin{pmatrix}
            \mathcal{U}_{+}^n & 0 \\ 0 & \mathcal{U}_{-}^n
            \end{pmatrix}.
\end{equation}
The block operators  $\mathcal{U}_{\pm}$ are explicitly found by using the
Eqs. (\ref{Eq:5q87}), (\ref{Eq:5q91}), (\ref{Eq:5q92}), (\ref{Eq:5q93}), (\ref{Eq:5q94}), (\ref{Eq:5q88}) and (\ref{Eq:5q89}) as follows:
 \begin{equation}
\mathcal{U_+}= \frac{-e^{\frac{i \pi }{8}}}{2\sqrt{2}} \left(
\begin{array}{cccc}
 0 &  \sqrt{3} & 0 & \sqrt{5} \\
 \sqrt{3}~e^{\frac{i \pi }{4}} & 0 & \sqrt{5}~e^{\frac{i \pi }{4}} & 0 \\
 0 &  \sqrt{5} & 0 & - \sqrt{3} \\
 -\sqrt{5}~e^{\frac{i \pi }{4}} & 0 & \sqrt{3}~e^{\frac{i \pi }{4}}  & 0 \\
\end{array}
\right)\mbox{and}
 \end{equation}
 \begin{equation}
\mathcal{U_-}= \frac{ e^{\frac{i \pi }{8}}}{4} \left(
\begin{array}{ccc}
 1 & 0 &  \sqrt{15} \\
 0 & 4~e^{\frac{i \pi }{4}} & 0 \\
  \sqrt{15} & 0 & -1 \\
\end{array}
\right).
 \end{equation}
The eigenvalues of the $\mathcal{U_+} \left( \mathcal{U_-}\right)$  are $\left\{(-1)^{1/4}, -(-1)^{1/4}, (-1)^{3/4}, -(-1)^{3/4}\right\}\left(\left\{(-1)^{3/8}, (-1)^{1/8}, -(-1)^{1/8}\right\}\right)$  and the eigenvectors are  $\left\lbrace\left[\sqrt{\frac{3}{5}},\sqrt{\frac{3}{5}},\frac{(-1)^{3/8}}{2}  \sqrt{\frac{5}{2}},-\frac{(-1)^{3/8}}{2}  \sqrt{\frac{5}{2}}\right]^T, \left[-2(-1)^{1/8} \sqrt{\frac{2}{5}},2 (-1)^{1/8} \sqrt{\frac{2}{5}},0,0\right]^T, \left[1,1,-\frac{(-1)^{3/8}}{2}  \sqrt{\frac{3}{2}},\frac{(-1)^{3/8}}{2}
\sqrt{\frac{3}{2}}\right]^T,\left[0,0,1,1\right]^T  \right\rbrace $\\$  \left(\left\lbrace\left[-\sqrt{\frac{3}{5}},\sqrt{\frac{5}{3}},0\right]^T, \left[0,0,1\right]^T, \left[1,1,0\right]^T\right\rbrace\right)$. The $n$th time evolution of the  blocks $\mathcal{U_{\pm}}$ is given as follows:
 \begin{equation}
\mathcal{U}_{+}^n= {\frac{e^{\frac{i n \pi }{4}}}{16}\left[
\begin{array}{cccc}
 {a_n}  \left(3+5 e^{\frac{i n \pi }{2}}\right)  & 2\sqrt{{6}}~b_n{e^{\frac{7i  \pi }{8}}}
  & 4{i \sqrt{15} e^{\frac{3i n \pi }{4}}}  \sin\left(\frac{n \pi }{4}\right)\cos\left(\frac{n \pi }{2}\right) & 2\sqrt{{10}}~b_n~{e^{(\frac{ i n \pi }{2}+\frac{3i  \pi }{8})}}   \\
 -2\sqrt{{6}}~b_n~e^{\frac{i  \pi }{8}}   & 8a_n   &-2\sqrt{{10}}~b_n~ {e^{\frac{i  \pi }{8}}}    & 0 \\
 {4i \sqrt{15} e^{\frac{3i n \pi }{4}}}  \sin\left(\frac{n \pi }{4}\right)\cos\left(\frac{n \pi }{2}\right) &2\sqrt{{10}}~b_n{e^{\frac{7i  \pi }{8}}}
 & a_n \left(5+3 e^{\frac{i n \pi
}{2}}\right)  & -2\sqrt{{6}}~b_n~{e^{(\frac{ i n \pi }{2}+\frac{3i  \pi }{8})}}    \\
 -2\sqrt{{10}}~b_n~{e^{(\frac{ i n \pi }{2}+\frac{5i  \pi }{8})}}     & 0 & 2\sqrt{{6}}~b_n~{e^{(\frac{ i n \pi }{2}+\frac{i  \pi }{8})}}
  & {8a_n~e^{\frac{i n \pi }{2}}}  \\
\end{array}
\right]}
 \end{equation}
 \begin{equation}
 \mbox{and}~~~~
 \mathcal{U}_{-}^n =\frac{e^{\frac{i n \pi }{8}}}{8}{\left[
\begin{array}{ccc}
 5 + 3~e^{i n \pi } & 0 &  \sqrt{15}\left( 1-e^{{ i n \pi }}\right) \\
 0 & 8 ~ e^{\frac{ i n \pi }{4}} & 0 \\
 \sqrt{15}\left( 1-e^{{ i n \pi }}\right) & 0 & 3 + 5~ e^{i n \pi } \\
\end{array}
\right]},
 \end{equation}
where $a_n$=$1+e^{i n \pi }$ and $b_n$=$1-e^{i n \pi }$. The time periodicity  of $\mathcal{U}_{+}^n$ and  $\mathcal{U}_{-}^n$  can be  shown to be  $8$ and $16$ respectively. Consequently,  the unitary operator $\mathcal{U}^n$ for $6$ qubits shows periodicity with a period of  $16$ i.e. $\mathcal{U}^n$=$\mathcal{U}^{n+16}.$
 \subsection{Initial state $\otimes^6\ket{0}$= $\ket{\theta_0 =0,\phi_0 =0}$}
The state $\ket{\psi_n}$ can be calculated by $n$ implementations  of  the unitary operator $\mathcal{U}$  on this initial state $\otimes^6\ket{0}$ as follows:
\begin{eqnarray}\nonumber
\ket{\psi_n} =\mathcal{U}^n\ket{000000}&=& \frac{1}{\sqrt{2}}\left(\mathcal{U}_{+}^n\ket{\phi_0^+}+\mathcal{U}_{-}^n\ket{\phi_0^-}\right)\\
&=& \frac{1}{\sqrt{2}}\left(\bar{a}_1\ket{\phi_0^+}+\bar{a}_2\ket{\phi_1^+}+\bar{a}_3\ket{\phi_2^+}+\bar{a}_4\ket{\phi_3^+}+\bar{a}_5\ket{\phi_0^-}+\bar{a}_6\ket{\phi_2^-}\right),
\end{eqnarray}
where the coefficients are,
\begin{eqnarray}\nonumber
\bar{a}_1&=&\frac{e^{\frac{i n \pi }{4}}}{16}\left[ \left(3+5~ e^{\frac{i n \pi }{2}}\right) \left(1+e^{i n \pi }\right)\right],~ \bar{a}_2 = \frac{\sqrt{{6}}}{8}\left[\left(-1+e^{i n \pi }\right)e^{(\frac{i \pi }{8}+\frac{i n\pi }{4})} \right],~~ \bar{a}_3=\frac{i \sqrt{15}~ e^{i n \pi }}{8}  \left[\sin\left(\frac{n \pi }{4}\right)-\sin\left(\frac{3 n \pi }{4}\right)\right],\\ \nonumber \bar{a}_4&=&\frac{\sqrt{10}}{8}\left[\left(-1+e^{i n \pi }\right) e^{\left(\frac{5i \pi }{8}+\frac{3 i n \pi}{4} \right)}\right], ~~~\bar{a}_5=\frac{e^{\frac{in \pi}{8}}}{8} \left(5+3~e^{{i n \pi }}\right)~ \mbox{and} ~~~\bar{a}_6= \frac{\sqrt{15}}{8}\left[e^{\frac{i n \pi }{8}}\left(1-e^{{i n \pi }}\right)\right].
\end{eqnarray}\\
By using the state $\ket{\psi_n}$ of the system, the one-and two-qubit reduced density matrices (RDM)
$\rho_1(n)=\text{tr}_{2,3,4,5,6} (|\psi_n \rangle \langle \psi_n |)$ and $\rho_{12}(n)=\text{tr}_{3,4,5,6} (|\psi_n \rangle \langle
\psi_n |)$ are obtained. The linear entropy, entanglement entropy (Von Neumann entropy), and concurrence are used as measures of entanglement in quantum systems.
\subsubsection{\bf{The linear entropy}}
 The linear entropy serves as
a measure of the degree of mixedness or impurity within a quantum state. It quantifies the entanglement in a bipartite (two-party)
quantum system \cite{wootters98} and it is defined as $1-\text{tr}\left[\rho_1^2(n)\right]$. The  single qubit RDM, $\rho_1(n)$, is given as follows:
\begin{equation}\label{von2}
\rho_1(n)=\frac{1}{4}\left(
\begin{array}{cc}
 2+a_n & w_n \\
 w_n^* & 2-a_n \\
\end{array}
\right),
\end{equation}\\
where the coefficients $a_n$ and $w_n$ are given as,
\begin{eqnarray}\nonumber
a_n&=&\frac{1}{4} \cos\left(\frac{n \pi }{8}\right) \cos\left(\frac{n \pi }{2}\right) \left[2-4 \cos\left(\frac{n \pi }{4}\right)+5
\cos\left(\frac{n \pi }{2}\right)+5 \cos(n \pi )\right] \mbox{and}\\ \nonumber
w_n&=&{(-1)^{1/8} e^{-\frac{13}{8} i n \pi }} \left(-1+e^{i n \pi }\right)
\left[(5+5~ i) \sqrt{2}+5 \left(-1+(-1)^{1/4}\right) e^{\frac{i n \pi }{8}}+10 (-1)^{3/4}~ e^{\frac{i n \pi }{2}}+5 ~i \left((-1+i)+\sqrt{2}\right) e^{\frac{9 i n \pi }{8}}\right.\\ \nonumber && \left. -2~ e^{\frac{3 i n \pi }{4}}-10 (-1)^{1/4}~ e^{i n \pi }-\right.
\left.10~ i~ e^{\frac{5 i n \pi }{4}}-2~ (-1)^{3/4}~ e^{\frac{3 i n \pi }{2}}++\left((1+5~ i)-(2+3~ i) \sqrt{2}\right)
\left(e^{\frac{5 i n \pi }{8}}+e^{\frac{-3 i n \pi }{8}}\right)\right.\\ \nonumber && \left.+10~
e^{\frac{- i n \pi }{4}}+5 \left(-i+(-1)^{3/4}\right) ~e^{\frac{- i n \pi }{8}}+10~ i ~e^{\frac{ i n \pi }{4}}\right]\Big{/}{64} .
\end{eqnarray}
The  expression for the linear entropy can be calculated as follows:
\begin{eqnarray}\label{Eq:Ap5Q4} \nonumber
S_{(\theta_0=0,\phi_0=0)}^{(6)}(n)&=& S_{(0,0)}^{(6)}(n)=1-\mbox{tr}\left[\rho_1^2(n)\right]\\ \nonumber
&=& \frac{1}{2}\left\lbrace 1-\frac{1}{64} \cos^2\left(\frac{n \pi }{8}\right) \cos^2\left(\frac{n \pi }{2}\right) \left(2-4 \cos\left(\frac{n
\pi }{4}\right)+5 \cos\left(\frac{n \pi }{2}\right)+5 \cos(n \pi )\right)^2+\sin^2\left(\frac{n \pi }{2}\right) \left[-456\right.\right.\\ \nonumber && \left.\left.+84 \sqrt{2}\left(1+\cos(n\pi)\right) +240 \sqrt{2} \left(-\cos\left(\frac{n \pi }{4}\right)+\sin\left(\frac{n \pi }{4}\right)\right)+40 \left(6-5 \sqrt{2}\right)
\cos\left(\frac{n \pi }{2}\right)+88 \cos (n\pi )\right.\right.\\  && \left.\left.+ 272 \sqrt{2} \left(\cos\left(\frac{3 n \pi }{4}\right)+\sin\left(\frac{3n \pi }{4}\right)\right)+ 480  \sin\left(\frac{n \pi }{2}\right)\right]\right\rbrace.
\end{eqnarray}
By observing  Eq.~(\ref{Eq:Ap5Q4}) and Fig. $1$ from the main text, we can see that the entanglement dynamics of the system exhibit a periodic pattern. This periodicity is expressed as, $S_{(0,0)}^{(6)}(n) = S_{(0,0)}^{(6)}(n+8)$. We also find that the values of linear entropy for consecutive odd \mbox{and} even values of $n$  are identical.
\subsubsection{\bf{The  Von Neumann entropy}}
The Von Neumann entropy, also known as entanglement entropy, quantifies information within a system and the  amount of correlations between quantum systems \cite{nielsenbook}. It is a quantum counterpart to the classical Shannon entropy and is defined as follows:
\begin{equation}\label{von}
S_{VN}=-\mbox{tr}\left[\rho \log\rho\right]=-\sum_{i=1}^d (\lambda_i\log\lambda_i),
\end{equation}
 where $\lambda_i$ are the eigenvalues  and  $d$ is the dimension of RDM matrix $\rho(n)$. For single  qubit ($d=2$),  Eq.~(\ref{von}) can be expressed as,
\begin{equation}\label{von1}
S_{VN}=-(\lambda_1\log\lambda_1 +\lambda_2\log\lambda_2 ),
\end{equation}
where $\lambda_1$ and $\lambda_2$ are the eigenvalues of  single qubit RDM, $\rho_1(n)$, in the  Eq.~(\ref{von2}). The eigenvalues are:
\begin{eqnarray}\nonumber
 &&\left\lbrace\frac{1}{2}\mp \frac{1}{16}\left[17+8 \left(2+\sqrt{2}\right) \cos\left(\frac{n \pi }{4}\right)+15 \left(\cos\left(\frac{n
\pi }{2}\right)- \sin\left(\frac{n
\pi }{2}\right)\right)-8 \sqrt{2} \left(\sin\left(\frac{n \pi }{4}\right)+  \sin\left(\frac{3 n \pi }{4}\right)\right) \right.\right.\\  && \left. \left. + ~8 \left(2-\sqrt{2}\right) \cos\left(\frac{3 n \pi }{4}\right)\right]^\frac{1}{2}\right\rbrace.
\end{eqnarray}\\
Using these eigenvalues in the Eq.~(\ref{von1}), we can  calculate the entanglement entropy. Similar to linear entropy, the von Neumann entropy for this state  also shows periodic behavior of the same period, which is shown in Fig. $1$ from the main text.
\subsubsection{\bf{Concurrence}}
The concurrence ($C$) is a measure  of entanglement present in a two-qubit quantum state. If $C=0$ the state is separable, and if $C=1$, the state is maximally entangled \cite{Wootters01}. The concurrence is given as follows:
\begin{equation}
\label{Eq:Ap5Q3}
\mathcal{C}(\rho_{12})=\text{max}\left(0, \sqrt{\lambda_1}-\sqrt{\lambda_2}-\sqrt{\lambda_3}-\sqrt{\lambda_4} \right),
\end{equation}
where $\lambda_l$ are eigenvalues in decreasing order of
$(\sigma_y \otimes \sigma_y)\rho_{12} (\sigma_y \otimes \sigma_y) \rho_{12}^*$, where $\rho_{12}^*$ is complex conjugation in
the standard $\sigma_z$ basis. The two-qubit RDM is  given as follows:
\begin{equation}
\rho_{12}(n)={\frac{1}{4}\left(
\begin{array}{cccc}
 b_{m} & c_{m} & c_{m} & d_{m}^* \\
 c_{m}^* & e_{m} & e_{m} & a_{m} \\
 c_{m}^* & e_{m} & e_{m} & a_{m} \\
 d_{m} & a_{m}^* & a_{m}^* & f_{m} \\
\end{array}
\right)},
\end{equation}
 where  the coefficients are  expressed as  follows:
\begin{eqnarray}\nonumber
b_{m}&=&\frac{1}{16} \left[22+6 \cos\left(\frac{n \pi }{8}\right)+10 \cos\left(\frac{3 n \pi }{8}\right)+4 \cos\left(\frac{n\pi }{2}\right)+10 \cos\left(\frac{5 n \pi }{8}\right)+6\cos\left(\frac{7 n \pi }{8}\right)+6 \cos(n \pi )\right],  \\ \nonumber
f_{m}&=&\frac{1}{4} \left[58+94 \cos\left(\frac{n \pi }{8}\right)+78 \cos\left(\frac{n \pi }{4}\right)+62 \cos\left(\frac{3
n \pi }{8}\right)+56 \cos\left(\frac{n \pi }{2}\right)+46\cos\left(\frac{5 n \pi }{8}\right)+46 \cos\left(\frac{3 n \pi }{4}\right)\right.\\ \nonumber  && \left.+26 \cos(n \pi )+46\cos\left(\frac{7 n \pi }{8}\right)\right] \sin^2\left(\frac{n \pi }{16}\right),~~e_{m}=\frac{1}{8} \left[3 \sin\left(\frac{n \pi }{4}\right)+\sin\left(\frac{3 n \pi }{4}\right)\right]^2,\\ \nonumber
d_{m}&=&\frac{1}{16} \left[4 \cos\left(\frac{n \pi }{2}\right)-6 \cos(n \pi )-{2~i}
 \left(i+\sin\left(\frac{n \pi }{8}\right)+ \sin\left(\frac{3 n \pi }{8}\right)-4~ i \sin\left(\frac{n \pi }{2}\right)- \sin\left(\frac{5
n \pi }{8}\right)-\sin\left(\frac{7 n \pi }{8}\right)\right)\right],\\ \nonumber
a_{m}&=&(-1)^{3/8} e^{-\frac{13}{8} i n \pi } \left(-1+e^{i n \pi }\right)
\left[6 \sqrt{2}+e^{\frac{i n \pi }{8}} \left((2-2~ i)-3 \sqrt{2}+6 ~i \sqrt{2} e^{\frac{3 i n \pi }{8}}+2 \left(-2+\sqrt{4-3~ i}\right)\left( e^{\frac{i
n \pi }{2}}+ e^{\frac{3i
n \pi }{2}}\right)\right.\right.\\ \nonumber
&&\left.\left.-(4-4~ i)\left( e^{\frac{5 i n \pi }{8}}-e^{\frac{13 i n \pi }{8}}\right)-6 \sqrt{2} e^{\frac{7 i n \pi }{8}}+\left(4-6 (-1)^{1/4}\right) e^{i n \pi }-(4+4~ i) \left(e^{\frac{9 i n
\pi }{8}}-e^{\frac{17 i n \pi }{8}}\right)+2~ i \sqrt{2} e^{\frac{11 i n \pi }{8}}\right.\right.\\ \nonumber
&&\left.\left. +i(-2+2~
i)-3~ i \sqrt{2} \right)\right]\Big{/}{64 \sqrt{2}}~~ \mbox{and}\\ \nonumber
c_{m}&=&(-1)^{3/8} e^{-\frac{13}{8} i n \pi } \left(-1+e^{i n \pi }\right)
\left[6 \sqrt{2}+\left((-2+2~ i)+3 \sqrt{2}\right) e^{\frac{i n \pi }{8}}+6~ i \sqrt{2} e^{\frac{i n \pi }{2}}+\left(4-(3-i) \sqrt{2}\right) e^{\frac{5
i n \pi }{8}}-\right.\\ \nonumber
&&\left.(4-4 ~i) e^{\frac{3 i n \pi }{4}}-6 \sqrt{2} e^{i n \pi }+\left(-4+6 (-1)^{1/4}\right) e^{\frac{9 i n \pi }{8}}-(4+4~ i) e^{\frac{5 i n \pi }{4}}+2~ i \sqrt{2} e^{\frac{3 i n \pi }{2}}+\left(4-(3-i) \sqrt{2}\right) e^{\frac{13 i n \pi }{8}}\right.\\ \nonumber
&&\left.+(4-4~ i)
e^{\frac{7 i n \pi }{4}}+i \left((-2+2 ~i)+3 \sqrt{2}\right) e^{\frac{17 i n \pi }{8}}+(4+4~ i) e^{\frac{9 i n \pi }{4}}\right]\Big{/}{64 \sqrt{2}}.\\ \nonumber
\end{eqnarray}
All these coefficients exhibits periodic behavior i.e $\left({b}_{m}(n), {a}_{m}(n), {d}_{m}(n), {c}_{m}(n), {f}_{m}(n), {e}_{m}(n)\right)$=$\left({b}_{m}(n+16), {a}_{m}(n+32)\right.$\\$\left., {d}_{m}(n+16), {c}_{m}(n+32)), {f}_{m}(n+16), {e}_{m}(n+8)\right)$. The concurrence and   the eigenvalues of $(\sigma_y \otimes \sigma_y)\rho_{12} (\sigma_y \otimes \sigma_y) \rho_{12}^*$ are also periodic in nature with period $8$. However,
due to a fairly large expression of eigenvalues and concurrence, it becomes complicated to understand its behavior with time. To show its periodic nature, we have tabulated the values of the concurrence and   the eigenvalues of $(\sigma_y \otimes \sigma_y)\rho_{12} (\sigma_y \otimes \sigma_y) \rho_{12}^*$   in
Table \ref{Table:supp1} for  the time evolution   $n$ = $0$ to $15$, instead of writing the large expressions. By using these eigenvalues in the Eq.~(\ref{Eq:Ap5Q3}), we can easily obtain the concurrence values, as detailed in Table  \ref{Table:supp1}. The corresponding  values are plotted in  the Fig. $2$  from the main text.
\begin{table}[hbtp]
  \centering
 \caption{The eigenvalues and concurrence ($C(n)$) for the initial state  $\otimes^6\ket{0}$ at  evolution steps ($n$)   for the system of $6$ qubits.}
 \renewcommand{\arraystretch}{2.4}
\begin{tabular}{|p{0.35cm}|p{5.6cm}|p{2.25cm}||p{0.35cm}|p{5.6cm}|p{2.25cm} | }
 \hline
$n$ &\hspace{1.9cm} Eigenvalues&\hspace{.25cm}Concurrence  &$n$ &\hspace{1.9cm} Eigenvalues& \hspace{.25cm}Concurrence \\
\hline
0 &$(0, 0, 0, 0)$& 0&8&$(0,0,0,0)$&0
 \\
\hline
1 & $(0.2678348192, 0.0625, 0.0446651808, 0)$& $0.05618621785$& 9&$(0.2678348192, 0.0625, 0.0446651808, 0)$&$0.05618621785$
\\
\hline

2 & $(0.2678348192, 0.0625, 0.0446651808, 0)$&$0.05618621785$&10 &$(0.2678348192, 0.0625, 0.0446651808, 0)$&$0.05618621785$
\\
\hline
3 & $(0.25, 0.25, 0, 0)$&$0$ &11 & $(0.25, 0.25, 0, 0)$&$0$
\\
\hline
4 & $(0.25, 0.25, 0, 0)$&$0$ &12 & $(0.25, 0.25, 0, 0)$&$0$
\\
\hline
5 &$(0.2678348192, 0.0625, 0.0446651808, 0)$& $0.05618621785$ &13& $(0.2678348192, 0.0625, 0.0446651808, 0)$& $0.05618621785$
\\
\hline
6 &$(0.2678348192, 0.0625, 0.0446651808, 0)$&$0.05618621785$ & 14&$(0.2678348192, 0.0625, 0.0446651808, 0)$&$0.05618621785$
\\
\hline
7 &$(0,0,0, 0)$ & $0$ & 15 & $(0,0,0, 0)$& $0$
\\
\hline
\end{tabular}
  \label{Table:supp1}
\end{table}\\
\subsection{Initial state $\ket{++++++}$= $\ket{\theta_0 =\pi/2,\phi_0 =-\pi/2}$}
We have previously explored  the detailed evolution of the state $\ket{000000}$. Now, let's examine the case of a six-qubit for the initial state $\ket{++++++}_y$, where $|+\rangle_y=\frac{1}{\sqrt{2}}(|0\rangle+i|1\rangle)$ is an eigenstate of $\sigma_y$ with eigenvalue $+1$. This state can  be expressed as, $\otimes ^6 {\ket{+}}_y=\frac{1}{4\sqrt{2}} \ket{\phi_0^+}+\frac{i\sqrt{3}}{4}\ket{\phi_1^+}- \frac{\sqrt{15}}{4\sqrt{2}} \ket{\phi_2^+}-\frac{i\sqrt{5}}{4}\ket{\phi_3^+}$. Its evolution is thus, restricted to positive parity subspace. The state $\ket{\psi_n}$ can be obtained  by   iteratively applying  unitary operator $\mathcal{U}$  $n$ times  on this initial state as follows:
\begin{eqnarray}\label{Eq:45} \nonumber
\ket{\psi_n}&=&\mathcal{U}_{+}^n\ket{++++++}\\
&=&e^{{\frac{ in \pi }{4}}}\left({\alpha}_{n}^\prime \ket{\phi_0^+}+{\beta}_{n}^\prime  \ket{\phi_1^+}+{\gamma}_{n}^\prime  \ket{\phi_2^+}+\zeta_n  \ket{\phi_3^+}\right).
\end{eqnarray}
Here, the coefficients are  given as follows:
\begin{eqnarray} \nonumber
 {\alpha}_{n}^\prime &=&\frac{e^{-\frac{3}{4} i n \pi }} {16\sqrt{2}}\left[3 \left(-1+e^{\frac{3i \pi }{8}}\right)+5 \left(1+e^{\frac{7i \pi }{8}}\right) e^{\frac{i
n \pi }{2}}-3 \left(1+e^{\frac{3i \pi }{8}}\right) e^{i n \pi }-5 \left(-1+e^{\frac{7i \pi }{8}}\right) e^{\frac{3 i n \pi }{2}}\right],\\ \nonumber
{\gamma}_{n}^\prime &=&\frac{\sqrt{{15}}i~e^{-\frac{3}{4} i n \pi }}{16\sqrt{2}}    \left[1-e^{\frac{3i\pi }{8}}+\left(i+e^{\frac{3i\pi }{8}}\right) e^{\frac{i n \pi }{2}}\right] \left[i+(1+i) e^{\frac{i n \pi }{2}}+e^{i n \pi }\right], \\ \nonumber
~{\beta}_{n}^\prime &=&\frac{\sqrt{3}~ e^{-\frac{3}{4} i n \pi }}{8}  \left[i-e^{\frac{i\text{  }\pi }{8}}+\left(i+e^{\frac{i\text{  }\pi }{8}}\right)
e^{i n \pi }\right]
~ \mbox{and}~~~ \zeta_n =\frac{\sqrt{5}~ e^{-\frac{1}{4} i n \pi }}{8}  \left[-i+e^{\frac{5i\text{  }\pi }{8}}-\left(i+e^{\frac{5i\text{  }\pi
}{8}}\right) e^{i n \pi }\right]. \\ \nonumber
\end{eqnarray}
\subsubsection{\bf{The linear entropy}}
Using the Eq. (\ref{Eq:45}), the single qubit RDM, $\rho_1(n)$, is then given as follows:
\begin{equation}
\rho_1(n)={\frac{1}{2}\left[
\begin{array}{cc}
 1 & Z_n \\
 Z_n^* & 1 \\
\end{array}
\right]},
\end{equation}
where the coefficient $Z_n$ is given as follows:
\begin{equation}\nonumber
 Z_n=\frac{-i}{16} \left[3(1+\cos(n\pi)) +\sqrt{2}\left(1-\cos(n\pi)\right)+10 \cos\left(\frac{n \pi }{2}\right)\right].
\end{equation}
The eigenvalues of $\rho_1(n)$ are $\frac{1}{16} \left[8\mp\sqrt{18+30\cos\left(\frac{n \pi }{2}\right)+16 \cos(n \pi )}\right]$. Thus, the expression for the linear entropy is given as follows:
\begin{eqnarray}\label{Eq:Ap6Q1}
S_{(\pi/2,-\pi/2)}^{(6)}(n)&=& \frac{1}{2}-\frac{1}{512}
\left[3(1+\cos(n\pi)) +\sqrt{2}\left(1-\cos(n\pi)\right)+10 \cos\left(\frac{n \pi }{2}\right)\right]^2.
\end{eqnarray}
From the Eq. ({\ref{Eq:Ap6Q1}}), we can see that the entanglement dynamics is periodic in nature with $S_{(\pi/2,-\pi/2)}^{(6)}(n)=S_{(\pi/2,-\pi/2)}^{(6)}(n+4)$ and  is shown in the Fig. $3$ from the main text. By using these eigenvalues in Eq.~(\ref{von1}), we can easily calculate the entanglement entropy for this state. Similar to linear entropy, the Von Neumann entropy for this state shows periodic behavior of the same period, which can be seen in Fig. $3$ from the main text.
\subsubsection{\bf{Concurrence}}
The two-qubit RDM, $\rho_{12}(n)$, using the  Eq. (\ref{Eq:45}), for this state  is given as follows:
\begin{equation}
\rho_{12}(n)={\frac{1}{2}\left(
\begin{array}{cccc}
 h_1 &h_4  & h_4 & h_2 \\
 h_4^* & h_3 & h_3 & h_6 \\
 h_4^* & h_3 & h_3 & h_6 \\
 h_2^* & h_6^* & h_6^* & h_1 \\
\end{array}
\right)},
\end{equation}
where  the coefficients are,
\begin{eqnarray} \nonumber
h_1&=&\frac{1}{8} \left[5- \left(\cos\left(\frac{n \pi }{2}\right)+\sin\left(\frac{n \pi }{2}\right)\right)\right],~h_2=\frac{-1}{8} \left[1+ 3\cos\left(\frac{n \pi }{2}\right)-\sin\left(\frac{n \pi }{2}\right)\right],~h_3=\frac{1}{8} \left[3+\cos\left(\frac{n \pi }{2}\right)+\sin\left(\frac{n \pi }{2}\right)\right]\\ \nonumber
h_4&=&\frac{-(1+i)}{64} \left[3(1+ i)(1+\cos(n\pi))+4 \sqrt{2}(1-\cos(n\pi))+20\left(1+ i\right) \cos\left(\frac{n
\pi }{2}\right)\right] \mbox{and} \\ \nonumber
h_6&=&\frac{(1-i)}{64} \left[3(1+ i)(1+\cos(n\pi))+4 \sqrt{2}(1-\cos(n\pi))+20\left(1+ i\right) \cos\left(\frac{n
\pi }{2}\right)\right]. \\ \nonumber
\end{eqnarray}
All these coefficients have periodic nature  of period $4$ i.e $\left({h}_{1}(n), {h}_{2}(n), {h}_{3}(n), {h}_{4}(n), {h}_{6}(n)\right)=\left({h}_{1}(n+4), {h}_{2}(n+4), {h}_{3}(n+4), {h}_{4}(n+4)), {h}_{6}(n+4)\right)$. The  concurrence and  eigenvalues of $(\sigma_y \otimes \sigma_y)\rho_{12} (\sigma_y \otimes \sigma_y) \rho_{12}^*$ for $n$ ranging from 0 to 8, arranged in decreasing order, are presented in
Table \ref{Table:LinearEntropy131sup}. The concurrence values repeat after every $n\rightarrow n+4$, which shows a periodicity of $4$. Their periodic nature  is shown in the Fig. $4$ from the main text.
\begin{table*}[hbtp]
  \centering
 \caption{The eigenvalues and concurrence ($C(n)$) for the initial state  $\otimes^6\ket{+}$ at  evolution steps ($n$)   for the system of $6$ qubits.}
 \renewcommand{\arraystretch}{2.4}
\begin{tabular}{|p{0.35cm}|p{5.6cm}|p{2.25cm}||p{0.35cm}|p{5.6cm}|p{2.25cm} | }
 \hline
$n$ &\hspace{1.9cm} Eigenvalues&\hspace{.25cm}Concurrence  &$n$ & \hspace{1.9cm}Eigenvalues& \hspace{.25cm}Concurrence \\
\hline
0 &$(0, 0, 0, 0)$& 0&4&$(0,0,0,0)$&0
 \\
\hline
1 & $(0.2678348192, 0.0625, 0.0446651808, 0)$& $0.05618621785$& 5&$(0.2678348192, 0.0625, 0.0446651808, 0)$&$0.05618621785$
\\
\hline
2 & $(0.25, 0.046875, 0.046875, 0)$&$0.06698729812$&6 &$(0.25, 0.046875, 0.046875, 0.)$&$0.06698729812$
\\
\hline
3 & $(0.2678348192, 0.0625, 0.0446651808, 0)$&$0.05618621785$ &7 & $(0.2678348192, 0.0625, 0.0446651808, 0)$&$0.05618621785$
\\
\hline
\end{tabular}
  \label{Table:LinearEntropy131sup}
\end{table*}
\\

\section{Exact analytical solution for eight-qubits case}
Here, the  eigenbasis for this case  is given as follows:\\
\begin{eqnarray}
\ket{\phi_0^{\pm}}&=&\frac{1}{\sqrt{2}}(\ket{w_0} \pm \ket{\overline{w_0}}),\\
\ket{\phi_1^{\pm}}&=&\frac{1}{\sqrt{2}}(\ket{w_1} \mp \ket{\overline{w_1}}),\\
\ket{\phi_2^{\pm}}&=&\frac{1}{\sqrt{2}}(\ket{w_2} \pm  \ket{\overline{w_2}}),\\
\ket{\phi_3^{\pm}}&=&\frac{1}{\sqrt{2}}(\ket{w_3} \mp  \ket{\overline{w_3}}),\\
\ket{\phi_4^{+}}&=&\frac{1}{\sqrt{70}}\sum_{\mathcal{P}}\ket{00001111},
\end{eqnarray}
where $\ket{w_0}=\ket{00000000}$,~$\ket{\overline{w_0}}=\ket{11111111}$, $\ket{w_1}=\frac{1}{\sqrt{8}}\sum_\mathcal{P} \ket{00000001}_\mathcal{P}$, $\ket{\overline {w_1}}=\frac{1}{\sqrt{8}}\sum_\mathcal{P}\ket{01111111}_\mathcal{P}$,~$\ket{w_2}=\frac{1}{\sqrt{28}}\sum_\mathcal{P} \ket{00000011}_\mathcal{P}$,$\ket{\overline{w_2}}=\frac{1}{\sqrt{28}}\sum_\mathcal{P}\ket{00111111}_\mathcal{P}$,
$\ket{w_3}=\frac{1}{\sqrt{56}}\sum_\mathcal{P} \ket{00000111}_\mathcal{P}$ and $\ket{\overline{w_3}}=\frac{1}{\sqrt{56}} \sum_\mathcal{P} \ket{00011111}_\mathcal{P}$. The unitary operator $\mathcal{U}$ in this basis is block diagonal in $\mathcal{U_+}$ and $\mathcal{U_-}$ having dimensions $5\times 5$ and  $4\times4$ respectively such that:
\begin{equation}
\mathcal{U}_{+}=\frac{1}{8}\left(
\begin{array}{ccccc}
 i & 0 & 2~i \sqrt{7} & 0 & i~ \sqrt{35} \\
 0 & -6 ~e^{\frac{i \pi }{4}} & 0 & -2 \sqrt{7}~ e^{\frac{i \pi }{4}} & 0 \\
 -2~i \sqrt{7} & 0 & -4~i & 0 & 2~i \sqrt{5} \\
 0 & -2 \sqrt{7}~ e^{\frac{i \pi }{4}} & 0 & 6~ e^{\frac{i \pi }{4}} & 0 \\
 i~ \sqrt{35} & 0 & -2~i \sqrt{5} & 0 & 3 ~i \\
\end{array}
\right)~ \mbox{and}
\end{equation}
\begin{equation}
\mathcal{U}_{-}=\frac{-1}{2\sqrt{2}}\left(
\begin{array}{cccc}
 0 & i & 0 &  i ~\sqrt{7} \\
 -e^{\frac{i \pi }{4}} & 0 & -\sqrt{7}~ e^{\frac{i \pi }{4}} & 0 \\
 0 & - i ~\sqrt{7} & 0 & i \\
 - \sqrt{7}~ e^{\frac{i \pi }{4}} & 0 & e^{\frac{i \pi }{4}} & 0 \\
\end{array}
\right).
\end{equation}
The eigenvalues of $\mathcal{U_{+}}$ $\left(\mathcal{U_{-}}\right)$ are $\left\lbrace i,e^{\frac{7i \pi }{6}},e^{\frac{- i \pi }{6}},e^{\frac{5 i \pi }{4}},e^{\frac{ i \pi }{4}}\right\rbrace$ $\left(\left\lbrace e^{\frac{7i \pi }{8}},e^{\frac{15i \pi }{8}},e^{\frac{11i \pi }{8}},e^{\frac{3i \pi }{8}}\right\rbrace\right)$ and the eigenvectors are\\
 $\left\lbrace\left[ \sqrt{\frac{5}{7}},-\sqrt{\frac{7}{5}},-\sqrt{\frac{7}{5}},0,0\right]^T, \left[0,0,0,\sqrt{7},-\frac{1}{\sqrt{7}}\right]^T,\left[0,-2
i \sqrt{\frac{3}{5}},2 i \sqrt{\frac{3}{5}},0,0 \right]^T,\left[0,0,0,1,1 \right]^T ,\left[1,1,1,0,0\right]^T\right\rbrace $\\ $\left(\left\lbrace\left[2 (-1)^{5/8} \sqrt{\frac{2}{7}},-2 (-1)^{5/8} \sqrt{\frac{2}{7}},0,0 \right]^T,\left[\frac{1}{\sqrt{7}},\frac{1}{\sqrt{7}},-\sqrt{7},-\sqrt{7} \right]^T,\left[0,0,2
(-1)^{1/8} \sqrt{2},-2 (-1)^{1/8} \sqrt{2} \right]^T,\left[1,1,1,1 \right]^T\right\rbrace \right)$.\\

 The $n$th time evolution of $\mathcal{U}_{+}$ and $\mathcal{U}_{-}$ is given as follows:
\begin{equation}
 \mathcal{U}_{+}^n=\frac{ e^{\frac{i n \pi }{2}}}{24}{\left[
\begin{array}{ccccc}
  \left(10+14 \cos\left(\frac{2 n \pi }{3}\right)\right) & 0 & 4 \sqrt{{21}}  \sin\left(\frac{2 n \pi }{3}\right) & 0 & -{4\sqrt{35}\sin^2\left(\frac{ n \pi }{3}\right)}
\\
 0 & 3{e^{\frac{-i n \pi }{4}}}  \left(1+7 e^{i n \pi }\right) & 0 & {-3\sqrt{7} e^{\frac{-i n \pi }{4}}} \left(1-e^{i n \pi }\right)
& 0 \\
- 4 \sqrt{{21}}  \sin\left(\frac{2 n \pi }{3}\right) & 0 & 24\cos\left(\frac{2 n \pi }{3}\right) & 0 & 4\sqrt{15}  \sin\left(\frac{2 n \pi }{3}\right) \\
 0 & {-3\sqrt{7} e^{\frac{-i n \pi }{4}}} \left(1-e^{i n \pi }\right) & 0 & 3{e^{\frac{-i n \pi }{4}} } \left(7+e^{i n \pi }\right)
& 0 \\
 -{4\sqrt{35}\sin^2\left(\frac{ n \pi }{3}\right)}  & 0 & -4\sqrt{15} \sin\left(\frac{2n\pi}{3}\right) & 0 & \left(14+10 \cos\left(\frac{2 n \pi }{3}\right)\right)
\\
\end{array}
\right]}
\end{equation}
\begin{equation}
 \mbox{and}~~\mathcal{U}_{-}^n=\frac{e^{\frac{7 i n \pi }{8}}}{8}{\left[
\begin{array}{cccc}
   4a_n & -\sqrt{2}b_n e^{\frac{5 i  \pi }{8}}
& 0 &  \sqrt{{14}}b_n e^{\frac{5i  \pi }{8}}  \\
 \sqrt{2}b_n e^{\frac{3 i  \pi }{8}} & \frac{a_n}{2}  \left(7~e^{\frac{-i n
\pi }{2}}+1\right) &  \sqrt{14}b_n e^{(\frac{- i n \pi }{2}+\frac{7 i  \pi }{8})}   &
-i \sqrt{7}\bar{c_n} e^{\frac{ i n \pi }{4}} \\
 0 & -\sqrt{14}b_n e^{(\frac{- i n \pi }{2}+\frac{ i  \pi }{8})} & 4a_n e^{\frac{- i n \pi }{2}}  & \sqrt{2}b_n e^{(\frac{- i n \pi }{2}+\frac{ i  \pi }{8})} \\
 \sqrt{{14}}b_n e^{\frac{3i  \pi }{8}}    & -i \sqrt{7}\bar{c_n} e^{\frac{ i n \pi }{4}} & -\sqrt{2}b_n e^{(\frac{- i n \pi }{2}+\frac{ 7i  \pi }{8})}  & \frac{ a_n}{2} \left(e^{\frac{-i n \pi }{2}}+7 \right)  \\
\end{array}
\right]},
\end{equation}
where $a_n$=$1+e^{i n \pi }$ and $b_n=e^{i n \pi }-1$. The time periodicity  of $\mathcal{U}_{+}^n\left(\mathcal{U}_{-}^n\right)$ is $24(16)$. Hence the time periodicity of $\mathcal{U}^n$ is $48$.
\subsection{Initial state $\otimes^8\ket{0}$= $\ket{\theta_0 =0,\phi_0 =0}$}
The state $\ket{\psi_n}$ can be calculated by acting $n$th time evolution of unitary operator $\mathcal{U}$ on the initial state $\otimes^8\ket{0}$ and is given as follows:
 \begin{eqnarray}
\ket{\psi_n}&=&\mathcal{U}^{n}\ket{00000000}= \frac{1}{\sqrt{2}}(\mathcal{U}_{+}^n\ket{\phi_0^+}+\mathcal{U}_{-}^n\ket{\phi_0^-}) \\  \nonumber
&=&\frac{1}{\sqrt{2}}\left( {b}_1^\prime\ket{\phi_0^+}+{b}_2^\prime\ket{\phi_2^+}+{b}_3^\prime\ket{\phi_4^+}+{b}_4^\prime\ket{\phi_0^-}+{b}_5^\prime\ket{\phi_1^-}+{b}_6^\prime\ket{\phi_3^-}\right),
\end{eqnarray}
where the coefficients are:
\begin{eqnarray}\nonumber
{b}_1^\prime&=&\frac{ e^{\frac{ i n \pi }{2}}}{12} \left[5+7 \cos\left(\frac{2 n \pi }{3}\right)\right],\;\; {b}_2^\prime=\frac{i~\sqrt{{7}} e^{\frac{ i n \pi }{6}}}{4\sqrt{{3}}}   \left[-1+ e^{\frac{4 i n \pi }{3}}\right], ~{b}_3^\prime=-\frac{{\sqrt{35}~e^{\frac{ i n \pi }{6}}}}{24}   \left(-1+e^{\frac{ 2i n \pi }{3}}\right)^2,~ {b}_4^\prime =\frac{e^{\frac{ 7i n \pi }{8}}}{2}  (1+e^{{ i n \pi }}),\\ \nonumber {b}_5^\prime &=&\frac{(-1+e^{{ i n \pi }})}{{4 \sqrt{2}}} \left[i~ \cos\left(\frac{\pi }{8}-\frac{7 n \pi }{8}\right)+\sin\left(\frac{\pi}{8}-\frac{7 n \pi }{8}\right)\right]~\mbox{and}~~ {b}_6^\prime=\frac{\sqrt{{7}} (-1+e^{{ i n \pi }})}{4\sqrt{{2}} }  \left[i~ \cos\left(\frac{\pi }{8}-\frac{7 n \pi }{8}\right)+\sin\left(\frac{\pi
}{8}-\frac{7 n \pi }{8}\right)\right].
\end{eqnarray}
\subsubsection{\bf{The linear entropy}}
 The single qubit RDM, $\rho_1(n)$, is given as follows:
\begin{equation}
\rho_1(n)=\frac{1}{4}\left[
\begin{array}{cc}
 2+\frac{1}{12} \left(5+7 \cos\left(\frac{2 n \pi }{3}\right)\right) \left(\cos\left(\frac{3 n \pi }{8}\right)+\cos\left(\frac{5
n \pi }{8}\right)\right) & {W_n} \\
 {W_n}^* & 2-\frac{1}{12} \left(5+7 \cos\left(\frac{2 n \pi }{3}\right)\right) \left(\cos\left(\frac{3 n \pi }{8}\right)+\cos\left(\frac{5
n \pi }{8}\right)\right) \\
\end{array}
\right],
\end{equation}
\begin{eqnarray}\nonumber
\mbox{where} ~W_n &=&\frac{1}{24} \sin\left(\frac{n \pi }{2}\right) \left[-3 i \cos\left(\frac{1}{8} (3+4 n) \pi \right)+7 \sqrt{3} \cos\left(\frac{1}{24}
(9+5 n) \pi \right)-7 \sqrt{3} \cos\left(\frac{1}{24} (9+37 n) \pi \right)\right.\\ \nonumber &&\left.+7 \sin\left(\frac{1}{24} (9+5 n) \pi \right)-20 \sin\left(\frac{1}{8}
(3+7 n) \pi \right)+7 \sin\left(\frac{1}{24} (9+37 n) \pi \right)-3 i \sin\left(\frac{1}{8} (\pi +4 n \pi )\right)\right]. \\ \nonumber
\end{eqnarray}
The eigenvalues of $\rho_1(n)$ are given as follows:
\begin{eqnarray}\nonumber
\lambda &=&\frac{1}{2}\mp\frac{1}{96}
 \left\lbrace \left(5+7 \cos\left(\frac{2 n \pi }{3}\right)\right)^2\left[ \cos^2\left(\frac{3 n \pi }{8}\right) +8 \cos\left(\frac{3
n \pi }{8}\right)  \cos\left(\frac{5 n \pi }{8}\right)+4 \cos^2\left(\frac{5n \pi }{8}\right)\right]+\sin^2\left(\frac{n \pi }{2}\right)\right.\\ \nonumber &&\left.
 \left[9 \cos^2\left(\frac{1}{8} (3+4 n) \pi \right)+147 \cos^2\left(\frac{1}{24} (9+5 n)
\pi \right)+147 \cos^2\left(\frac{1}{24} (9+13 n) \pi \right)+
\sin\left(\frac{1}{24} (9+5 n) \pi \right)\right.\right.\\ \nonumber &&  \left.\left.\left(-98 \sqrt{3} \cos\left(\frac{1}{24} (9+13n) \pi \right) +49 \sin\left(\frac{1}{24}
(9+5 n) \pi \right)-280\sin\left(\frac{1}{8} (3+7 n) \pi \right)\right)+49 \sqrt{3} \sin\left(\frac{1}{12} (9+5 n) \pi \right)\right.\right.\\ \nonumber && \left.\left.+
280 \sqrt{3} \cos\left(\frac{1}{24} (9+37 n) \pi \right) \sin\left(\frac{1}{8} (3+7 n) \pi \right)+400 \sin^2\left(\frac{1}{8}
(3+7 n) \pi \right)-
\sin\left(\frac{1}{24} (9+13 n) \pi \right)\right.\right.\\ \nonumber && \left. \left.\left(98 \sin\left(\frac{1}{24} (9+5 n) \pi \right)-280 \sin\left(\frac{1}{8} (3+7
n) \pi \right)-49 \sin\left(\frac{1}{24} (9+13 n) \pi \right)\right)+
14 \cos\left(\frac{1}{24} (9+5 n) \pi \right)\right.\right.\\ \nonumber && \left.\left. \left(-21 \cos\left(\frac{1}{24} (9+37 n) \pi \right)+\sqrt{3} \left(-20 \sin\left(\frac{1}{8}
(3+7 n) \pi \right)-7 \sin\left(\frac{1}{24} (9+13 n) \pi \right)\right)\right)-49 \sqrt{3} \sin\left(\frac{1}{12} (9+13 n) \pi \right)\right.\right.\\ \nonumber && \left.\left. +
\sin\left(\frac{1}{8} (\pi +4 n \pi )\right)\left(18 \cos\left(\frac{1}{8} (3+4 n) \pi \right) +9 \sin\left(\frac{1}{8}
(\pi +4 n \pi )\right)\right)\right]\right\rbrace^\frac{1}{2}.
\end{eqnarray}
The expression for the linear entropy is given as follows:
 \begin{eqnarray}  \nonumber \label{Eq:Ap58Q1}
S_{(0,0)}^{8}(n)&=&\frac{1}{2}
\left\lbrace 1-\frac{1}{576} \left[5+7 \cos\left(\frac{2 n \pi }{3}\right)\right]^2 \left[\cos\left(\frac{3 n \pi }{8}\right)+\cos\left(\frac{11
n \pi }{8}\right)\right]^2+\sin^2\left(\frac{n \pi }{2}\right) \left[-\left(18-9\sqrt{2}\right) \cos(n \pi ) \right. \right.\\ \nonumber && \left.\left. +560 \left(\cos\left(\frac{2 n \pi }{3}\right)+\sqrt{3} \sin\left(\frac{2 n \pi }{3}\right)\right)+196 \left(\cos\left(\frac{4n \pi }{3}\right) - \sqrt{3} \sin\left(\frac{4n \pi }{3}\right)\right)+378 \sin\left(\frac{1}{12} (3+5 n) \pi \right)\right.\right.\\ \nonumber && \left.\left.
-378 \sqrt{3} \sin\left(\frac{1}{12} (9+5 n) \pi \right)+378 \sin\left(\frac{1}{12}(3+13 n) \pi \right)+378 \sqrt{3} \sin\left(\frac{1}{12} (9+13 n) \pi \right)+9 \left(-90+\sqrt{2}\right)\right.\right.\\  && \left.\left. -792 \sin\left(\frac{1}{4} (\pi +7 n \pi )\right)\right]\right\rbrace.
 \end{eqnarray}
From the Eq.~(\ref{Eq:Ap58Q1}), we can see that the entanglement dynamics is periodic in nature with
$S_{(0,0)}^{(8)}(n)$=$S_{(0,0)}^{(8)}(n+24)$. We also find that the linear entropy for consecutive odd and even values of $n$
remain the same. This is shown in Fig. $1$ from the main text. Substituting these eigenvalues in Eq.(\ref{von1}), we can calculate the entanglement entropy for this  state. Similar to linear entropy, it also shows the periodic nature of the same period, which is  shown in the Fig. $1$ from the main text.
\subsubsection{\bf{Concurrence}}
 The two-qubit RDM, $\rho_{12}(n)$, is given as follows:
\begin{equation}
 \rho_{12}(n)={\frac{1}{4}\left[
\begin{array}{cccc}
 g_n^\prime & y_n^\prime & y_n^\prime & o_n^* \\
 y_n^* & t_n^\prime & t_n^\prime & q_n \\
 y_n^* & t_n^\prime & t_n^\prime & q_n \\
 o_n & q_n^* & q_n^* & e_n^\prime \\
\end{array}
\right]},
\end{equation}
where  the coefficients are:
\begin{eqnarray}\nonumber \label{Eq:Ap8Q3}
g_n^\prime&=&\frac{1}{72} \left[102+21 \cos\left(\frac{7 n \pi }{24}\right)+30 \cos\left(\frac{3 n \pi }{8}\right)+20 \cos\left(\frac{2
n \pi }{3}\right)+21 \cos\left(\frac{17 n \pi }{24}\right)+18 \cos(n \pi )+21 \cos\left(\frac{25 n \pi }{24}\right) \right.\\ \nonumber &&\left. +4 \cos\left(\frac{4
n \pi }{3}\right)+ 30 \cos\left(\frac{5 n \pi }{8}\right)+21 \cos\left(\frac{ n \pi }{24}\right)\right], ~~t_n^\prime=\frac{1}{36} \left[21-12 \cos\left(\frac{2 n \pi }{3}\right)-9 \cos(n \pi )\right],\\ \nonumber
e_n^\prime&=&\frac{1}{72} \left[102-21 \cos\left(\frac{7 n \pi }{24}\right)-30 \cos\left(\frac{3 n \pi }{8}\right)+24 \cos\left(\frac{2
n \pi }{3}\right)-21 \cos\left(\frac{17 n \pi }{24}\right)+18 \cos (n\pi )-21 \cos\left(\frac{25 n \pi }{24}\right)\right.\\ \nonumber &&\left. -30 \cos\left(\frac{5 n \pi }{8}\right)-21 \cos\left(\frac{ n \pi }{24}\right)\right], \\ \nonumber
o_n&=&\frac{1}{72} \left[18-3~ i \sqrt{3} \cos\left(\frac{7 n \pi }{24}\right)-3~ i \sqrt{3} \cos\left(\frac{17 n \pi }{24}\right)-18
\cos (n\pi )+3~ i \sqrt{3} \cos\left(\frac{25 n \pi }{24}\right)+3~ i \sqrt{3} \cos\left(\frac{ n \pi }{24}\right)\right.\\ \nonumber &&\left. -20 \sqrt{3}
\sin\left(\frac{2 n \pi }{3}\right)+4 \sqrt{3} \sin\left(\frac{4 n \pi }{3}\right)\right],\\ \nonumber
y_n&=&\frac{1}{48} \left[\frac{1}{8} e^{-\frac{1}{24} i (9+ n) \pi } \left(7+e^{\frac{2 i n \pi }{3}} \left(10\left(1-e^{i n \pi }\right)-7 e^{\frac{i n \pi }{3}}+7 e^{\frac{2 i n \pi }{3}}-7 e^{\frac{5 i n \pi }{3}}\right)\right)+e^{\frac{i \pi }{8}-\frac{7 i n \pi }{8}} \left(4 \sqrt{3} \cos\left(\frac{n \pi }{6}\right)\right.\right.\\  \nonumber
&&\left.\left. -4 \sqrt{3} \cos\left(\frac{7 n \pi
}{6}\right)+(4-4 i) \sqrt{2} ~e^{\frac{7 i n \pi }{4}} \left(2 \cos\left(\frac{n \pi }{6}\right)+\cos\left(\frac{n \pi }{2}\right)\right)
\sin^2\left(\frac{n \pi }{6}\right) \left(3 \sqrt{3} \cos\left(\frac{n \pi }{3}\right)-5 \sin\left(\frac{n \pi }{3}\right)\right)\right.\right.\\ \nonumber
&&\left.\left.- 15 \sin\left(\frac{n \pi }{6}\right) \sin\left(\frac{n \pi }{3}\right)\left(\sin\left(\frac{n \pi }{3}\right)-\sin\left(\frac{n \pi }{6}\right) \sin\left(\frac{2 n \pi }{3}\right)\right)\right)\right]\mbox{and}  \\ \nonumber
q_n&=&\frac{1}{192}
\left[-6~e^{\frac{i \pi }{4}}+32~e^{\frac{i n \pi }{8}} \left(2 \cos\left(\frac{n \pi }{6}\right)+\cos\left(\frac{n \pi }{2}\right)\right)
\sin^2\left(\frac{n \pi }{6}\right) \left(3 \sqrt{3} \cos\left(\frac{n \pi }{3}\right)-5 \sin\left(\frac{n \pi }{3}\right)\right)+2~e^{\frac{i \pi }{4}+\frac{15 i n \pi }{8}}\right.\\ \nonumber
&&\left. \left(3 \cos\left(\frac{n \pi }{8}\right)-i \left(8 \sqrt{3}\left( \cos\left(\frac{n
\pi }{6}\right)- \cos\left(\frac{7 n \pi }{6}\right)\right)-3 \sin\left(\frac{n \pi }{8}\right)-4 \left(\sin\left(\frac{n \pi }{6}\right)-\sin\left(\frac{7 n \pi }{6}\right)\right)\right) -20
\sin\left(\frac{n \pi }{2}\right)\right)\right]. \\ \nonumber
\end{eqnarray}
All  these coefficients are periodic in nature i.e. $\left({g}_{n}^\prime(n), {t}_{n}^\prime(n), {o}_{n}(n), {y}_{44}(n), {e}_{n}^\prime(n)\right)$=$\left({g}_{n}^\prime(n+48), {t}_{n}^\prime(n+6), {o}_{n}(n+48),\right.$\\$\left.{y}_{n}(n+48)), {e}_{n}^\prime(n+48)\right)$. The concurrence and  eigenvalues $\lambda_l$ (in decreasing order) of $(\sigma_y \otimes \sigma_y)\rho_{12} (\sigma_y \otimes \sigma_y) \rho_{12}^*$ for  $n$ = $0$ to $47$  are presented in Table \ref{Table:LinearEntropy1211sup}. The values of concurrence are plotted  in the Fig. $2$ from the main text.
 \begin{table}[htbp]
  \centering
 \caption{The eigenvalues and concurrence ($C(n)$) for the initial state $\otimes^8\ket{0}$ at  evolution steps ($n$) for the system of $8$ qubits.}
 \renewcommand{\arraystretch}{2.4}
\begin{tabular}{|p{0.35cm}|p{6.3cm}|p{2.25cm}||p{0.35cm}|p{6.3cm}|p{2.25cm} | }
 \hline
$n$ &\hspace{2.2cm} Eigenvalues &\hspace{.25cm}Concurrence  &$n$ & \hspace{2.2cm}Eigenvalues& \hspace{.25cm}Concurrence \\
\hline
0 &$(0, 0, 0, 0)$& 0&24 &$(0,0,0,0)$&0
 \\
\hline
1 & $(0.254993942654, 0.0625, 0.0575060573456, 0)$& $0.015165042945$& 25&$(0.254993942654, 0.0625, 0.0575060573456, 0)$&$0.015165042945$
\\
\hline
2 &$(0.254993942654, 0.0625, 0.0575060573456, 0)$&$0.015165042945$ &26 & $(0.254993942654, 0.0625, 0.0575060573456, 0)$&$0.015165042945$
\\
\hline
3 & $(0.264971936445, 0.0625, 0.0553405635547, 0)$&$0.029508497188$&27 &$(0.264971936445, 0.0625, 0.0553405635547, 0)$&$0.029508497188$
\\
\hline
4 &$(0.264971936445, 0.0625, 0.0553405635547, 0)$&$0.029508497188$ & 28& $(0.264971936445, 0.0625, 0.0553405635547, 0)$&$0.029508497188$
\\
\hline
5 & $(0.125, 0.125, 0, 0)$&$0$ &29 & $(0.125, 0.125, 0, 0)$&$0$
\\
\hline
6 & $(0.125, 0.125, 0, 0)$&$0$ &30 &$(0.125, 0.125, 0, 0)$&$0$
\\
\hline

7 &$(0.244782496611, 0.0625, 0.05990500339, 0)$&$0$ & 31&$(0.244782496611, 0.0625, 0.05990500339, 0)$&$0$
\\
\hline
8 &$(0.2496744438147, 0.0625, 0.062337274936, 0)$&$0$ & 32&$(0.2496744438147, 0.0625, 0.062337274936, 0)$&$0$
\\
\hline
9 &$(0.254993942654, 0.0625, 0.0575060573456, 0)$& $0.015165042945$ &33& $(0.254993942654, 0.0625, 0.0575060573456, 0)$& $0.015165042945$
\\
\hline
 10& $(0.254993942654, 0.0625, 0.0575060573456, 0)$& $0.015165042945$&34 & $(0.254993942654, 0.0625, 0.0575060573456, 0)$& $0.015165042945$
\\
\hline
11 &$(0.25,0.25,0,0)$&$0$ & 35&$(0.25,0.25,0,0)$&$0$
\\
\hline
12 &$(0.25,0.25,0,0)$&$0$ &36 &$(0.25,0.25,0,0)$&$0$
\\
\hline
13 &$(0.254993942654, 0.0625, 0.0575060573456, 0)$ & $0.015165042945$ & 37 & $(0.254993942654, 0.0625, 0.0575060573456, 0)$& $0.015165042945$
\\
\hline
14 &$(0.254993942654, 0.0625, 0.0575060573456, 0)$& $0.015165042945$& 38 & $(0.254993942654, 0.0625, 0.0575060573456, 0)$ & $0.015165042945$
\\
\hline
15 &$(0.244782496611, 0.0625, 0.05990500339, 0)$&$0$ &39 & $(0.244782496611, 0.0625, 0.05990500339, 0)$&$0$
\\
\hline
16 &$(0.244782496611, 0.0625, 0.05990500339, 0)$ &$0$&40 & $(0.244782496611, 0.0625, 0.05990500339, 0)$ &$0$
\\
\hline
17 & $(0.125, 0.125, 0, 0)$ &$0$&41 &$(0.125, 0.125, 0, 0)$ &$0$
\\
\hline
18 & $(0.125, 0.125, 0, 0)$&$0$& 42&$(0.125, 0.125, 0, 0)$ &$0$
\\
\hline
19 & $(0.264971936445, 0.0625, 0.0553405635547, 0)$&$0.029508497188$&43 & $(0.264971936445, 0.0625, 0.0553405635547, 0)$&$0.029508497188$
\\
\hline
20 &$(0.264971936445, 0.0625, 0.0553405635547, 0)$&$0.029508497188$ &44 &$(0.264971936445, 0.0625, 0.0553405635547, 0)$&$0.029508497188$
\\
\hline
21 &$(0.254993942654, 0.0625, 0.0575060573456, 0)$ & $0.015165042945$ & 45& $(0.254993942654, 0.0625, 0.0575060573456, 0)$ & $0.015165042945$
\\
\hline
22 & $(0.254993942654, 0.0625, 0.0575060573456, 0)$& $0.015165042945$  & 46& $(0.254993942654, 0.0625, 0.0575060573456, 0)$ & $0.015165042945$
\\
\hline
23& $(0,0,0,0)$&$0$&47 & $(0,0,0,0)$ &$0$
\\
\hline
\end{tabular}
  \label{Table:LinearEntropy1211sup}
\end{table}
\subsection{Initial state $\otimes^8\ket{+}$= $\ket{\theta_0 =\pi/2,\phi_0 =-\pi/2}$}
The initial state can be  expressed as $\otimes ^8 {\ket{+}}_y=\frac{1}{8\sqrt{2}} \ket{\phi_0^+}+\frac{i}{4}\ket{\phi_1^+}- \frac{\sqrt{14}}{8} \ket{\phi_2^+}-\frac{i\sqrt{7}}{4}\ket{\phi_3^+}+\frac{\sqrt{70}}{16}\ket{\phi_4^+}$. The state $\ket{\psi_n}$ can be obtained by $n$th time evolution of unitary operator $\mathcal{U}$ on this initial state. Thus,
\begin{eqnarray}\nonumber
\ket{\psi_n}&=&\mathcal{U}_{+}^n\ket{++++++++}\\
&=& {c}_{1}^\prime \ket{\phi_0^+}+ {c}_{2}^\prime \ket{\phi_1^+} +  {c}_{3}^\prime \ket{\phi_2^+}+ {c}_{4}^\prime\ket{\phi_3^+}+ {c}_{5}^\prime\ket{\phi_4^+},
\end{eqnarray}
where $ {c}_{1}^\prime=-\frac{e^{\frac{i n \pi }{2}}}{{24
\sqrt{2}}} \left[-10+7 \cos\left(\frac{2 n \pi }{3}\right)+7 \sqrt{3} \sin\left(\frac{2 n \pi }{3}\right)\right]$, ${c}_{2}^\prime=\frac{i}{4} ~ e^{\frac{i n \pi }{4}}$,
$ {c}_{3}^\prime=\frac{1}{12} \sqrt{\frac{7}{2}}~ e^{\frac{i n \pi }{2}} \left[-3 \cos\left(\frac{2 n \pi }{3}\right)+\sqrt{3} \sin\left(\frac{2
n \pi }{3}\right]\right)$, $ {c}_{4}^\prime=-\frac{\sqrt{7}~i}{4}~   e^{\frac{i n \pi }{4}}$
\mbox{and} $ {c}_{5}^\prime=\frac{1}{24} \sqrt{\frac{35}{2}}~ e^{\frac{i n \pi }{2}} \left[2+\cos\left(\frac{2 n \pi }{3}\right)+\sqrt{3} \sin\left(\frac{2
n \pi }{3}\right)\right]$.\\
\subsubsection{\bf{The linear entropy}}
 The  $\rho_1(n)$, for this state is given as follows:
\begin{equation}
 \rho_1(n)={\frac{1}{2}\left[
\begin{array}{cc}
 1 & -\frac{i}{12}  \cos\left(\frac{n \pi }{4}\right) \left(5+7 \cos\left(\frac{2 n \pi }{3}\right)\right) \\
 \frac{i}{12}  \cos\left(\frac{n \pi }{4}\right) \left(5+7 \cos\left(\frac{2 n \pi }{3}\right)\right) & 1 \\
\end{array}
\right]}.
\end{equation}
The eigenvalues of $\rho_1(n)$ are $ \left\lbrace \frac{1}{2}\pm\frac{1}{48}\left[10 \cos\left(\frac{n \pi }{4}\right)+7 \left(\cos\left(\frac{5 n \pi }{12}\right)+\cos\left(\frac{11
n \pi }{12}\right)\right)\right]\right\rbrace$. The linear entropy is then given as follows:
\begin{equation}\label{Eq:8Ap5}
S_{(\pi/2,-\pi/2)}^{(8)}(n)=\frac{1}{2}\left[1-\frac{1}{144} \cos^2\left(\frac{n \pi }{4}\right) \left(5+7 \cos\left(\frac{2 n \pi }{3}\right)\right)^2\right].
\end{equation}
From the Eq. ({\ref{Eq:8Ap5}), we can see that the entanglement dynamics is periodic in nature with  $S_{(\pi/2,-\pi/2)}^{(8)}(n)=S_{(\pi/2,-\pi/2)}^{(8)}(n+12)$. This is shown in Fig. $3$ from the main text. By using these eigenvalues in  the Eq.~(\ref{von1}), we can easily calculate the entanglement entropy for this initial state. Similar to linear entropy, the von Neumann entropy for this state shows periodic behavior of the same period, which is  shown in Fig. $3$ from the main text.
\subsubsection{\bf{Concurrence}}
 Here, $\rho_{12}(n)$,  is given as follows:
\begin{equation}
\rho_{12}={\frac{1}{2}\left(
\begin{array}{cccc}
 a &d  & d & b \\
 d^* & c & c & f^* \\
  d^* &c& c & f^* \\
 b &  f & f  & a \\
\end{array}
\right)},
\end{equation}
where all the coefficients are,
\begin{eqnarray} \nonumber
a&=&\frac{1}{12} \left[7- \cos\left(\frac{2 n \pi }{3}\right)- \sqrt{3} \sin\left(\frac{2
n \pi }{3}\right)\right], ~~ b=\frac{-1}{12} \left[3+3 \cos\left(\frac{2 n \pi }{3}\right)- \sqrt{3} \sin\left(\frac{2
n \pi }{3}\right)\right],\\ \nonumber
d&=&\frac{1}{24} \left[-i \cos\left(\frac{n \pi }{4}\right) \left(5+7 \cos\left(\frac{2 n \pi }{3}\right)\right)-\sqrt{3} \sin\left(\frac{n
\pi }{4}\right) \sin\left(\frac{2 n \pi }{3}\right)\right],~~c=\frac{1}{12} \left[5+ \cos\left(\frac{2 n \pi }{3}\right)+\sqrt{3} \sin\left(\frac{2
n \pi }{3}\right)\right]\\ \nonumber
\mbox{and} ~~~f&=&\frac{1}{24} \left[i \cos\left(\frac{n \pi }{4}\right) \left(5+7 \cos\left(\frac{2 n \pi }{3}\right)\right)+\sqrt{3} \sin\left(\frac{n
\pi }{4}\right) \sin\left(\frac{2 n \pi }{3}\right)\right].\\  \nonumber
\end{eqnarray}
All these coefficients have periodic nature such that $\left({a}_{m}(n), {b}_{m}(n), {c}_{m}(n), {d}_{m}(n), {f}_{m}(n)\right)$=$\left({a}_{m}(n+3), {b}_{m}(n+3), {c}_{m}(n+3),\right.$\\$\left.{d}_{m}(n+24)), {f}_{m}(n+24)\right)$. The concurrence and  eigenvalues $\lambda_l$ (in decreasing order) of $(\sigma_y \otimes \sigma_y)\rho_{12} (\sigma_y \otimes \sigma_y) \rho_{12}^*$ for  $n$=$0$ to $23$  are tabulated in the table \ref{Table:LinearEntropy1222sup}. The concurrence values are plotted in the Fig. $4$ from the main text.
 \begin{table*}[hbtp]
  \centering
 \caption{The eigenvalues and concurrence ($C(n)$) for the initial state  $\otimes^8\ket{+}$ at  evolution steps ($n$) for the system of $8$ qubits.}
 \renewcommand{\arraystretch}{2.4}
\begin{tabular}{|p{0.35cm}|p{6.3cm}|p{2.25cm}||p{0.35cm}|p{6.3cm}|p{2.25cm} | }
 \hline
$n$ &\hspace{2.2cm} Eigenvalues &\hspace{.25cm}Concurrence  &$n$ & \hspace{2.2cm}Eigenvalues& \hspace{.25cm}Concurrence \\
\hline
0 &$(0, 0, 0, 0)$& 0&12&$(0,0,0,0)$&0
 \\
\hline
1 & $(0.254993942654, 0.0625, 0.0575060573456, 0)$& $0.015165042945$& 13&$(0.254993942654, 0.0625, 0.0575060573456, 0)$&$0.015165042945$
\\
\hline

2 & $(0.264971936445, 0.0625, 0.0553405635547, 0)$&$0.029508497188$&14 &$(0.264971936445, 0.0625, 0.0553405635547, 0)$&$0.029508497188$
\\
\hline
3 & $(0.125, 0.125, 0, 0)$&$0$ &15 & $(0.125, 0.125, 0, 0)$&$0$
\\
\hline
4 &$(0.244782496611, 0.0625, 0.05990500339, 0)$&$0$ & 16&$(0.244782496611, 0.0625, 0.05990500339, 0)$&$0$
\\
\hline
5 &$(0.254993942654, 0.0625, 0.0575060573456, 0)$& $0.015165042945$ &17& $(0.254993942654, 0.0625, 0.0575060573456, 0)$& $0.015165042945$
\\
\hline
6 &$(0.25,0.25,0,0)$&$0$ & 18&$(0.25,0.25,0,0)$&$0$
\\
\hline
7 &$(0.254993942654, 0.0625, 0.0575060573456, 0)$ & $0.015165042945$ & 19 & $(0.254993942654, 0.0625, 0.0575060573456, 0)$& $0.015165042945$
\\
\hline
8 &$(0.244782496611, 0.0625, 0.05990500339, 0)$&$0$ &20 & $(0.244782496611, 0.0625, 0.05990500339, 0)$&$0$
\\
\hline
9 & $(0.125, 0.125, 0, 0)$ &$0$&21 &$(0.125, 0.125, 0, 0)$ &$0$
\\
\hline
10 &$(0.264971936445, 0.0625, 0.0553405635547, 0)$&$0.029508497188$ &22 &$(0.264971936445, 0.0625, 0.0553405635547, 0)$&$0.029508497188$
\\
\hline
11 &$(0.254993942654, 0.0625, 0.0575060573456, 0)$ & $0.015165042945$ & 23& $(0.254993942654, 0.0625, 0.0575060573456, 0)$ & $0.015165042945$
\\
\hline
\end{tabular}
  \label{Table:LinearEntropy1222sup}
\end{table*}
\section{Exact analytical solution for ten-qubits case}
Now, we consider a $10$ qubit system that is confined in an eleven-dimensional symmetric subspace. The eigenbasis  in which the unitary operator becomes block diagonal is given as follows:
\begin{eqnarray}
\ket{\phi_0^{\pm}}&=&\frac{1}{\sqrt{2}}(\ket{w_0} \mp \ket{\overline{w_0}}),\\
\ket{\phi_1^{\pm}}&=&\frac{1}{\sqrt{2}}(\ket{w_1} \pm  \ket{\overline{w_1}}),\\
\ket{\phi_2^{\pm}}&=&\frac{1}{\sqrt{2}}(\ket{w_2} \mp  \ket{\overline{w_2}}),\\
\ket{\phi_3^{\pm}}&=&\frac{1}{\sqrt{2}}(\ket{w_3} \pm  \ket{\overline{w_3}}),\\
\ket{\phi_4^{\pm}}&=&\frac{1}{\sqrt{2}}(\ket{w_4} \mp  \ket{\overline{w_4}}),\\
\ket{\phi_5^{+}}&=&\frac{1}{\sqrt{252}}\sum_\mathcal{P}\ket{0000011111},
\end{eqnarray}
where $\ket{w_0}=\ket{0000000000}$, $\ket{\overline{w_0}}=\ket{1111111111}$, $\ket{w_1}=\frac{1}{\sqrt{10}} \sum_\mathcal{P} \ket{0000000001}_\mathcal{P}$, $\ket{\overline {w_1}}=\frac{1}{\sqrt{10}} \sum_\mathcal{P} \ket{0111111111}_\mathcal{P}$, $\ket{w_2}=\frac{1}{\sqrt{45}} \sum_\mathcal{P} \ket{0000000011}_\mathcal{P}$, $\ket{\overline{w_2}}=\frac{1}{\sqrt{45}} \sum_\mathcal{P} \ket{0011111111}_\mathcal{P}$, $\ket{w_3}=\frac{1}{\sqrt{120}}\sum_\mathcal{P} \ket{0000000111}_\mathcal{P}$, $\ket{\overline{w_3}}=\frac{1}{\sqrt{120}} \sum_\mathcal{P} \ket{0001111111}_\mathcal{P}$, $\ket{w_4}=\frac{1}{\sqrt{210}}\sum_\mathcal{P} \ket{0000001111}_\mathcal{P}$  and  $\ket{\overline{w_4}}=\frac{1}{\sqrt{210}} \sum_\mathcal{P} \ket{0000111111}_\mathcal{P}$. As we discussed earlier in this set of basis, $\mathcal{U}$ is block diagonal in $\mathcal{U}_{+}$ and $\mathcal{U}_{-}$ having dimensions $6\times6$ and $5\times5$ respectively. The block $\mathcal{U}_{+}$ is written in positive parity basis $\left\lbrace\phi_0^+,\phi_1^+,\phi_2^+,\phi_3^+,\phi_4^+,\phi_5^+ \right\rbrace $ as follows:
\begin{equation}
\mathcal{U}_{+}=-\frac{e^{\frac{3i \pi }{8}}}{8\sqrt{2}} \left(
\begin{array}{cccccc}
 0 &  \sqrt{5} & 0 &  2\sqrt{15} & 0 & 3\text{  }\sqrt{7} \\
 -e^{\frac{i \pi }{4}} \sqrt{5} & 0 & -9 ~e^{\frac{i \pi }{4}} & 0 & -e^{\frac{i \pi }{4}} \sqrt{42} & 0 \\
 0 & 9  & 0 & 2\text{  }\sqrt{3} & 0 & - \sqrt{35} \\
 2~ e^{\frac{i \pi }{4}}\sqrt{15} & 0 & 2~ e^{\frac{i \pi }{4}}\sqrt{3} & 0 & 2~ e^{\frac{i \pi }{4}} \sqrt{14} & 0 \\
 0 &  \sqrt{42} & 0 & - 2\sqrt{14} & 0 &  \sqrt{30} \\
 -3~ e^{\frac{i \pi }{4}} \sqrt{7} & 0 & e^{\frac{i \pi }{4}} \sqrt{35} & 0 & -e^{\frac{i \pi }{4}} \sqrt{30} & 0 \\
\end{array}
\right),
\end{equation}
whereas, the block $\mathcal{U}_{-}$  written in negative parity basis $\left\lbrace\phi_0^-,\phi_1^-,\phi_2^-,\phi_3^-,\phi_4^-\right\rbrace$ is as follows:
\begin{equation}
\mathcal{U}_{-}=\frac{e^{\frac{3i \pi }{8}}}{16} \left(
\begin{array}{ccccc}
 1 & 0 & 3\text{  }\sqrt{5} & 0 & \sqrt{210} \\
 0 & -8~ e^{\frac{i \pi }{4}} & 0 & -8 \sqrt{3}~ e^{\frac{i \pi }{4}}  & 0 \\
 3\text{  }\sqrt{5} & 0 & 13  & 0 & - \sqrt{42} \\
 0 & 8\sqrt{3}~ e^{\frac{i \pi }{4}}  & 0 & -8~ e^{\frac{i \pi }{4}} & 0 \\
  \sqrt{210} & 0 & -\sqrt{42} & 0 & 2 \\
\end{array}
\right).
\end{equation}
The eigenvalues for $\mathcal{U_{+}}\left(\mathcal{U_{-}}\right)$ are ${\{-1,-1,1,1,i,-i\}}\left(\left\{-(-1)^{3/8},(-1)^{3/8},-(-1)^{23/24},(-1)^{3/8},-(-1)^{7/24}\right\}\right)$\\ and the eigenvectors are $\left\lbrace\left[(-1)^{3/8} \sqrt{\frac{2}{7}},\sqrt{\frac{5}{42}},-(-1)^{3/8} \sqrt{\frac{2}{7}},\sqrt{\frac{5}{42}},-\sqrt{\frac{15}{14}},-\sqrt{\frac{15}{14}} \right]^T,\left[-\sqrt{\frac{5}{7}},-\frac{8(-1)^{5/8}}{\sqrt{21}},-\sqrt{\frac{5}{7}},\frac{8 (-1)^{5/8}}{\sqrt{21}},0,0\right]^T,\right.$\\ $\left. \left[-(-1)^{3/8} \sqrt{\frac{10}{7}},3 \sqrt{\frac{3}{14}},(1-i) (-1)^{5/8} \sqrt{\frac{5}{7}},3 \sqrt{\frac{3}{14}},-\sqrt{\frac{3}{14}},-\sqrt{\frac{3}{14}}\right]^T,
\left[0,0,0,0,-\frac{4
(-1)^{1/8}}{\sqrt{7}},\frac{4 (-1)^{1/8}}{\sqrt{7}} \right]^T,\left[0,1,0,1,1,1 \right]^T,\right.$\\ $\left.\left[1,0,1,0,0,0 \right]^T \right\rbrace \left(\left\lbrace\left[0.684653,-0.0809597,-0.0616442~ i,0,0.728869,0\right]^T,\left[0,0,\frac{1}{\sqrt{2}},0,\frac{1}{\sqrt{2}}\right]^T, \left[-0.306186,0.866322,0,\right.\right.\right.$\\$\left.\left.\left. 0.287612,0\right]^T,\left[0,0,0,\frac{i}{\sqrt{2}} ,0,0.\, \frac{-i}{\sqrt{2}}\right]^T,\left[-0.661438,-0.48483-0.0638078~ i,0,0.621313,0\right]^T \right\rbrace\right)$.  The $n$th time evolution  of $\mathcal{U}_{\pm}$ is  given  as follows:
\begin{equation}
\mathcal{U}_{+}^n=\left(
\begin{array}{cccccc}
 \frac{a_n}{64} \left(17+15 e^{\frac{i n \pi }{2}}\right)  & \frac{(-1)^{3/8}b_n}{16}  \sqrt{\frac{5}{2}}  & \frac{3\sqrt{5}a_n}{64}  \left(-1+e^{\frac{i n \pi }{2}}\right)  & \frac{(-1)^{7/8}}{8}  \sqrt{\frac{15}{2}}
e^{\frac{i n \pi }{2}}  & \frac{a_n}{32} \sqrt{\frac{105}{2}} \left(1-e^{\frac{i n \pi }{2}}\right)  & \frac{3(-1)^{3/8}\sqrt{{7}}b_n}{16\sqrt{2}}    \\
 -\frac{(-1)^{5/8}b_n}{16}  \sqrt{\frac{5}{2}}  & \frac{a_n}{2}  & -\frac{9 (-1)^{5/8}~b_n }{16 \sqrt{2}} & 0 & -\frac{(-1)^{5/8} \sqrt{21} ~b_n }{16} & 0 \\
 \frac{-3\sqrt{5}~a_n}{64}  \left(1-e^{\frac{i n \pi }{2}}\right)  & \frac{9 (-1)^{3/8} b_n}{16 \sqrt{2}}
& \frac{a_n}{64} \left(29+3 e^{\frac{i n \pi }{2}}\right)  & \frac{(-1)^{7/8}~b_n}{8}  \sqrt{\frac{3}{2}} e^{\frac{i n \pi
}{2}}  & \frac{a_n}{32} \sqrt{\frac{21}{2}} \left(1-e^{\frac{i n \pi }{2}}\right)  & -\frac{(-1)^{3/8} ~\sqrt{{35}}b_n}{16\sqrt{2}}
  \\
 \frac{-(-1)^{1/8}~\sqrt{{15}}b_n}{8\sqrt{2}}   e^{\frac{i n \pi }{2}} & 0 & -\frac{(-1)^{1/8}~b_n}{8}  \sqrt{\frac{3}{2}}
e^{\frac{i n \pi }{2}}  & e^{i n \pi } \cos\left(\frac{n \pi }{2}\right) & \frac{(-1)^{1/8} \sqrt{7}~b_n}{8}  e^{\frac{i
n \pi }{2}}  & 0 \\
 \frac{a_n}{32} \sqrt{\frac{105}{2}} \left(1-e^{\frac{i n \pi }{2}}\right)  & \frac{(-1)^{3/8} \sqrt{21}~b_n}{16}   & \frac{a_n}{32} \sqrt{\frac{21}{2}} \left(1-e^{\frac{i n \pi }{2}}\right)  & -\frac{(-1)^{7/8} \sqrt{7}~b_n}{8}
e^{\frac{i n \pi }{2}} & \frac{a_n}{32} \left(9+7 e^{\frac{i n \pi }{2}}\right)  & \frac{(-1)^{3/8} \sqrt{15}~b_n}{16}
  \\
 -\frac{3(-1)^{5/8} ~b_n}{16} \sqrt{\frac{7}{2}}  & 0 & \frac{(-1)^{5/8}~b_n}{16}  \sqrt{\frac{35}{2}}
& 0 & -\frac{(-1)^{5/8} \sqrt{15}~b_n}{16}   & \frac{a_n}{2}  \\
\end{array}
\right)
\end{equation}
\begin{equation}
\mbox{and}~~\mathcal{U}_{-}^n=\left(
\begin{array}{ccccc}
 f_n & 0 & -0.209632~b_n(-1)^\frac{3n}{8} . & 0 & -0.452856~b_n (-1)^\frac{3n}{8}
 \\
 0 & 0.5 c_n (-1)^\frac{31n}{24}  & 0 & (-0.5 i)~d_n(-1)^\frac{31n}{24}  & 0 \\
 -0.209632~ b_n(-1)^\frac{3n}{8}  & 0 & e_n & 0 & 0.2025232~b_n(-1)^\frac{3n}{8}
\\
 0 & (0.5 i)~d_n(-1)^\frac{31n}{24}  & 0 & 0.5~c_n(-1)^{31 n/24}  & 0 \\
 -0.452856~ b_n (-1)^\frac{3n}{8}  & 0 & 0.2025232~b_n(-1)^\frac{3n}{8} & 0 & g_n
\\
\end{array}
\right),
\end{equation}
where $a_n$=$1+e^{i n \pi }$,~$f_n$=$(-1)^\frac{3n}{8}  \left(0.53125 +0.46875 (-1)^n\right)$,~$g_n$=$(-1)^\frac{3n}{8} \left(0.5625+0.4375 (-1)^n\right)$, $b_n=e^{i n \pi }-1$,~ $c_n$=$1+(-1)^{2 n/3}$, $d_n$=$(-1)^{2 n/3}-1$ and $e_n$=$(-1)^\frac{3n}{8} \left(0.90625+0.09375(-1)^n\right)$. The time periodicity of   $\mathcal{U}_{+}^n\left(\mathcal{U}_{-}^n\right)$ is $16(48)$. Hence, the time periodicity of $\mathcal{U}^n$ is $48$.
\subsection{Initial state $\otimes^{10}\ket{0}$= $\ket{\theta_0 =0,\phi_0 =0}$}
The state $\ket{\psi_n}$ can be obtained after the $n$ implementations
of the unitary operator $\mathcal{U}$ on this initial state. Thus,
\begin{eqnarray}
\ket{\psi_n}&=&\mathcal{U}^{n}\ket{0000000000}= \frac{1}{\sqrt{2}}(\mathcal{U}_{+}^n\ket{\phi_0^+}+\mathcal{U}_{-}^n\ket{\phi_0^-}) \\  \nonumber
&=&b_1\ket{\phi_0^+}+b_2\ket{\phi_1^+}+b_3\ket{\phi_2^+}+b_4\ket{\phi_3^+}+b_5\ket{\phi_4^+}+b_6\ket{\phi_5^+}+b_7\ket{\phi_0^-}+b_8\ket{\phi_4^-} +b_9\ket{\phi_5^-},
\end{eqnarray}
where the coefficients are:
\begin{eqnarray}\nonumber
b_1&=&\frac{1}{64} \left(17+15 e^{\frac{i n \pi }{2}}\right) \left(1+e^{i n \pi }\right),~ b_2=\frac{\sqrt{5}(-1)^{5/8}}{16\sqrt{2}} \left[1-e^{i n \pi }\right],~ {b_3}=\frac{3\sqrt{5} }{64} \left(-1+e^{\frac{i n \pi }{2}}\right) \left(1+e^{i n \pi }\right),\\ \nonumber b_4&=&-\frac{(-1)^{1/8}}{8}  \sqrt{\frac{15}{2}}\left[ e^{\frac{i n \pi }{2}} \left(-1+e^{i n \pi }\right)\right],~b_5=-\frac{1}{32} \sqrt{\frac{105}{2}} \left[\left(-1+e^{\frac{i n \pi }{2}}\right) \left(1+e^{i n \pi }\right)\right],~b_6=\frac{3(-1)^{5/8}}{16}  \sqrt{\frac{7}{2}} \left[\left(1-e^{i n \pi }\right)\right]\\ \nonumber b_7 &=&(-1)^{3 n/8} \left(0.53125\, +0.46875 (-1)^n\right),~b_8=0.209632(-1)^{3 n/8} \left(1-(-1)^n\right)~ \mbox{and}~ b_9=0.452856 (-1)^{3 n/8} \left(1- (-1)^n\right).
\end{eqnarray}
\subsubsection{\bf{The linear entropy}}
Here, $\rho_1(n)$ is given as follows:
\begin{equation}
\rho_1(n)={\frac{1}{4}\left(
\begin{array}{cc}
 2+t_n
& z_n \\
 z_n^* & 2-t_n \\
\end{array}
\right)},
\end{equation}
where  the coefficients $t_n$ and $z_n$ are given as follows:
\begin{eqnarray} \nonumber
 t_n&=&3.872 \cos\left(\frac{n \pi }{8}\right) \left[-0.5+\cos^2\left(\frac{n \pi }{4}\right)\right] \left[0.23012+\cos\left(\frac{n
\pi }{4}\right)\right]
\left[\left(-0.08358+ \cos\left(\frac{n \pi }{4}\right)\right)^2+0.2736091\sin^2\left(\frac{n \pi }{4}\right)\right]\mbox{and}\\ \nonumber
z_n&=&e^{-\frac{3}{2} i n \pi }
\left[(-0.07655642372~(1 -i)+
e^{\frac{7 i n \pi }{4}} \left(-(0.31818455215 + 0.335096930284~ i) \left(\cos\left(\frac{n \pi }{8}\right)-\cos\left(\frac{7n \pi }{8}\right)\right)\right.\right. \\ \nonumber && \left.\left. - (0.11488397871 + 0.155714071363 ~i) \left(\cos\left(\frac{3n \pi }{8}\right)-\cos\left(\frac{5n \pi }{8}\right)\right)+0.07655642372  \left(\left(1-i\right)\cos\left(\frac{n \pi}{4}\right)\right)\right.\right. \\ \nonumber && \left.\left.+(0.1557140713628 + 0.114883978711~ i)\left(\sin\left(\frac{n \pi }{8}\right)+\sin\left(\frac{7n \pi }{8}\right)\right)+0.07655642372  \left(\left(1+i\right)\sin\left(\frac{n \pi
}{4}\right)\right)\right.\right. \\ \nonumber && \left.\left.+ (0.3350969303 + 0.31818455215431 ~i) \left(\sin\left(\frac{3n \pi }{8}\right)+\sin\left(\frac{5n \pi }{8}\right)\right)\right)\right].
\end{eqnarray}
The eigenvalues of $\rho_1(n)$ are given as follows:
\begin{eqnarray}\nonumber
\lambda_1 &=& 0.5\pm 0.5 \cos(n\pi) \left[
  0.2509765625 + 0.426776695296637 \cos(\frac{n\pi}{4}) +
   0.073223304703364 \cos(\frac{3n\pi}{4}) - 0.176776695296637 \right.\\ \nonumber
&& \left.\left(\sin(\frac{n\pi}{4})+\cos(\frac{3n\pi}{4})\right) +
   0.2490234375 \left(\cos(\frac{n\pi}{2}) - \sin(\frac{n\pi}{2})\right)\right]^{\frac{1}{2}}.
\end{eqnarray}\\
Thus, the  expression for linear entropy  is given by,
\begin{eqnarray}\label{Eq:ap10Q11} \nonumber
S_{(0,0)}^{(10)}(n)&=& 0.37451171875-(0.21338834765) \cos\left(\frac{n \pi }{4}\right)-0.036611652352 \cos\left(\frac{3
n \pi }{4}\right)+0.088388347649\sin\left(\frac{ n \pi }{4}\right)\\ && +  ~0.0883883476484 \sin\left(\frac{3
n \pi }{4}\right)-0.12451171875\left( \cos\left(\frac{n \pi }{2}\right)-\sin\left(\frac{n \pi }{2}\right)\right).
\end{eqnarray}
Using Eq.~(\ref{Eq:ap10Q11}), we can evaluate the entanglement dynamics for this state. We found  that it is periodic in nature having periodicity $8$ i.e. $S_{(0,0)}^{(10)}(n)=S_{(0,0)}^{(10)}(n+8)$. Notably, the linear entropy remains unchanged for consecutive odd and even values of $n$, which is shown in the Fig. $1$ from the main text. Using these eigenvalues in Eq.~(\ref{von1}), we can easily calculate the entanglement entropy for this initial state. Similar to linear entropy, the von Neumann entropy for this state shows periodic behavior of the same period, which is shown in Fig. $1$ from the main text.
\subsubsection{\bf{Concurrence}}
 The two qubit RDM, $\rho_{12}(n)$, for this state is given as follows:
\begin{equation}
\rho_{12}(n)={\frac{1}{4}\left(
\begin{array}{cccc}
 w_1 & y_1 & y_1 &  x_1^ *\\
 y_1^ *& u_1 & u_1 & v_1 \\
 y_1^ * & u_1 &  u_1& v_1 \\
 x_1 & v_1^ *& v_1^ *& s_1\\
\end{array}
\right)},
\end{equation}
where  the coefficients are:
\begin{eqnarray}\nonumber
w_1&=&1.375+0.46875 \left(\cos\left(\frac{n \pi }{8}\right)+\cos\left(\frac{7 n \pi }{8}\right)\right)+0.53125 \left(\cos\left(\frac{3 n \pi }{8}\right)+\cos\left(\frac{5 n \pi }{8}\right)\right)+0.25\cos\left(\frac{n
\pi }{2}\right)+0.375\cos(n \pi ),\\ \nonumber
s_1&=&1.375-0.46875 \left(\cos\left(\frac{n \pi }{8}\right)+\cos\left(\frac{7 n \pi }{8}\right)\right)-0.53125 \left(\cos\left(\frac{3 n \pi }{8}\right)+\cos\left(\frac{5 n \pi }{8}\right)\right)+0.25\cos\left(\frac{n
\pi }{2}\right)+0.375\cos(n \pi ),\\ \nonumber
x_1&=&0.125 +0.25 \cos\left(\frac{n \pi }{2}\right)-0.375 \cos(n \pi )-(0.03125
i)\left[ \sin\left(\frac{n \pi }{8}\right)+ \sin\left(\frac{3 n \pi }{8}\right)\right]-0.5 \sin\left(\frac{n \pi }{2}\right)\\ \nonumber &&+(0.03125i) \left[\sin\left(\frac{5n \pi }{8}\right)+ \sin\left(\frac{7
n \pi }{8}\right)\right],\\ \nonumber
u_1&=& \left[4.5-0.25 \cos\left(\frac{n \pi }{8}\right)+6.5 \cos\left(\frac{n \pi }{4}\right)-0.25\cos\left(\frac{3 n
\pi }{8}\right)+4 \cos\left(\frac{n \pi }{2}\right)-0.25 \cos\left(\frac{5 n \pi }{8}\right)+2.5 \cos\left(\frac{3 n \pi }{4}\right)\right.\\ \nonumber &&
\left.-0.25
\cos\left(\frac{7 n \pi }{8}\right)+0.5 \cos(n \pi )\right] \sin^2\left(\frac{n \pi }{8}\right),\\ \nonumber
y_1 &=&-(0.2165342655 + 0.005979428631 i)  \left(\cos\left(\frac{n \pi }{8}\right)-\cos\left(\frac{7n \pi }{8}\right)\right)-(0.101650286722 + 0.014435617696 i)
\\ \nonumber && \left(\cos\left(\frac{3n \pi }{8}\right)-\cos\left(\frac{5n \pi }{8}\right)\right)+(0.08969142946 - 0.0144356176955 i) \left(\sin\left(\frac{ n \pi }{8}\right)+\sin\left(\frac{ 7n \pi }{8}\right)\right)\\ \nonumber &&+(0.24540550083 - 0.005979428631 i) \left(\sin\left(\frac{ 3n \pi }{8}\right)+\sin\left(\frac{ 5n \pi }{8}\right)\right) \mbox{and}\\ \nonumber
v_1&=&-(0.2165342655 -0.005979428631 i)  \left(\cos\left(\frac{n \pi }{8}\right)-\cos\left(\frac{7n \pi }{8}\right)\right)-(0.101650286722 - 0.014435617696 i)
\\ \nonumber && \left(\cos\left(\frac{3n \pi }{8}\right)-\cos\left(\frac{5n \pi }{8}\right)\right)+(0.08969142946 + 0.0144356176955 i) \left(\sin\left(\frac{ n \pi }{8}\right)+\sin\left(\frac{ 7n \pi }{8}\right)\right)\\ \nonumber &&+(0.24540550083 + 0.005979428631 i) \left(\sin\left(\frac{ 3n \pi }{8}\right)+\sin\left(\frac{ 5n \pi }{8}\right)\right).
\end{eqnarray}
All these coefficients are periodic in nature i.e. $\left(\bar{w}_1(n), \bar{s}_1(n), \bar{x}_1(n), \bar{u}_1(n), \bar{y}_1(n), \bar{v}_1(n)\right)$=$\left(\bar{w}_1(n+4), \bar{s}_1(n+4), \bar{x}_1(n+8), \right.$\\$\left.\bar{u}_1(n+4), \bar{y}_1(n+4), \bar{v}_1(n+4)\right)$. The eigenvalues $\lambda_l$ (in decreasing order) of $(\sigma_y \otimes \sigma_y)\rho_{12} (\sigma_y \otimes \sigma_y) \rho_{12}^*$   and concurrence values  for  $n$ ranging from $0$ to $15$  are represented in Table \ref{Table:LinearEntropy103sup}, which shows a period of $8$. The concurrence values are plotted in the Fig. $2$ from the main text.

\begin{table*}[hbtp!]
  \centering
 \caption{The eigenvalues and concurrence ($C(n)$) for the initial state $\otimes^{10}\ket{0}$ at  evolution steps ($n$)   for the system of $10$ qubits.}
 \renewcommand{\arraystretch}{2.4}
\begin{tabular}{|p{0.35cm}|p{6cm}|p{2.25cm}||p{0.35cm}|p{6cm}|p{2.25cm} | }
 \hline
$n$ &\hspace{2cm} Eigenvalues& \hspace{.25cm}Concurrence  &$n$ &\hspace{2cm} Eigenvalues& \hspace{.25cm}Concurrence \\
\hline
0 &$(0, 0, 0, 0)$& 0&8&$(0,0,0,0)$&0
 \\
\hline
1 & $(0.25128814735, 0.0625, 0.061211852656, 0)$& $0.003876200145$& 9&$(0.25128814735, 0.0625, 0.061211852656, 0)$&$0.003876200145$
\\
\hline

2 & $(0.25128814735, 0.0625, 0.061211852656, 0)$&$0.003876200145$&10 &$(0.25128814735, 0.0625, 0.061211852656, 0)$&$0.003876200145$
\\
\hline
3 & $(0.25, 0.25, 0, 0)$&$0$ &11 & $(0.25, 0.25, 0, 0)$&$0$
\\
\hline
4 & $(0.25, 0.25, 0, 0)$&$0$ &12 & $(0.25, 0.25, 0, 0)$&$0$
\\
\hline
5 &$(0.25128814735, 0.0625, 0.061211852656, 0)$& $0.003876200145$ &13& $(0.25128814735, 0.0625, 0.061211852656, 0)$& $0.003876200145$
\\
\hline
6 &$(0.25128814735, 0.0625, 0.061211852656, 0)$&$0.003876200145$ & 14&$(0.25128814735, 0.0625, 0.061211852656, 0)$&$0.003876200145$
\\
\hline
7 &$(0,0,0, 0)$ & $0$ & 15 & $(0,0,0, 0)$& $0$
\\
\hline
\end{tabular}
  \label{Table:LinearEntropy103sup}
\end{table*}
\subsection{Initial state $\otimes^{10}\ket{+}$= $\ket{\theta_0 =\pi/2,\phi_0 =-\pi/2}$}
The  state can be written as, $\otimes ^{10} {\ket{+}}_y=\frac{1}{16\sqrt{2}}\left( \ket{\phi_0^+}+{i \sqrt{10}}\ket{\phi_1^+}- {3\sqrt{5}} \ket{\phi_2^+}-{2i\sqrt{30}}\ket{\phi_3^+}+{\sqrt{210}}\ket{\phi_4^+}+{3 i\sqrt{14}}\ket{\phi_5^+}\right)$. The state $\ket{\psi_n}$ can be obtained by operating unitary operator $n$ times on this initial state. Thus,
\begin{eqnarray}
\ket{\psi_n}&=& \mathcal{U}_{+}^n\ket{++++++++++}\\  \nonumber
            &=&q_{1n} \ket{\phi_0^+}+w_{1n} \ket{\phi_1^+}+ r_{1n} \ket{\phi_2^+}+t_{1n} \ket{\phi_3^+}+y_{1n} \ket{\phi_4^+}+o_{1n} \ket{\phi_5^+},
\end{eqnarray}
where  the coefficients are:
\begin{eqnarray} \nonumber
 q_{1n}&=&\frac{-17 \left(-1+(-1)^{7/8}\right)-15 \left(1+(-1)^{3/8}\right) e^{\frac{i n \pi }{2}}+17 \left(1+(-1)^{7/8}\right) e^{i n \pi
}+15 \left(-1+(-1)^{3/8}\right) e^{\frac{3 i n \pi }{2}}}{64 \sqrt{2}},\\ \nonumber
 r_{1n}&=&-\frac{3}{64} \sqrt{\frac{5}{2}} \left[1-(-1)^{7/8}+\left(1+(-1)^{3/8}\right) e^{\frac{i n \pi }{2}}+\left(1+(-1)^{7/8}\right) e^{i
n \pi }-\left(-1+(-1)^{3/8}\right) e^{\frac{3 i n \pi }{2}}\right],\\  \nonumber
w_{1n}&=&\frac{ \sqrt{5}}{32} \left[i+(-1)^{5/8}-\left(-i+(-1)^{5/8}\right) e^{i n \pi }\right], ~~t_{1n}=\frac{\sqrt{15}~ e^{\frac{i n \pi }{2}}}{16}  \left[-i-(-1)^{1/8}+\left(-i+(-1)^{1/8}\right) e^{i n \pi }\right]\\ \nonumber
y_{1n}&=&\frac{\sqrt{105}}{64}  \left[1-(-1)^{7/8}+\left(1+(-1)^{3/8}\right) e^{\frac{i n \pi }{2}}+\left(1+(-1)^{7/8}\right) e^{i n \pi
}-\left(-1+(-1)^{3/8}\right) e^{\frac{3 i n \pi }{2}}\right]\mbox{and } \\ \nonumber
 o_{1n}&=&\frac{3\sqrt{7}}{32}  \left[i+(-1)^{5/8}-\left(-i+(-1)^{5/8}\right) e^{i n \pi }\right].
\end{eqnarray}
\subsubsection{\bf{The linear entropy}}
The $\rho_1(n)$  for this state is given by,
\begin{equation}
  \rho_1(n)={\frac{1}{2}\left[
\begin{array}{cc}
 1 & V \\
V^* & 1 \\
\end{array}
\right]},
\end{equation}
where $V$=$-\frac{i}{64} \left[17 \left(1+\cos(n \pi )\right)+\sqrt{2} \left(1-\cos(n \pi )\right) +30 \cos\left(\frac{ n \pi
}{2}\right)\right]$. The eigenvalues of $\rho_1(n)$  are $ \left[\frac{1}{2}\pm\frac{1}{64} \left(258\right.\right.$\\$\left.\left.+510 \cos\left(\frac{n \pi }{2}\right)+256 \cos(n \pi )\right)^\frac{1}{2}\right]$. Thus, the linear entropy is given as follows:
\begin{eqnarray}\label{eq:ap100Q5}
S_{(\pi/2,-\pi/2)}^{(10)}(n)&=&\frac{1}{2}\left\lbrace 1-\frac{1}{4096}\left[17 \left(1+\cos(n \pi )\right)+\sqrt{2} \left(1-\cos(n \pi )\right) +30 \cos\left(\frac{ n \pi
}{2}\right)\right]^2\right\rbrace.
\end{eqnarray}
From the Eq. (\ref{eq:ap100Q5}), we can see that the entanglement dynamics is
periodic in nature having periodicity $4$. Using these eigenvalues in Eq.~(\ref{von1}), we can easily calculate the entanglement entropy for this initial state. Similar to linear entropy, the von Neumann entropy  shows periodic behavior with the same period, which is shown in Fig. $3$ from the main text.
\subsubsection{\bf{Concurrence}}
 The $\rho_{12}(n)$ is given as follows,
\begin{equation}
\rho_{12}(n)={\frac{1}{2}\left[
\begin{array}{cccc}
  g_n& h_n & h_n & j_n \\
 h_n^* & l_n & l_n & k_n \\
h_n^* & l_n & l_n & k_n \\
 j_n^* & k_n^* & k_n^* & g_n \\
\end{array}
\right]},
\end{equation}
where all the coefficients are:
\begin{eqnarray} \nonumber
g_n&=&\frac{1}{8} \left[5- \left(\cos\left(\frac{n \pi }{2}\right)- \sin\left(\frac{n \pi }{2}\right)\right)\right],~j_n=\frac{-1}{8} \left[1+3 \cos\left(\frac{n \pi }{2}\right)+\sin\left(\frac{n \pi }{2}\right)\right],~l_n=\frac{1}{8} \left[3+ \left(\cos\left(\frac{n \pi }{2}\right)- \sin\left(\frac{n \pi }{2}\right)\right)\right],\\ \nonumber
h_n&=&\left(-\frac{1}{512}-\frac{i}{512}\right) \left[ 4\sqrt{2}(1 - \cos(n \pi) + (1 + i) (60 \cos\left(\frac{n\pi}{2}\right) + 34 (1 + \cos(n \pi)))\right]~ \mbox{and}\\ \nonumber
k_n&=&\left(-\frac{1}{512}-\frac{i}{512}\right) \left[ 4i\sqrt{2}(1 - \cos(n \pi) + (1 + i) (60 \cos\left(\frac{n\pi}{2}\right) + 34 (1 + \cos(n \pi)))\right].
\end{eqnarray}
All these coefficients have periodic nature with period $4$ i.e $\left({g}_{n}(n), {h}_{n}(n), {j}_{n}(n), {k}_{n}(n), {l}_{n}(n)\right)$=$\left({g}_{n}(n+4), {h}_{n}(n+4),\right.$\\$\left.{j}_{n}(n+4), {k}_{n}(n+4)), {l}_{n}(n+4)\right)$. The  concurrence and eigenvalues $\lambda_l$ (in decreasing order) of $(\sigma_y \otimes \sigma_y)\rho_{12} (\sigma_y \otimes \sigma_y) \rho_{12}^*$   for $n$ ranging from $0$ to $7$ are presented in Table \ref{Table:LinearEntropy153sup}. It  is shown in the Fig. $4$ from the main text.
\begin{table*}[!htp]
  \centering
 \caption{The eigenvalues and concurrence ($C(n)$) for the initial state $\otimes^{10}\ket{+}$ at  evolution steps ($n$)   for the system of $10$ qubits.}
 \renewcommand{\arraystretch}{2.4}
\begin{tabular}{|p{0.35cm}|p{6cm}|p{2.25cm}||p{0.35cm}|p{6cm}|p{2.25cm} | }
 \hline
$n$ &\hspace{2cm} Eigenvalues& \hspace{.25cm} Concurrence  &$n$ &\hspace{2cm} Eigenvalues& \hspace{.25cm} Concurrence \\
\hline
0 &$(0, 0, 0, 0)$& 0&4&$(0,0,0,0)$&0
 \\
\hline
1 & $(0.25128814735, 0.0625, 0.061211852656, 0)$& $0.003876200145$& 5&$(0.25128814735, 0.0625, 0.061211852656, 0)$& $0.003876200145$
\\
\hline
2 & $(0.25, 0.0615234375, 0.0615234375, 0)$&$0.003921629176$&6 &$(0.25, 0.0615234375, 0.0615234375, 0)$&$0.003921629176$
\\
\hline
3 & $(0.25128814735, 0.0625, 0.061211852656, 0)$& $0.003876200145$ &7 & $(0.25128814735, 0.0625, 0.061211852656, 0)$& $0.003876200145$
\\
\hline
\end{tabular}
  \label{Table:LinearEntropy153sup}
\end{table*}\\

\section{Exact analytical solution for twelve-qubits case}
Now, we consider a $12$ qubit system that is confined in a thirteen-dimensional symmetric subspace. The eigenbasis  in which the unitary operator becomes block diagonal is given as follows:
\begin{eqnarray}
\ket{\phi_0^{\pm}}&=&\frac{1}{\sqrt{2}}(\ket{w_0} \mp \ket{\overline{w_0}}),\\
\ket{\phi_1^{\pm}}&=&\frac{1}{\sqrt{2}}(\ket{w_1} \pm  \ket{\overline{w_1}}),\\
\ket{\phi_2^{\pm}}&=&\frac{1}{\sqrt{2}}(\ket{w_2} \mp  \ket{\overline{w_2}}),\\
\ket{\phi_3^{\pm}}&=&\frac{1}{\sqrt{2}}(\ket{w_3} \pm  \ket{\overline{w_3}}),\\
\ket{\phi_4^{\pm}}&=&\frac{1}{\sqrt{2}}(\ket{w_4} \mp  \ket{\overline{w_4}}),\\
\ket{\phi_5^{\pm}}&=&\frac{1}{\sqrt{2}}(\ket{w_5} \mp  \ket{\overline{w_5}}),  \\
\mbox{and} ~~\ket{\phi_6^{+}}&=&\frac{1}{\sqrt{924}}\sum_\mathcal{P}\ket{000000111111},
\end{eqnarray}
here $\ket{w_0}$=$\ket{000000000000}$, $\ket{\overline{w_0}}$=$\ket{111111111111}$, $\ket{w_1}$=$\frac{1}{\sqrt{12}} \sum_\mathcal{P} \ket{000000000001}_\mathcal{P}$, $\ket{\overline {w_1}}$=$\frac{1}{\sqrt{12}} \sum_\mathcal{P} \ket{011111111111}_\mathcal{P}$,\\ $\ket{w_2}$=$\frac{1}{\sqrt{66}} \sum_\mathcal{P} \ket{000000000011}_\mathcal{P}$, $\ket{\overline{w_2}}$=$\frac{1}{\sqrt{66}} \sum_\mathcal{P} \ket{001111111111}_\mathcal{P}$, $\ket{w_3}$=$\frac{1}{\sqrt{220}}\sum_\mathcal{P} \ket{000000000111}_\mathcal{P}$,\\$\ket{\overline{w_3}}$=$\frac{1}{\sqrt{220}} \sum_\mathcal{P} \ket{000111111111}_\mathcal{P}$, $\ket{w_4}$=$\frac{1}{\sqrt{495}}\sum_\mathcal{P} \ket{000000001111}_\mathcal{P}$, $\ket{\overline{w_4}}$=$\frac{1}{\sqrt{495}} \sum_\mathcal{P} \ket{000011111111}_\mathcal{P}$,\\$\ket{w_5}$=$\frac{1}{\sqrt{792}}\sum_\mathcal{P} \ket{000000011111}_\mathcal{P}$ and  $\ket{\overline{w_5}}$=$\frac{1}{\sqrt{792}} \sum_\mathcal{P} \ket{000001111111}_\mathcal{P}$. As we discussed earlier in this set of basis $\mathcal{U}$ is block diagonal in $\mathcal{U}_{+}$ and $\mathcal{U}_{-}$ having dimensions $6\times6$ and $5\times5$ respectively. The block $\mathcal{U}_{+}$ is written in positive parity basis $\left\lbrace\phi_0^+,\phi_1^+,\phi_2^+,\phi_3^+,\phi_4^+,\phi_5^+,\phi_6^+ \right\rbrace $ as follows:
\begin{equation}
\mathcal{U}_{+}=\frac{1}{32}\left(
\begin{array}{ccccccc}
  e^{-\frac{i \pi }{4}} & 0 &  \sqrt{66}~ e^{-\frac{i \pi }{4}} & 0 & 3 \sqrt{55}~ e^{-\frac{i \pi }{4}} & 0 &  \sqrt{462}~ e^{-\frac{i \pi }{4}} \\
 0 & -10 i & 0 & -2i \sqrt{165} & 0 & -2 i \sqrt{66} & 0 \\
  \sqrt{66}~ e^{\frac{3 i \pi }{4}} & 0 & 13~ e^{\frac{3 i \pi }{4}} & 0 &  \sqrt{30}~ e^{\frac{3 i \pi }{4}} & 0 & -6\sqrt{7}~ e^{\frac{3 i \pi }{4}}
\\
 0 & -2i \sqrt{165} & 0 & -2i & 0 & 6 i \sqrt{10} & 0 \\
 3 \sqrt{55}~ e^{-\frac{i \pi }{4}} & 0 &  \sqrt{30}~ e^{-\frac{i \pi }{4}} & 0 & -17~ e^{-\frac{i \pi }{4}} & 0 &  \sqrt{210}~ e^{-\frac{i \pi }{4}}
\\
 0 & -2 i \sqrt{66} & 0 & 6 i \sqrt{10} & 0 & -20 i & 0 \\
  \sqrt{462}~ e^{\frac{3 i \pi }{4}} & 0 & -6 \sqrt{7}~ e^{\frac{3 i \pi }{4}} & 0 &  \sqrt{210}~ e^{\frac{3 i \pi }{4}} & 0 & -10 ~e^{\frac{3 i \pi
}{4}} \\
\end{array}
\right),
\end{equation}
whereas, the block $\mathcal{U}_{-}$  written in negative parity basis $\left\lbrace\phi_0^-,\phi_1^-,\phi_2^-,\phi_3^-,\phi_4^-,\phi_5^-\right\rbrace$ is as follows: \\
\begin{equation}
\mathcal{U}_{-}=\frac{1}{16}\left(
\begin{array}{cccccc}
 0 & - \sqrt{3}~ e^{-\frac{i \pi }{4}} & 0 & - \sqrt{55}~ e^{-\frac{i \pi }{4}} & 0 & -3 \sqrt{22}~ e^{-\frac{i \pi }{4}} \\
 i \sqrt{3} & 0 & 2 i \sqrt{22} & 0 & i \sqrt{165} & 0 \\
 0 & -2 \sqrt{22}~ e^{\frac{3 i \pi }{4}} & 0 & -2 \sqrt{30}~ e^{\frac{3 i \pi }{4}} & 0 & 4 \sqrt{3}~ e^{\frac{3 i \pi }{4}} \\
 i \sqrt{55} & 0 & 2 i \sqrt{15} & 0 & -9 i & 0 \\
 0 & - \sqrt{165}~ e^{-\frac{i \pi }{4}} & 0 & 9~ e^{-\frac{i \pi }{4}} & 0 & - \sqrt{10}~ e^{-\frac{i \pi }{4}} \\
 3 i \sqrt{22} & 0 & -4i \sqrt{3} & 0 &  i \sqrt{10} & 0 \\
\end{array}
\right).
\end{equation}
The eigenvalues for $\mathcal{U_{+}}\left(\mathcal{U_{-}}\right)$ are ${\left\{e^{\frac{17 i n \pi }{12}},e^{\frac{i n \pi }{2}},e^{\frac{i n \pi }{12}},e^{\frac{3 i n \pi }{4}},e^{\frac{3 i n \pi }{4}},e^{\frac{3
i n \pi }{2}},e^{\frac{3 i n \pi }{2}}\right\}}\left(\left\{(-1)^{5/8},(-1)^{5/8},-(-1)^{1/8},(-1)^{1/8},\right.\right.$\\$\left.\left.-(-1)^{5/8},-(-1)^{5/8}\right\}\right)$ and the eigenvectors are $\left\lbrace\left[0.586301969978~i,0,0.25,0,0.3952847075211~i,0,0.6614378277661 \right]^T,\right.$\\ $\left.\left[0,-0.5863019699779289,0,0.6846531968814575,0,0.4330127018922193,0\right]^T,\right.$\\ $\left.
\left[-0.5863019699779286~ i,0,0.25,0,-0.3952847075210475~i,0,0.6614378277661478 \right]^T,\right.$\\ $\left.\left[-0.5590169943749475,0,0,0,0.82915619758885,0,0 \right]^T,\right.$\\ $\left.\left[-0.02351922113063-0.076007268961~ i,0,0.925891611177,0,0.034884642434 + 0.1127369986151~ i,-0.349954134882 \right]^T ,\right.$\\ $\left.\left[0,0.7864166677879,0,0.60872499475+0.01230981550379~ i,0,0.1023344758186\, -0.01946352728421~i,0 \right]^T,\right.$\\ $\left.\left[0,0.11788408575671\, +0.10196506311775~i,0,-0.45183482623614+0.08731766337638~i,0,0.87402939521,0 \right]^T\right\rbrace $\\$\left(\left\lbrace\left[\frac{8(-1)^{1/8}}{3}  \sqrt{\frac{2}{11}},-\frac{1}{3}\sqrt{\frac{5}{11}},0,0,-\frac{8(-1)^{1/8}}{3}  \sqrt{\frac{2}{11}},-\frac{1}{3}\sqrt{\frac{5}{11}}\right]^T,\left[\frac{1}{\sqrt{66}},-2
(-1)^{7/8} \sqrt{\frac{5}{33}},-\sqrt{\frac{11}{6}},-\sqrt{\frac{11}{6}},\frac{1}{\sqrt{66}},2 (-1)^{7/8} \sqrt{\frac{5}{33}}\right]^T,\right.\right.$\\$\left.\left.\left[0,0,-\frac{4 (-1)^{5/8}}{\sqrt{3}},\frac{4 (-1)^{5/8}}{\sqrt{3}},0,0\right]^T,\left[\frac{1}{3}\sqrt{\frac{5}{2}},\frac{2(-1)^{7/8}}{3} ,-\sqrt{\frac{5}{2}},-\sqrt{\frac{5}{2}},\frac{1}{3}\sqrt{\frac{5}{2}},-\frac{2(-1)^{7/8}}{3}
\right]^T,\left[0,1,0,0,0,1\right]^T,\left[1,0,1,1,1,0\right]^T \right\rbrace\right)$. \\

 The $n$th time evolution  of $\mathcal{U}_{\pm}$ is  given  as follows :
\begin{equation}
\mathcal{U}_{+}^n=\left(
\begin{array}{ccccccc}
 j_n & 0 & -\bar{s_n}  & 0 & j_n^{\prime} & 0 & -\bar{j_n} \\
 0 & y_n & 0 & \bar{g_n}  & 0 & \bar{h_n} & 0 \\
 \bar{s_n}  & 0 & h_n^{\prime}  & 0 & -\bar{r_n}
 & 0 &h_n \\
 0 &\bar{g_n}  & 0 & y_n^{\prime}
& 0 & (-0.296463531b_n)~ e^{\frac{i n \pi }{2}} & 0 \\
j_n^{\prime}  & 0 & \bar{r_n} & 0 & t_n^{\prime} & 0 & (-0.261456 i~r_n) ~e^{\frac{i n \pi }{12}}  \\
 0 &\bar{h_n}   & 0 & (-0.296463531b_n) ~e^{\frac{i n \pi }{2}}
& 0 & r_n^{\prime} & 0 \\
\bar{j_n}  & 0 & h_n& 0 & (0.261456 i~r_n)~ e^{\frac{i n \pi
}{12 }} & 0 & s_n^{\prime}  \\
\end{array}
\right)\mbox{and}
\end{equation}
\begin{equation}
\mathcal{U}_{-}^n= \frac{e^{\frac{5 i n \pi }{8}}}{32} \left(
\begin{array}{cccccc}
 16a_n & - e^{\frac{i \pi }{8}} \sqrt{3} b_n & 0 & - e^{\frac{i \pi }{8}} \sqrt{55}b_n  & 0 & -6b_n e^{\frac{i \pi }{8}} \sqrt{\frac{11}{2}}
 \\
  e^{\frac{7 i  \pi }{8}} \sqrt{3} b_n & \frac{a_n}{2}  \left(11e^{\frac{-i n \pi }{2}}+21 \right) & - 2b_n  \sqrt{{22}} e^{(\frac{-i n \pi }{2}+\frac{3i \pi }{8})}
&  i\bar{c_n} \sqrt{165} e^{\frac{ i n \pi }{4}}
&  \sqrt{165}b_ne^{\frac{7 i  \pi }{8}}   & - i\bar{c_n} \sqrt{{66}} e^{\frac{ i n \pi
}{4}}  \\
 0 & 2b_n  \sqrt{{22}} e^{(\frac{-i n \pi }{2}+\frac{5i \pi }{8})}  & 16a_n e^{\frac{-i n \pi }{2}}  & 2b_n e^{(\frac{-i n \pi }{2}+\frac{5i \pi }{8})} \sqrt{{30}} & 0 & -4\sqrt{3}b_n e^{(\frac{-i
n \pi }{2}+\frac{5i
 \pi }{8})}  \\
   e^{\frac{7 i  \pi }{8}} \sqrt{55} b_n &  i \bar{c_n}\sqrt{165} e^{\frac{ i n \pi }{4}}  & -2b_n e^{(\frac{-i n \pi }{2}+\frac{3i \pi }{8})} \sqrt{{30}}  & \frac{a_n}{2}  \left(15e^{\frac{-i n \pi }{2}}+17 \right)  & -{9b_n}
e^{\frac{7 i  \pi }{8}} & - 6i\bar{c_n} \sqrt{\frac{5}{2}} e^{\frac{ i n \pi }{2}} \\
 0 & -  \sqrt{165}b_n e^{\frac{ i  \pi }{8}}  & 0 & {9}b_n e^{\frac{i  \pi }{8}}
 & 16a_n & - \sqrt{{10}}b_n e^{\frac{
i  \pi }{8}}  \\
3b_n  e^{\frac{7 i  \pi }{8}}\sqrt{{22}}  & - i\bar{c_n} \sqrt{{66}} e^{\frac{
i n \pi }{4}}  & 4\sqrt{3}b_n e^{(\frac{-i
n \pi }{2}+\frac{3i
 \pi }{8})}  & -{3} i \sqrt{{10}}\bar{c_n} e^{\frac{ i n \pi }{4}}  &   \sqrt{{10}}b_n e^{\frac{7i
\pi }{8}}  &  a_n \left(3e^{\frac{-i n \pi }{2}}+13 \right)  \\
\end{array}
\right)
\end{equation}
where $a_n$=$1+e^{i n \pi }$, $b_n=-1+e^{i n \pi }$, $t_n$=$1+ e^{\frac{-2 i n \pi }{3}} $, $r_n$=$1- e^{\frac{-2 i n \pi }{3}}$, $y_n$=$e^{\frac{i n \pi }{2}}\left(0.34375 +0.65625~ e^{ i n \pi }\right)$,\\$h_n$=$(0.165359 ) e^{\frac{i n \pi}{12 }}-(0.330719) e^{\frac{3 i n \pi }{4}}$,~ $j_n$=$(0.3125) e^{\frac{3 i n \pi }{4}}+(0.34375t_n ) e^{\frac{i n \pi }{12}}$, ~$j_n^{\prime}$=$ -(0.4635124) e^{\frac{3 i n \pi }{4}}+(0.231756 t_n) e^{\frac{i n \pi }{12}}$,\\$h_n^{\prime}$=$ (0.875) e^{\frac{3 i n \pi }{4}}+(0.0625 t_n) e^{\frac{i
n \pi }{12}}$, $y_n^{\prime}$=$(0.46875) e^{\frac{i n \pi }{2}}+(0.53125) e^{\frac{3 i n \pi }{2}}$, $t_n^{\prime}$=$(0.15625t_n) e^{\frac{i n \pi }{12 }}+(0.6875)
e^{\frac{3 i n \pi }{4}}$, $r_n^{\prime}$=$(0.1875) e^{\frac{i n \pi }{2}}+(0.8125) e^{\frac{3 i n \pi }{2}}$, $s_n^{\prime}$=$(0.125) e^{\frac{3 i n \pi }{4}}+(0.43745t_n) e^{\frac{ i n \pi }{12}}$, $\bar{s}_n$= $(0.1465755 i~r_n) e^{\frac{i n \pi }{12}}$, $\bar{j}_n$=$(0.387802 i~r_n) e^{\frac{i n \pi }{12}}$,\\$\bar{g}_n$=$(0.40141352b_n) e^{\frac{i n \pi }{2}} $, $\bar{h}_n$=$(0.2538762b_n) e^{\frac{i n \pi }{2}}$, $\bar{r}_n$=$(-0.0988212 i~r_n) e^{\frac{in \pi }{12}}$ and $\bar{c}_n$=$\sin\left(\frac{n \pi }{4}\right)-\sin\left(\frac{3 n \pi }{4}\right)$. The time periodicity of each element of $\mathcal{U}_{+}^n\left(\mathcal{U}_{-}^n\right)$ is $24(16)$. Hence, the time periodicity of $\mathcal{U}^n$ is $48$.
\subsection{Initial state $\otimes^{12}\ket{0}$= $\ket{\theta_0 =0,\phi_0 =0}$}
The state $\ket{\psi_n}$ can be obtained after the  $n$ implementations of the unitary operator $\mathcal{U}$ on this initial state. Thus,
\begin{eqnarray}
\ket{\psi_n}&=&\mathcal{U}^{n}\ket{000000000000}= \frac{1}{\sqrt{2}}(\mathcal{U}_{+}^n\ket{\phi_0^+}+\mathcal{U}_{-}^n\ket{\phi_0^-}) \\  \nonumber
&=&b_1^{\prime}\ket{\phi_0^+}+b_2^{\prime}\ket{\phi_2^+}+b_3^{\prime}\ket{\phi_4^+}+b_4^{\prime}\ket{\phi_6^+}+b_5^{\prime}\ket{\phi_0^-}+b_6^{\prime}\ket{\phi_1^-}+b_7^{\prime}\ket{\phi_3^-}+b_8^{\prime}\ket{\phi_5^-} ,
\end{eqnarray}
where the coefficients are:
\begin{eqnarray}\nonumber
b_1^{\prime}&=&(0.3125)~ e^{\frac{3 i n \pi }{4}}+(0.34375\, )~ e^{\frac{i n \pi }{12}}\left(1+ e^{\frac{-2 i n \pi }{3}} \right), ~b_2^{\prime}=(0.1465755~ i)~ e^{\frac{i n \pi }{12}}\left(1- e^{\frac{-2 i n \pi }{3}} \right), \\ \nonumber{b_3^{\prime}}&=&-(0.4635124)~ e^{\frac{3 i n \pi }{4}}+(0.231756)~ e^{\frac{i n \pi }{12}} \left(1+ e^{\frac{-2i n \pi }{3}} \right),~ b_4^{\prime}=(0.387802 ~i)~ e^{\frac{i n \pi }{12}} \left(1- e^{\frac{-2 i n \pi }{3}} \right),~b_5^{\prime}=\frac{e^{\frac{5 i n \pi }{8}}}{2}  \left(1+e^{i n \pi }\right),\\ \nonumber
b_6^{\prime}&=&\frac{(-1)^{7/8} \sqrt{3}}{32}  e^{\frac{5 i n \pi }{8}} \left(-1+e^{i n \pi }\right),~ b_7^{\prime} =\frac{(-1)^{7/8} \sqrt{55}~e^{\frac{5 i n \pi }{8}}}{32}   \left(-1+e^{i n \pi }\right) ~\mbox{and}~~b_8^{\prime}=\frac{3 (-1)^{7/8}~ e^{\frac{5 i n \pi }{8}}}{16} \sqrt{\frac{11}{2}}  \left(-1+e^{i n \pi }\right) .
\end{eqnarray}
\subsubsection{\bf{The linear entropy}}
The $\rho_1(n)$ is given as follows:
\begin{equation}
\rho_1(n)={\frac{1}{4}\left(
\begin{array}{cc}
 2+t_n^{\prime}
& x_n \\
 x_n^* & 2-t_n^{\prime} \\
\end{array}
\right)},
\end{equation}
where the  coefficients $t_n$ and $x_n$ are given as follows:
\begin{eqnarray}\nonumber
t_n&=&\frac{5~e^{-\frac{1}{2} i n \pi } \left(1+e^{i n \pi }\right)}{16}  \left[\cos\left(\frac{3 n \pi }{8}\right)+\frac{11}{10}\left(
\cos\left(\frac{25 n \pi }{24}\right)+ \cos\left(\frac{7 n \pi }{24}\right)\right)\right]~~\mbox{and}\\ \nonumber
x_n&=&(-0.288712354) \left(\cos\left(\frac{n \pi }{8}\right)-\cos\left(\frac{7 n \pi }{8}\right)\right)+
(0.119588573) \left(\sin\left(\frac{n \pi }{8}\right)+\sin\left(\frac{7 n \pi }{8}\right)\right)+\left(1-\cos(n\pi)\right)\\&&\nonumber\left(0.0448684
\cos\left(\frac{13 n \pi }{24}\right)+(0.272715211) \cos\left(\frac{29
n \pi }{24}\right)+(0.3408091711)
\sin\left(\frac{13 n \pi }{24}\right)-0.20926174123 \sin\left(\frac{29
n \pi }{24}\right)\right).
\end{eqnarray}
The eigenvalues of $\rho_1(n)$ are given as follows:
\begin{eqnarray}\nonumber
\lambda &=&0.5 \pm
\left[0.041748046875+ 0.040916791185\cos\left(\frac{n\pi}{12}\right)+0.035634186961\left(\cos\frac{n\pi}{4}\right)+0.06243896484375\cos\left(\frac{n\pi}{3}\right)\right.\\ \nonumber &&\left.+0.026199786034\cos\left(\frac{5n\pi}{12}\right)+0.01542619052857\cos\left(\frac{7n\pi}{12}\right)+0.02081298828125\cos\left(\frac{2n\pi}{3}\right)\right.\\ \nonumber &&\left.
+0.0061138599142\cos\left(\frac{3n\pi}{4}\right)+0.00070918537814\cos\left(\frac{11n\pi}{12}\right)-0.0053867977527\left(\sin\left(\frac{n\pi}{12}\right)+\sin\left(\frac{11n\pi}{12}\right)\right)\right.\\ \nonumber &&\left.-0.0147601635233\left(\sin\left(\frac{n\pi}{4}\right)+\sin\left(\frac{3n\pi}{4}\right)\right)-0.036049153161\left(\sin\left(\frac{n\pi}{3}\right)+\sin\left(\frac{2n\pi}{3}\right)\right)-0.020103802903\right.\\ \nonumber &&\left.\left(\sin\left(\frac{5n\pi}{12}\right)+\sin\left(\frac{7n\pi}{12}\right)\right)\right]^\frac{1}{2}.
\end{eqnarray}
Thus, the expression for linear entropy of $\rho_1(n)$ is given by,
\begin{eqnarray}\label{Eq:ap10Q1} \nonumber
S_{(0,0)}^{(12)}(n)&=& \frac{1}{16} \left\lbrace 8-0.1953125 \cos (n \pi) \left(1 +\cos (n \pi) \right)^2 \left[\cos\left(\frac{3 n \pi }{8}\right)+1.1\cos\left(\frac{25
n \pi }{24}\right)+1.1 \cos\left(\frac{41 n \pi }{24}\right)\right]^2 \right.\\ \nonumber &&\left.-0.1667096466
\left[\cos\left(\frac{n \pi }{8}\right)-\cos\left(\frac{7 n \pi }{8}\right)-0.4142135623731 \left(\sin\left(\frac{n \pi }{8}\right)+\sin\left(\frac{7 n \pi }{8}\right)\right)+ (1-\cos(n\pi))\right.\right.\\ \nonumber  &&
\left.\left.\left( -0.944591414368  \cos\left(\frac{29 n \pi }{24}\right)-0.15540858564 \cos\left(\frac{13 n \pi }{24}\right)-1.180445403469\sin\left(\frac{13
n \pi }{24}\right)\right.\right.\right.\\  &&
\left.\left.\left.+0.724810484858 \sin\left(\frac{29 n \pi }{24}\right)\right)\right]^2\right\rbrace.
\end{eqnarray}
With the help of Eq. (\ref{Eq:ap10Q1}), we can evaluate the entanglement dynamics for this state, which we found periodic in nature having periodicity $24$ i.e. $S_{(0,0)}^{(12)}(n)=S_{(0,0)}^{(12)}(n+24)$. Notably, the linear entropy remains unchanged for consecutive odd and even values of $n$, which is shown in Fig. $1$ from the main text.
Using these eigenvalues in Eq.~(\ref{von1}), we can easily calculate the entanglement entropy for this initial state. Similar to linear entropy, the von Neumann entropy for this state shows periodic behavior with the same period, which is shown in Fig. $1$ from the main text.
\subsubsection{\bf{Concurrence}}
 The  $\rho_{12}(n)$ for this state is given as follows:
\begin{equation}
\rho_{12}(n)={\frac{1}{4}\left(
\begin{array}{cccc}
 w_1^{\prime} & y_1^{\prime} & y_1^{\prime} &  x_1^{\prime}{ *}\\
 y_1^{\prime}{*}& u_1^{\prime} & u_1^{\prime} & v_1^{\prime} \\
 y_1^{\prime} {*} & u_1^{\prime} &  u_1^{\prime}& v_1^{\prime} \\
 x_1^{\prime} & v_1^{\prime}{ *}& v_1^{\prime}{*} & s_1^{\prime}\\
\end{array}
\right)},
\end{equation}
where  the coefficients are:
\begin{eqnarray}\nonumber
w_1^{\prime}&=&\frac{17}{12}+\frac{5}{16} \cos\left(\frac{n \pi }{8}\right)+\frac{11}{32}\left( \cos\left(\frac{13 n \pi }{24}\right)+\cos\left(\frac{29
n \pi }{24}\right)+\cos\left(\frac{11 n \pi }{24}\right)+\cos\left(\frac{5 n \pi }{24}\right)\right)+\frac{11}{48} \cos\left(\frac{2
n \pi }{3}\right)\\ \nonumber &&+\frac{5}{16} \cos\left(\frac{7 n \pi }{8}\right)+\frac{1}{4} \cos(n \pi )+\frac{5}{48} \cos\left(\frac{4 n \pi
}{3}\right),\\ \nonumber
s_1^{\prime}&=&\frac{17}{12}-\frac{5}{16}\left( \cos\left(\frac{n \pi }{8}\right)+\cos\left(\frac{7 n \pi }{8}\right)\right)-\frac{11}{32}\left(
\cos\left(\frac{13 n \pi }{24}\right)+\cos\left(\frac{29 n \pi }{24}\right)+\cos\left(\frac{11 n \pi }{24}\right)+\cos\left(\frac{5
n \pi }{24}\right)\right)\\ \nonumber &&+\frac{11}{48} \cos\left(\frac{2 n \pi }{3}\right) +\frac{1}{4} \cos(n \pi )+\frac{5}{48} \cos\left(\frac{4
n \pi }{3}\right),\\ \nonumber
x_1^{\prime}&=&(0.25)(1-\cos(n \pi ))-(\, 0.0180421959121758~ i) \left(\cos\left(\frac{13 n \pi }{24}\right)-\cos\left(\frac{5
n \pi }{24}\right)+\cos\left(\frac{11n \pi }{24}\right)-\cos\left(\frac{29 n \pi }{24}\right)\right)\\ \nonumber &&-0.39692831006786783 \sin\left(\frac{2
n \pi }{3}\right)+
0.18042195912175818 \sin\left(\frac{4 n \pi }{3}\right),\\ \nonumber
\end{eqnarray}
\begin{eqnarray}
u_1^{\prime}&=&\frac{7}{12}-\frac{11}{48} \cos\left(\frac{2 n \pi }{3}\right)-\frac{1}{4} \cos(n \pi )-\frac{5}{48} \cos\left(\frac{4
n \pi }{3}\right),\\ \nonumber
y_1^{\prime} &=&(-0.144356177 \left(\cos\left(\frac{n \pi }{8}\right)-\cos\left(\frac{7 n \pi }{8}\right)\right)+
0.05979428631\left(\sin\left(\frac{n \pi }{8}\right)+\sin\left(\frac{7 n \pi }{8}\right)\right)+(1-\cos(n \pi))\\ \nonumber &&\left[(0.02243419-0.00117749~
i) \cos\left(\frac{13 n \pi }{24}\right)+
(0.1363576054 +0.00715692~ i) \cos\left(\frac{29 n \pi }{24}\right)\right.\\ \nonumber &&\left.+(0.170404586 -0.0089439213~
i) \sin\left(\frac{13 n \pi }{24}\right)-(0.104630871 +0.005491697~ i) \sin\left(\frac{29
n \pi }{24}\right)\right]~~ \mbox{and}\\ \nonumber
v_1^{\prime}&=&(-0.144356177)
\left[ \cos\left(\frac{n \pi }{8}\right)- \cos\left(\frac{7 n \pi }{8}\right)-(0.414213563
)\left( \sin\left(\frac{n \pi }{8}\right)+\sin\left(\frac{7 n \pi }{8}\right)\right)+(1-\cos(n \pi))\right.\\ \nonumber &&\left.\left(-(0.155408586 +0.008156836~ i) \cos\left(\frac{13
n \pi }{24}\right)
-(0.944591415 -0.049578192~ i)\cos\left(\frac{29 n \pi }{24}\right)\right.\right. \\ \nonumber &&\left.\left.-(1.18044540, +0.0619573156~ i) \sin\left(\frac{13 n \pi }{24}\right)  +
(0.724810485 -0.0380426845~i)\sin\left(\frac{29 n \pi }{24}\right)\right)\right].
\end{eqnarray}
All these coefficients are periodic in nature such that $\left(w_1^{\prime}(n), s_1^{\prime}(n), x_1^{\prime}(n),u_1^{\prime}(n), y_1^{\prime}(n), v_1^{\prime}(n)\right)$=$\left(w_1^{\prime}(n)(n+4), s_1^{\prime}(n)(n+4),\right.$\\$\left.x_1^{\prime}(n)(n+8), u_1^{\prime}(n)(n+4), y_1^{\prime}(n)(n+4), v_1^{\prime}(n)(n+4)\right)$. The concurrence and  eigenvalue $\lambda_l$ (in decreasing order) of $(\sigma_y \otimes \sigma_y)\rho_{12} (\sigma_y \otimes \sigma_y) \rho_{12}^*$  for $n=0$ to $48$  are  tabulated in Table \ref{Table:table33Supp} . Its nature can be observed in the  Fig. $2$ from the main text.
 \begin{table*}[bp]
  \centering
 \caption{The eigenvalues and concurrence ($C(n)$) for the initial state $\otimes^{12}\ket{0}$ at  evolution steps ($n$) for a system of $12$ qubits.}
 \renewcommand{\arraystretch}{2.4}
\begin{tabular}{|p{0.35cm}|p{6.3cm}|p{2.25cm}||p{0.35cm}|p{6.3cm}|p{2.25cm} | }
 \hline
$n$ & \hspace{2.2cm}Eigenvalues&\hspace{.25cm}Concurrence  &$n$ &\hspace{2.2cm} Eigenvalues&\hspace{.25cm} Concurrence \\
\hline
0 &$(0, 0, 0, 0)$& 0&24 &$(0,0,0,0)$&0
 \\
\hline
1 & $(0.250324640853, 0.0625, 0.062175359147, 0)$& $0.000974662566$& 25&$(0.250324640853, 0.0625, 0.062175359147, 0)$&$0.000974662566$
\\
\hline
2 &$(0.250324640853, 0.0625, 0.062175359147, 0)$&$0.000974662566$ &26 & $(0.250324640853, 0.0625, 0.062175359147, 0)$&$0.000974662566$
\\
\hline
3 & $(0.250973723613, 0.0625, 0.0620145576372, 0)$&$0.001945554635$&27 &$(0.250973723613, 0.0625, 0.0620145576372, 0)$&$0.001945554635$
\\
\hline
4 &$(0.250973723613, 0.0625, 0.0620145576372, 0)$&$0.001945554635$ & 28& $(0.250973723613, 0.0625, 0.0620145576372, 0)$&$0.001945554635$
\\
\hline
5 & $(0.125, 0.125, 0, 0)$&$0$ &29 & $(0.125, 0.125, 0, 0)$&$0$
\\
\hline
6 & $(0.125, 0.125, 0, 0)$&$0$ &30 &$(0.125, 0.125, 0, 0)$&$0$
\\
\hline

7 &$(0.2496744438147, 0.0625, 0.062337274936, 0)$&$0$ & 31&$(0.2496744438147, 0.0625, 0.062337274936, 0)$&$0$
\\
\hline
8 &$(0.2496744438147, 0.0625, 0.062337274936, 0)$&$0$ & 32&$(0.2496744438147, 0.0625, 0.062337274936, 0)$&$0$
\\
\hline
9 &$(0.250324640853, 0.0625, 0.062175359147, 0)$& $0.000974662566$ &33& $(0.250324640853, 0.0625, 0.062175359147, 0)$& $0.000974662566$
\\
\hline
 10& $(0.250324640853, 0.0625, 0.062175359147, 0)$& $0.000974662566$&34 & $(0.250324640853, 0.0625, 0.062175359147, 0)$& $0.000974662566$
\\
\hline
11 &$(0.25,0.25,0,0)$&$0$ & 35&$(0.25,0.25,0,0)$&$0$
\\
\hline
12 &$(0.25,0.25,0,0)$&$0$ &36 &$(0.25,0.25,0,0)$&$0$
\\
\hline
13 &$(0.250324640853, 0.0625, 0.062175359147, 0)$ & $0.000974662566$ & 37 & $(0.250324640853, 0.0625, 0.062175359147, 0)$& $0.000974662566$
\\
\hline
14 &$(0.250324640853, 0.0625, 0.062175359147, 0)$& $0.000974662566$ & 38 & $(0.250324640853, 0.0625, 0.062175359147, 0)$ & $0.000974662566$
\\
\hline
15 &$(0.2496744438147, 0.0625, 0.062337274936, 0)$&$0$ &39 & $(0.2496744438147, 0.0625, 0.062337274936, 0)$&$0$
\\
\hline
16 &$(0.2496744438147, 0.0625, 0.062337274936, 0)$ &$0$&40 & $(0.2496744438147, 0.0625, 0.062337274936, 0)$ &$0$
\\
\hline
17 & $(0.125, 0.125, 0, 0)$ &$0$&41 &$(0.125, 0.125, 0, 0)$ &$0$
\\
\hline
18 & $(0.125, 0.125, 0, 0)$&$0$& 42&$(0.125, 0.125, 0, 0)$ &$0$
\\
\hline
19 & $(0.250973723613, 0.0625, 0.0620145576372, 0)$&$0.001945554635$&43 & $(0.250973723613, 0.0625, 0.0620145576372, 0)$&$0.001945554635$
\\
\hline
20 &$(0.250973723613, 0.0625, 0.0620145576372, 0)$&$0.001945554635$ &44 &$(0.250973723613, 0.0625, 0.0620145576372, 0)$&$0.001945554635$
\\
\hline
21 &$(0.250324640853, 0.0625, 0.062175359147, 0)$ & $0.000974662566$ & 45& $(0.250324640853, 0.0625, 0.062175359147, 0)$ & $0.000974662566$
\\
\hline
22 & $(0.250324640853, 0.0625, 0.062175359147, 0)$& $0.000974662566$  & 46& $(0.250324640853, 0.0625, 0.062175359147, 0)$ & $0.000974662566$
\\
\hline
23& $(0,0,0,0)$&$0$&47 & $(0,0,0,0)$ &$0$
\\
\hline

\end{tabular}
  \label{Table:table33Supp}
\end{table*}
\subsection{Initial state $\otimes^{12}\ket{+}$= $\ket{\theta_0 =\pi/2,\phi_0 =-\pi/2}$}
This state can be written as follows:
\begin{eqnarray}
  \otimes ^{12} {\ket{+}}_y=\frac{1}{32\sqrt{2}}\left( \ket{\phi_0^+}+{i \sqrt{12}}\ket{\phi_1^+}- {\sqrt{66}} \ket{\phi_2^+}-{2i\sqrt{55}}\ket{\phi_3^+}+{3\sqrt{55}}\ket{\phi_4^+}+{i\sqrt{792}}\ket{\phi_5^+}-\sqrt{462}\ket{\phi_6^+}\right).
\end{eqnarray}
The state $\ket{\psi_n}$  then can be obtained by operating unitary operator $n$ times on this initial state. Thus,
\begin{eqnarray}
\ket{\psi_n}&=& \mathcal{U}_{+}^n\ket{++++++++++}\\  \nonumber
            &=&(q_{1n} \ket{\phi_0^+}+r_{1n} \ket{\phi_1^+}+ w_{1n} \ket{\phi_2^+}+t_{1n} \ket{\phi_3^+}+y_{1n} \ket{\phi_4^+}+o_{1n} \ket{\phi_5^+}+l_{1n} \ket{\phi_6^+}),
\end{eqnarray}
where  the coefficients are:
\begin{eqnarray} \nonumber
 q_{1n}&=&(0.12153398+0.210503025~i)~ e^{\frac{i n \pi }{12}}-0.22097087~ e^{\frac{3 i n \pi }{4}}+(0.12153398-0.210503025~i)~e^{\frac{17 i n \pi }{12}},\\  \nonumber
 r_{1n}&=&0.0765465547~i ~e^{\frac{3 i n \pi }{2}}, ~~
w_{1n}=(-0.0897588+0.05182226235~ i)~ e^{\frac{i n \pi }{12}}-(0.0897588\, +0.0518222624~ i) e^{\frac{17i n \pi }{12}},\\ \nonumber
t_{1n}&=&-(0.3277527650531724 ~i)~ e^{\frac{3 i n \pi }{2}}\\ \nonumber
y_{1n}&=&(0.0819382\, +0.14192111~ i)~ e^{\frac{i n \pi }{12}}+(0.327752765)~ e^{\frac{3 i n \pi }{4}}+(0.0819382\,
-0.14192111 ~i)~ e^{\frac{17 i n \pi }{12}}, \\ \nonumber
 o_{1n}&=&0.6218671481916376~i ~e^{\frac{3 i n \pi }{2}}~~~~~\mbox{and }\\ \nonumber
 l_{1n}&=&(-0.23747943989954154+0.13710881855300194~i)~ e^{\frac{i n \pi }{12}}-(0.23747943989954154+0.13710881855300194~i) ~e^{\frac{17
i n \pi }{12}}.
\end{eqnarray}
\subsubsection{\bf{The linear entropy}}
The $\rho_1(n)$ is given as follows:
\begin{equation}
  \rho_1(n)={\frac{1}{2}\left[
\begin{array}{cc}
 1 & V_1 \\
V_1^{{*}} & 1 \\
\end{array}
\right]},
\end{equation}
where $V_1=-i\left[11\left(\cos\left(\frac{n \pi }{12}\right)+ \cos\left(\frac{17 n \pi }{12}\right)\right)+10\cos\left(\frac{3 n \pi }{4}\right)\right]/32$. The eigenvalues of the $\rho_1(n)$ are $\left[\frac{1}{2}\pm\frac{1}{64} \left(11 \cos\left(\frac{n \pi }{12}\right)\right.\right.$\\ $\left.\left.+10 \cos\left(\frac{3 n \pi }{4}\right)+11
\cos\left(\frac{17 n \pi }{12}\right)\right)\right]$. Thus, the linear entropy is given as follows:
\begin{eqnarray}\label{eq:ap10Q5}
S_{(\pi/2,-\pi/2)}^{(12)}(n)=\frac{1}{2}-\frac{1}{2048}\left[11 \cos\left(\frac{n \pi }{12}\right)+10 \cos\left(\frac{3 n \pi }{4}\right)+11 \cos\left(\frac{17
n \pi }{12}\right)\right]^2.
\end{eqnarray}
From the Eq. (\ref{eq:ap10Q5}), we can see that the entanglement dynamics is periodic in nature having periodicity $12$. By using these eigenvalues in Eq.~(\ref{von1}), we can easily calculate the entanglement entropy for this initial state. Similar to linear entropy, the von Neumann entropy for this state shows periodic behavior of the same period, which can be shown in Fig. $3$ from the main text.
\subsubsection{\bf{Concurrence}}
 The two-qubit $\rho_{12}(n)$ is given as follows,
\begin{equation}
\rho_{12}(n)={\frac{1}{2}\left[
\begin{array}{cccc}
  g_n^{\prime}& h_n^{\prime} & h_n^{\prime} & j_n^{\prime} \\
 h_n^{{\prime}*} & l_n^{\prime} & l_n^{\prime} & k_n^{\prime} \\
h_n^{{\prime}*} & l_n^{\prime} & l_n^{\prime} & k_n^{\prime} \\
 j_n^{{\prime}*}& k_n^{{\prime}*} & k_n^{{\prime}*} & g_n^{\prime} \\
\end{array}
\right]},
\end{equation}
where all the coefficients are:
\begin{eqnarray} \nonumber
g_n^{\prime}&=&\frac{7}{12}-\frac{5}{96} \cos\left(\frac{2 n \pi }{3}\right)-\frac{1}{32} \cos\left(\frac{4 n \pi }{3}\right)-0.090210979560879\sin\left(\frac{2
n \pi }{3}\right)+0.05412658773652736 \sin\left(\frac{4 n \pi }{3}\right),\\ \nonumber
j_n^{\prime}&=&-\frac{1}{4}-\frac{5}{32} \cos\left(\frac{2 n \pi }{3}\right)-\frac{3}{32} \cos\left(\frac{4 n \pi }{3}\right)+0.090210979560879
\sin\left(\frac{2 n \pi }{3}\right)-0.05412658773652738 \sin\left(\frac{4 n \pi }{3}\right),\\ \nonumber
l_n^{\prime}&=&\frac{5}{12}+\frac{5}{96} \cos\left(\frac{2 n \pi }{3}\right)+\frac{1}{32} \cos\left(\frac{4 n \pi }{3}\right)+0.090210979560879
\sin\left(\frac{2 n \pi }{3}\right)-0.05412658773652738 \sin\left(\frac{4 n \pi }{3}\right),\\ \nonumber
h_n^{\prime}&=&(-0.0090210979561-0.171875~ i) \cos\left(\frac{n \pi }{12}\right)-\frac{5~i}{32}  \cos\left(\frac{3 n \pi }{4}\right)+(0.0090210979561
-0.171875~ i) \cos\left(\frac{17 n \pi }{12}\right) ~~\mbox{and} \\ \nonumber
k_n^{\prime}&=&(0.0090210979561\, -0.171875~ i) \cos\left(\frac{n \pi }{12}\right)-\frac{5~i}{32}  \cos\left(\frac{3 n \pi }{4}\right)-(0.0090210979561+0.171875~i)
\cos\left(\frac{17 n \pi }{12}\right).
\end{eqnarray}
 \begin{table*}[t!]
  \centering
 \caption{The eigenvalues and concurrence ($C(n)$) for the initial state $\otimes^{12}\ket{+}$ at  evolution steps ($n$) for the system of $12$ qubits.}
 \renewcommand{\arraystretch}{2.4}
\begin{tabular}{|p{0.35cm}|p{6.3cm}|p{2.25cm}||p{0.35cm}|p{6.3cm}|p{2.25cm} | }
 \hline
$n$ &\hspace{2.2cm} Eigenvalues&\hspace{.25cm} Concurrence  &$n$ &\hspace{2.2cm} Eigenvalues& \hspace{.25cm} Concurrence \\
\hline
0 &$(0, 0, 0, 0)$& 0&12&$(0,0,0,0)$&0
 \\
\hline
1 & $(0.250324640853, 0.0625, 0.062175359147, 0)$& $0.000974662566$& 13&$(0.250324640853, 0.0625, 0.062175359147, 0)$&$0.000974662566$
\\
\hline
2 & $(0.250973723613, 0.0625, 0.0620145576372, 0)$&$0.001945554635$&14 &$(0.250973723613, 0.0625, 0.0620145576372, 0)$&$0.001945554635$
\\
\hline
3 & $(0.125, 0.125, 0, 0)$&$0$ &15 & $(0.125, 0.125, 0, 0)$&$0$
\\
\hline
4 &$(0.2496744438147, 0.0625, 0.062337274936, 0)$&$0$ & 16&$(0.2496744438147, 0.0625, 0.062337274936, 0)$&$0$
\\
\hline
5&$(0.250324640853, 0.0625, 0.062175359147, 0)$& $0.000974662566$ &17& $(0.250324640853, 0.0625, 0.062175359147, 0)$& $0.000974662566$
\\
\hline
6 &$(0.25,0.25,0,0)$&$0$ & 18&$(0.25,0.25,0,0)$&$0$
\\
\hline
7 &$(0.250324640853, 0.0625, 0.062175359147, 0)$ & $0.000974662566$ & 19 & $(0.250324640853, 0.0625, 0.062175359147, 0)$& $0.000974662566$
\\
\hline
8 &$(0.2496744438147, 0.0625, 0.062337274936, 0)$&$0$ &20 & $(0.2496744438147, 0.0625, 0.062337274936, 0)$&$0$
\\
\hline
9 & $(0.125, 0.125, 0, 0)$ &$0$&21&$(0.125, 0.125, 0, 0)$ &$0$
\\
\hline
10 & $(0.250973723613, 0.0625, 0.0620145576372, 0)$&$0.001945554635$&22 & $(0.250973723613, 0.0625, 0.0620145576372, 0)$&$0.001945554635$
\\
\hline
11 &$(0.250324640853, 0.0625, 0.062175359147, 0)$ & $0.000974662566$ & 23& $(0.250324640853, 0.0625, 0.062175359147, 0)$ & $0.000974662566$
\\
\hline
\end{tabular}
  \label{Table:LinearEntropy1278sup}
\end{table*}
All these coefficients have periodic nature i.e $\left({g}_{n}^{\prime}(n), {h}^{\prime}_{n}(n), {j}_{n}^{\prime}(n), {k}_{n}^{\prime}(n), {l}_{n}^{\prime}(n)\right)$=$\left({g}_{n}^{\prime}(n+12), {h}_{n}^{\prime}(n+12), {j}_{n}^{\prime}(n+12),\right.$\\$\left.{k}_{n}^{\prime}(n+12)), {l}_{n}^{\prime}(n+12)\right)$. The concurrence and  eigenvalues $\lambda_l$ (in decreasing order) of $(\sigma_y \otimes \sigma_y)\rho_{12} (\sigma_y \otimes \sigma_y) \rho_{12}^*$ for values of  $n$   ranging from $0$ to $23$ are  represented in Table \ref{Table:LinearEntropy1278sup}. The corresponding concurrence values are plotted in the Fig. $4$ from the main text.
\section{Invariance  of $\delta(n)$  under  the  global unitary transformation}\label{sec:example-section}
In this section, we study the invariance property of the quantity $\delta(n)$=$\sum_{i,j} ({A}_{i,j}{A}^*_{i,j})/2N$,
where  ${A}_{i,j}=\mathcal{U}^n_{i,j}-\mathcal{U}_{i,j}$, where  $A=\mathcal{U}^n-\mathcal{U}$ having dimension $N+1$.
Specifically, we show that  $\delta(n)$ is  invariant under  the unitary transformation $\tilde{A}=G {A} G^\dagger$ i.e.\\
\begin{equation}\label{eq:ap10Q8}
\sum_{m,k}~\tilde{A}_{m,k}\tilde{A}^*_{m,k}=\sum_{i,j}~{A}_{i,j}{A}^*_{i,j} ,
\end{equation}
where $G$ is a unitary matrix $GG^\dagger=I$. To prove  the above relation, let us  first evaluate each element of  $(G {A} G^\dagger)_{m,k}$ and  $(G {A} G^\dagger)^*_{m,k}$  as follows,
 \begin{eqnarray}
 (G {A} G^\dagger)_{m,k}&=&\sum_{l,n} G_{m,l} {A}_{l,n} G^*_{k,n}\label{Eq:Ap6Q9}  ~~~ \mbox{and} \\
(G {A} G^\dagger)^*_{m,k}&=&\sum_{i,j} G^*_{m,i} {A}^*_{i,j} G_{k,j}. \label{Eq:Ap6Q10}
\end{eqnarray}
Now, substituting the expression from the  Eq. (\ref{Eq:Ap6Q9}) and  Eq.  (\ref{Eq:Ap6Q10})  into the LHS of  Eq. (\ref{eq:ap10Q8})  and further simplifying  as follows:\\
 \begin{eqnarray}\nonumber
\sum_{i,j} (G {A} G^\dagger)_{m,k} (G \tilde{A} G^\dagger)^\dagger_{m,k}=&&\sum_{l,n} G_{m,l} {A}_{l,n} G^*_{k,n} \sum_{i,j} G^*_{m,i} {A}^*_{i,j} G_{k,j},\\ \nonumber
=&&\sum_{k,l,m,n,j,i}({A}_{l,n}{A}^*_{i,j}) (G_{m,l} G^*_{m,i})(G^*_{k,n}G_{k,j}),\\
=&&\sum_{l,n,i,j}{A}_{l,n}{A}^*_{i,j} \sum_{m} G^*_{m,l}G_{m,i}  \sum_{k} G^*_{k,n} G_{k,j}.\\ \nonumber
\end{eqnarray}
Now, using the unitary property of matrix $G$, we get $ \sum_{m} G^*_{m,l}G_{m,i} =\delta_{li}$. Thus, we get
 \begin{eqnarray}\nonumber
\sum_{i,j} (G {A} G^\dagger)_{i,j} (G {A} G^\dagger)^\dagger_{i,j}= &&\sum_{l,n,i,j}{A}_{l,n}{A}^*_{i,j}~ \delta_{li}~\delta_{nj},\\
 =&&\sum_{i,j}~{A}_{i,j}{A}^*_{i,j}.
 \end{eqnarray}
 Thus, proving the claimed invariance.
\bibliography{ref2024}